\documentclass[12pt]{article}
\usepackage{amssymb}
\usepackage{amsmath}
\usepackage{graphicx}
\usepackage{indentfirst}
\usepackage{cite}
\usepackage{appendix}

\allowdisplaybreaks

\begin{document}

\title{Production of $\psi (4040)$, $\psi (4160)$, and $\psi (4415)$ mesons in
strong interactions}
\author{Sheng-Nan Xu and Xiao-Ming Xu}
\date{}
\maketitle \vspace{-1cm}
\centerline{Department of Physics, Shanghai University, Baoshan,
Shanghai 200444, China}

\begin{abstract}
Using the inelastic scattering of charmed strange mesons by open-charm 
mesons in hadronic matter produced
in Pb-Pb collisions at the Large Hadron Collider, we study the
production of $\psi (4040)$, $\psi (4160)$, and $\psi (4415)$ mesons.
Master rate equations are established for inelastic scattering. The
scattering is caused by quark interchange in association with color 
interactions between all constituent pairs in different mesons. We consider
fifty-one reactions between charmed strange and open-charm mesons.
Unpolarized cross sections for the reactions are obtained from a 
temperature-dependent interquark potential. The temperature dependence of the
cross sections causes the contributions of the reactions to the
production of $\psi (4040)$, $\psi (4160)$, and $\psi (4415)$ to change
with decreasing temperature during the evolution of hadronic matter.
For central Pb-Pb collisions at $\sqrt{s_{NN}}=5.02$ TeV, 
the master rate equations reveal that the $\psi(4040)$ number density is 
larger than the $\psi(4160)$ number density which is larger than the
$\psi(4415)$ number density.
\end{abstract}

\noindent
Keywords: Inelastic meson-meson scattering, quark interchange, relativistic
constituent quark potential model, master rate equation.

\noindent
PACS: 13.75.Lb; 12.39.Jh; 12.39.Pn

\vspace{0.5cm}
\leftline{\bf I. INTRODUCTION}
\vspace{0.5cm}

Since the discovery of $\psi (4040)$, $\psi (4160)$, and $\psi (4415)$ mesons
produced in electron-positron collisions \cite{Augustin,Siegrist,Brandelik},
the three mesons have been of interest to hadron physicists
\cite{Zhukova,Ablikim1,Aaij,Ablikim2,Ablikim3,Ablikim4,Ablikim5}.
They are easily produced at electron-positron colliders. Via electromagnetic
interactions, the electron and the positron become a virtual photon that
splits into a charm quark and a charm
antiquark. This colorless $c\bar c$ pair of a small size evolves into a 
$c\bar c$ meson
directly if the $c\bar c$ relative momentum is small
or indirectly by radiating gluons if the relative momentum is large. 
The production of 
$\psi (4040)$, $\psi (4160)$, and $\psi (4415)$
in $e^+ e^-$ annihilation can be studied in nonrelativistic
quantum chromodynamics that includes color-singlet and color-octet 
contributions \cite{HLC} or from an electron-positron-photon vertex, a
photon propagator, and the direct connection between the photon
and the $\psi (4040)$ and $\psi (4160)$ meson fields \cite{PGK,BIO}.

Further interest in $\psi (4040)$, $\psi (4160)$, and $\psi (4415)$ mesons has
arisen in the context of ultrarelativistic heavy-ion collisions.
The history of ultrarelativistic heavy-ion collisions is divided into the
following stages: initial nucleus-nucleus collisions, thermalization of 
deconfined quark-gluon matter that has no temperatures, 
evolution of quark-gluon plasmas, 
hadronization of the quark-gluon plasma at the critical temperature 
$T_{\rm c}$, and evolution of hadronic matter until kinetic freeze-out.
Some species of hadrons produced in Pb-Pb collisions at the Large Hadron 
Collider (LHC) have been measured. We expect the production of $\psi (4040)$, 
$\psi (4160)$, and $\psi (4415)$ mesons in Pb-Pb collisions within quantum 
chromodynamics (QCD).
$\psi (4040)$, $\psi (4160)$, and $\psi (4415)$ mesons are generally 
identified with the $3 ^3S_1$, $2 ^3D_1$, and $4 ^3S_1$ states of a charm quark
and a charm antiquark, respectively \cite{GI,BGS,VFV,OSEF}. Because
$\psi (4040)$, $\psi (4160)$, and $\psi (4415)$ mesons are dissolved in 
hadronic matter when the temperature of hadronic matter 
is larger than $0.97T_{\rm c}$,
$0.95T_{\rm c}$, and $0.87T_{\rm c}$, respectively \cite{wxLXW}, they can only
be produced in hadronic matter. An open-charm meson contains only a charm quark
or a charm antiquark. Quark interchange between two open-charm
mesons in association with color interactions between two constituents may
produce charmonia. Therefore,
the production of $\psi (4040)$, $\psi (4160)$, and $\psi (4415)$ can be used
to probe hadronic matter that results from the quark-gluon plasma created 
in ultrarelativistic heavy-ion collisions \cite{wxLXW}.

The production of $\psi (4040)$, $\psi (4160)$, and $\psi (4415)$ mesons
in $e^+ e^-$ annihilation
relates to electromagnetic and strong interactions, whereas the one
in ultrarelativistic heavy-ion collisions invloves only strong interactions.
The mechanism of the latter is different from that of the former. 
Models differ according to the corresponding mechanisms.

Hadronic matter contains not only charmed mesons (for example, $D^+$, $D^-$,
$D^0$, $\bar{D}^0$, $D^{*+}$, $D^{*-}$, $D^{*0}$, and $\bar{D}^{*0}$ mesons)
but also charmed strange mesons (for example, $D_s^+$, $D_s^-$, $D_s^{*+}$,
and $D_s^{*-}$ mesons).
In this work, we study the production of $\psi (4040)$, $\psi (4160)$, 
and $\psi (4415)$ via quark interchange between a charmed meson and a charmed 
strange meson and between two charmed strange mesons in hadronic matter. This
includes establishing master rate equations with new source terms that involve
charmed strange mesons, calculating cross sections for the production of the 
charmonia in meson scattering by charmed strange mesons, and studying the
number densities of the charmonia yielded in
central Pb-Pb collisions at the center-of-mass energy per
nucleon-nucleon pair $\sqrt{s_{NN}}=5.02$ TeV at the LHC. 
We note that the charmonium production in the scattering of charmed
strange mesons by open-charm mesons has not been studied.

In a vacuum, $\psi (4040)$, $\psi (4160)$, and $\psi (4415)$ 
mesons decay to two open-charm mesons, where the masses of the three charmonia
are larger than the sum of the masses of the two open-charm mesons. 
The Schr\"odinger equation with the potential that will be presented in Sect.
III gives energy eigenvalues and quark-antiquark relative-motion wave 
functions. The sum of an eigenvalue, the quark mass, and the antiquark mass
gives a meson mass. The potential originates from perturbative QCD at short
distances and lattice QCD at intermediate and large distances. The confining
potential that corresponds to the lattice results depends on temperature, and
its value becomes smaller and smaller with increasing temperature.
It contributes to the eigenvalues and the meson masses. Corresponding to
meson quantum numbers, the Schr\"odinger equation indicates that 
quark-antiquark relative-motion wave functions of $\psi (4040)$, $\psi (4160)$,
and $\psi (4415)$ occupy large space and the ones of $D$, $D^*$, $D_s$, and 
$D_s^*$ occupy small space at zero
temperature. Hence, $\psi (4040)$, $\psi (4160)$, and $\psi (4415)$ mesons
are more sensitive to confinement than $D$, $D^*$, $D_s$, and $D_s^*$ mesons.
When the temperature increases from zero, confinement becomes weaker and 
weaker, and confinement contributions to $\psi (4040)$, $\psi (4160)$, and 
$\psi (4415)$ masses decrease faster than contributions to $D$, $D^*$, $D_s$,
and $D_s^*$ masses.
In hadronic matter, the masses of the three charmonia become smaller than the 
sum of the masses of the two open-charm mesons \cite{wxLXW}. 
Thus, the charmonium decays to the two open-charm 
mesons are forbidden by energy conservation. Hence, $\psi (4040)$, 
$\psi (4160)$, and $\psi (4415)$ mesons are stable in hadronic matter.

The remainder of this paper is organized as follows. In Sect.~II, we establish
master rate 
equations for $\psi (4040)$, $\psi (4160)$, and $\psi (4415)$ mesons. 
In Sect.~III, we provide cross-section formulas for inelastic meson-meson
scattering governed by quark interchange and introduce the 
temperature-dependent interquark potential.
In Sect.~IV, we present numerical cross sections for twenty-seven reactions, 
show number densities of the three mesons produced in central Pb-Pb collisions,
and provide relevant discussions. In Sect.~V, we summarize this paper.

\vspace{0.5cm}
\leftline{\bf II. MASTER RATE EQUATIONS}
\vspace{0.5cm}

We use the notation
$D=\left( \begin{array}{c}D^+\\ D^0 \end{array} \right)$,
$\bar{D}=\left( \begin{array}{c}\bar{D}^0\\ D^- \end{array} \right)$,
$K=\left( \begin{array}{c}K^+\\ K^0 \end{array} \right)$,
$\bar{K}=\left( \begin{array}{c}\bar{K}^0\\ K^- \end{array} \right)$,
$D^{\ast}=\left( \begin{array}{c}D^{\ast +}\\ D^{\ast 0} \end{array} \right)$,
$\bar{D}^{\ast}=\left( \begin{array}{c}\bar{D}^{\ast 0}\\ D^{\ast -}
\end{array} \right)$,
$K^{\ast}=\left( \begin{array}{c}K^{\ast +}\\ K^{\ast 0} \end{array} \right)$,
and $\bar{K}^{\ast}=\left( \begin{array}{c}\bar{K}^{\ast 0}\\ K^{\ast -}
\end{array} \right)$ 
for the isospin doublets. 
Hadronic matter produced in Pb-Pb collisions at LHC energies contains many
charmed mesons and charmed strange mesons. The production of $\psi(4040)$, 
$\psi(4160)$, and $\psi(4415)$ mesons from two charmed mesons
has been studied in Ref. \cite{lyLXW}.
Now we consider the production of $\psi(4040)$, 
$\psi(4160)$, and $\psi(4415)$ from a charmed strange meson and a 
charmed meson and from two charmed strange mesons as follows:
\begin{displaymath}
D_s^+ \bar{D} \to K^\ast \psi(4040),~D_s^+ \bar{D} \to K^\ast \psi(4160),
~D_s^+ \bar{D} \to K^\ast \psi(4415),
\end{displaymath}
\begin{displaymath}
D_s^+ \bar{D}^\ast \to K \psi(4040),~D_s^+ \bar{D}^\ast \to K \psi(4160),
~D_s^+ \bar{D}^\ast \to K \psi(4415),
\end{displaymath}
\begin{displaymath}
D_s^+ \bar{D}^\ast \to K^\ast \psi(4040),
~D_s^+ \bar{D}^\ast \to K^\ast \psi(4160),
~D_s^+ \bar{D}^\ast \to K^\ast \psi(4415),
\end{displaymath}
\begin{displaymath}
D_s^{\ast +} \bar{D} \to K \psi(4040),~D_s^{\ast +} \bar{D} \to K \psi(4160),
~D_s^{\ast +} \bar{D} \to K \psi(4415),
\end{displaymath}
\begin{displaymath}
D_s^{\ast +} \bar{D} \to K^\ast \psi(4040),
~D_s^{\ast +} \bar{D} \to K^\ast \psi(4160),
~D_s^{\ast +} \bar{D} \to K^\ast \psi(4415),
\end{displaymath}
\begin{displaymath}
D_s^{\ast +} \bar{D}^\ast \to K \psi(4040),
~D_s^{\ast +} \bar{D}^\ast \to K \psi(4160),
~D_s^{\ast +} \bar{D}^\ast \to K \psi(4415),
\end{displaymath}
\begin{displaymath}
D_s^{\ast +} \bar{D}^\ast \to K^\ast \psi(4040),
~D_s^{\ast +} \bar{D}^\ast \to K^\ast \psi(4160),
~D_s^{\ast +} \bar{D}^\ast \to K^\ast \psi(4415),
\end{displaymath}
\begin{displaymath}
D_s^+ D_s^{\ast -} \to \eta \psi(4040),~D_s^+ D_s^{\ast -} \to \eta \psi(4160),
~D_s^+ D_s^{\ast -} \to \eta \psi(4415),
\end{displaymath}
\begin{displaymath}
D_s^{\ast +} D_s^{\ast -} \to \eta \psi(4040),
~D_s^{\ast +} D_s^{\ast -} \to \eta \psi(4160),
~D_s^{\ast +} D_s^{\ast -} \to \eta \psi(4415).
\end{displaymath}
Applying charge conjugation to the above reactions, we obtain $D_s^- D$,
$D_s^- D^\ast$, $D_s^{\ast -} D$, $D_s^{\ast -} D^\ast$, and 
$D_s^- D_s^{\ast +}$ reactions. In total, in this study, we consider fifty-one
new reactions
to produce $\psi (4040)$, $\psi (4160)$, and $\psi (4415)$ mesons.

Denote the number densities of $\psi (4040)$, $\psi (4160)$, and $\psi (4415)$
mesons by $n_{\psi (4040)}$, $n_{\psi (4160)}$, and $n_{\psi (4415)}$, 
respectively. These number densities change according to the following rate 
equations,
\begin{equation} \label{rate}
\partial_{\mu}(n_{R}u^{\mu})=\varTheta_{R},
\end{equation}
where $\mu$ is the space-time index, $R$ represents $\psi (4040)$, 
$\psi (4160)$, or $\psi (4415)$, and $u^{\mu}$ is the 
four-velocity of a fluid element in hadronic matter. We use $v_{ij}$ for the
relative velocity of mesons $i$ and $j$ and $\sigma_{ij \to 
i^\prime j^\prime}$ for the isospin-averaged unpolarized cross section for
$ij \to i^\prime j^\prime$.
The source terms are given by
\begin{align}
\varTheta_R=
&\langle\sigma_{D\bar{D} \rightarrow \rho R}v_{D\bar D}\rangle n_Dn_{\bar D}
+\langle\sigma_{D\bar{D}^* \rightarrow \pi R}v_{D\bar{D}^*}\rangle n_D
n_{\bar{D}^*}
\notag \\
&+\langle\sigma_{D^*\bar{D} \rightarrow \pi R}v_{D^*\bar D}\rangle n_{D^*}
n_{\bar D}
+\langle\sigma_{D\bar{D}^* \rightarrow \rho R}v_{D\bar{D}^*}\rangle n_D
n_{\bar{D}^*} 
\notag \\
&+\langle\sigma_{D^*\bar{D} \rightarrow \rho R}v_{D^*\bar D}\rangle n_{D^*}
n_{\bar D}
+\langle\sigma_{D^*\bar{D}^* \rightarrow \pi R}v_{D^*\bar{D}^*}\rangle n_{D^*}
n_{\bar{D}^*}
\notag \\
&+\langle\sigma_{D^*\bar{D}^* \rightarrow \rho R}v_{D^*\bar{D}^*}\rangle
n_{D^*}n_{\bar{D}^*}
+\langle\sigma_{D_s^+\bar{D} \rightarrow K^* R}v_{D_s^+\bar D}\rangle 
n_{D_s^+}n_{\bar D}
\notag \\
&+\langle\sigma_{D_s^-D \rightarrow \bar{K}^* R}v_{D_s^-D}\rangle
n_{D_s^-}n_D
+\langle\sigma_{D_s^+\bar{D}^* \rightarrow K R}v_{D_s^+\bar{D}^*}\rangle 
n_{D_s^+}n_{\bar{D}^*}
\notag \\
&+\langle\sigma_{D_s^-D^* \rightarrow \bar{K} R}v_{D_s^-D^*}\rangle 
n_{D_s^-}n_{D^*}
+\langle\sigma_{D_s^+\bar{D}^* \rightarrow K^* R}v_{D_s^+\bar{D}^*}\rangle 
n_{D_s^+}n_{\bar{D}^*}
\notag \\
&+\langle\sigma_{D_s^-D^* \rightarrow \bar{K}^* R}v_{D_s^-D^*}\rangle 
n_{D_s^-}n_{D^*}
+\langle\sigma_{D_s^{*+}\bar{D} \rightarrow K R}v_{D_s^{*+}\bar D}\rangle 
n_{D_s^{*+}}n_{\bar D}
\notag \\
&+\langle\sigma_{D_s^{*-}D \rightarrow \bar{K} R}v_{D_s^{*-}D}\rangle 
n_{D_s^{*-}}n_D
+\langle\sigma_{D_s^{*+}\bar{D} \rightarrow K^* R}v_{D_s^{*+}\bar D}\rangle 
n_{D_s^{*+}}n_{\bar D}
\notag \\
&+\langle\sigma_{D_s^{*-}D \rightarrow \bar{K}^* R}v_{D_s^{*-}D}\rangle 
n_{D_s^{*-}}n_D
+\langle\sigma_{D_s^{*+}\bar{D}^* \rightarrow K R}v_{D_s^{*+}\bar{D}^*}\rangle 
n_{D_s^{*+}}n_{\bar{D}^*}
\notag \\
&+\langle\sigma_{D_s^{*-}D^* \rightarrow \bar{K} R}v_{D_s^{*-}D^*}\rangle 
n_{D_s^{*-}}n_{D^*}
+\langle\sigma_{D_s^{*+}\bar{D}^* \rightarrow K^* R}v_{D_s^{*+}\bar{D}^*}
\rangle n_{D_s^{*+}}n_{\bar{D}^*}
\notag \\
&+\langle\sigma_{D_s^{*-}D^* \rightarrow \bar{K}^* R}v_{D_s^{*-}D^*}\rangle 
n_{D_s^{*-}}n_{D^*}
+\langle\sigma_{D_s^+D_s^{*-} \rightarrow \eta R}v_{D_s^+D_s^{*-}}\rangle 
n_{D_s^+}n_{D_s^{*-}}
\notag \\
&+\langle\sigma_{D_s^{*+}D_s^- \rightarrow \eta R}v_{D_s^{*+}D_s^-}\rangle 
n_{D_s^{*+}}n_{D_s^-}
+\langle\sigma_{D_s^{*+}D_s^{*-} \rightarrow \eta R}v_{D_s^{*+}D_s^{*-}}\rangle
n_{D_s^{*+}}n_{D_s^{*-}}
\notag \\
&-\langle\sigma_{\rho R \rightarrow D\bar{D}}v_{\rho R}\rangle n_\rho n_R
-\langle\sigma_{\pi R \rightarrow D\bar{D}^*}v_{\pi R}\rangle n_\pi n_R
\notag \\
&-\langle\sigma_{\pi R \rightarrow D^*\bar{D}}v_{\pi R}\rangle n_\pi n_R
-\langle\sigma_{\rho R \rightarrow D\bar{D}^*}v_{\rho R}\rangle n_\rho n_R
\notag \\
&-\langle\sigma_{\rho R \rightarrow D^*\bar{D}}v_{\rho R}\rangle n_\rho n_R
-\langle\sigma_{\pi R \rightarrow D^*\bar{D}^*}v_{\pi R}\rangle n_\pi n_R
\notag \\
&-\langle\sigma_{\rho R \rightarrow D^*\bar{D}^*}v_{\rho R}\rangle n_\rho n_R
-\langle\sigma_{K^* R \rightarrow D_s^+\bar{D}}v_{K^* R}\rangle 
n_{K^*} n_R
\notag \\
&-\langle\sigma_{\bar{K}^* R \rightarrow D_s^-D}v_{\bar{K}^* R}\rangle
n_{\bar{K}^*} n_R
-\langle\sigma_{K R \rightarrow D_s^+\bar{D}^*}v_{K R}\rangle n_K n_R
\notag \\
&-\langle\sigma_{\bar{K} R \rightarrow D_s^-D^*}v_{\bar{K} R}\rangle 
n_{\bar{K}} n_R
-\langle\sigma_{K^* R \rightarrow D_s^+\bar{D}^*}v_{K^* R}\rangle n_{K^*}n_R
\notag \\
&-\langle\sigma_{\bar{K}^* R \rightarrow D_s^-D^*}v_{\bar{K}^* R}\rangle 
n_{\bar{K}^*} n_R
-\langle\sigma_{K R \rightarrow D_s^{*+}\bar{D}}v_{K R}\rangle n_Kn_R
\notag \\
&-\langle\sigma_{\bar{K} R \rightarrow D_s^{*-}D}v_{\bar{K} R}\rangle 
n_{\bar{K}} n_R
-\langle\sigma_{K^* R \rightarrow D_s^{*+}\bar{D}}v_{K^* R}\rangle n_{K^*}n_R
\notag \\
&-\langle\sigma_{\bar{K}^* R \rightarrow D_s^{*-}D}v_{\bar{K}^* R}\rangle 
n_{\bar{K}^*} n_R
-\langle\sigma_{K R \rightarrow D_s^{*+}\bar{D}^*}v_{K R}\rangle n_K n_R
\notag \\
&-\langle\sigma_{\bar{K} R \rightarrow D_s^{*-}D^*}v_{\bar{K} R}\rangle 
n_{\bar{K}} n_R
-\langle\sigma_{K^* R \rightarrow D_s^{*+}\bar{D}^*}v_{K^* R} 
\rangle n_{K^*} n_R
\notag \\
&-\langle\sigma_{\bar{K}^* R \rightarrow D_s^{*-}D^*}v_{\bar{K}^* R}\rangle 
n_{\bar{K}^*} n_R
-\langle\sigma_{\eta R \rightarrow D_s^+D_s^{*-}}v_{\eta R}\rangle n_\eta n_R
\notag \\
&-\langle\sigma_{\eta R \rightarrow D_s^{*+}D_s^-}v_{\eta R}\rangle n_\eta n_R
-\langle\sigma_{\eta R \rightarrow D_s^{*+}D_s^{*-}}v_{\eta R}\rangle
n_\eta n_R ,
\notag \\
\end{align}
where $n_D$, $n_{\bar D}$, $n_{D^*}$, $n_{\bar{D}^*}$, $n_{D_s^+}$, 
$n_{D_s^-}$, $n_{D_s^{*+}}$, $n_{D_s^{*-}}$, $n_\pi$, $n_\rho$, $n_K$, 
$n_{\bar K}$, $n_{K^\ast}$, $n_{\bar{K}^\ast}$, and $n_\eta$
are the number densities of
$D$, $\bar D$, $D^*$, $\bar{D}^*$, $D_s^+$, $D_s^-$, $D_s^{*+}$, $D_s^{*-}$,
$\pi$, $\rho$, $K$, $\bar K$, $K^\ast$, $\bar{K}^\ast$, and $\eta$ mesons, 
respectively;
$\langle \sigma_{ij\to i^\prime j^\prime} v_{ij}\rangle$ indicates
the average cross section weighted by the relative velocity,
\begin{equation}
\langle \sigma_{ij\to i^\prime j^\prime} v_{ij}\rangle
=\frac{\int\frac{d^3 k_i}{(2\pi)^3}f_i(k_i)
\frac{d^3 k_j}{(2\pi)^3}f_j(k_j)\sigma_{ij \to i^\prime j^\prime}
(\sqrt{s}, T)v_{ij}}
{\int\frac{d^3 k_i}{(2\pi)^3}f_i(k_i)\int\frac{d^3 k_j}{(2\pi)^3}
f_j(k_j)},
\end{equation}
where $\sqrt s$ is the center-of-mass energy of mesons $i$ and $j$, $T$
the temperature, and
$f_i(k_i)$ the momentum distribution function of meson $i$ with the 
four-momentum $k_i$ in the rest frame of hadronic matter.
The first seven terms on the right-hand side of Eq. (2) have been taken into 
account in Ref. \cite{lyLXW}. While the first twenty-four terms produce
$\psi (4040)$, $\psi (4160)$, and $\psi (4415)$ mesons, the last twenty-four
terms break them.

The master rate equations involve the temperature and the transverse velocity
of hadronic matter, which are given by the relativistic hydrodynamic equation,
\begin{equation}
\partial_{\mu}T^{\mu\nu}=0,
\end{equation}
where $T^{\mu\nu}$ is the energy-momentum tensor \cite{GMKR},
\begin{equation}
T^{\mu\nu}=(\epsilon+P)u^{\mu}u^{\nu}-Pg^{\mu\nu}
+\eta [\nabla^\mu u^\nu +\nabla^\nu u^\mu -\frac{2}{3}
(g^{\mu \nu}-u^\mu u^\nu)\nabla \cdot u],
\end{equation}
where $\nabla^\mu =\partial^\mu -u^\mu u \cdot \partial$; $\epsilon$, $P$,
$g^{\mu \nu}$, and $\eta$ are the energy density, the pressure, the metric, 
and the shear viscosity, respectively.

For a large volume of particles, if cross sections for particle-particle
scattering are very large, the mean free path of particles is very short, and
the particles form a perfect fluid \cite{GMKR,KH}. Thus, the 
first
two terms on the right-hand side of Eq. (5) produce the hydrodynamic equation,
\begin{equation}
\partial_{\mu}[(\epsilon+P)u^{\mu}u^{\nu}-Pg^{\mu\nu}]=0.
\end{equation}
If the cross sections are not very large, this matter, which the particles 
form, is not a perfect fluid.
Then, viscosities such as the shear viscosity, which is proportional to the
inverse of the cross sections, need to be taken into account in studying matter
evolution. The influence of the bulk viscosity may be neglected \cite{NEP}.
By including the shear viscosity in the energy-momentum tensor in Eq. (5),
we establish Eq. (4) \cite{GMKR}.

If matter is in thermal equilibrium, hydrodynamics can be applied to
this matter. To establish thermal equilibrium, particles need to frequently 
collide
with each other. Since pions are the dominant hadron species in hadronic
matter, we study collision cases of pions. The average transverse momenta
of charged pions and charged kaons produced in central Pb-Pb collisions at 
$\sqrt{s_{NN}}=5.02$ TeV are 0.5682 GeV/$c$ and 0.9177 GeV/$c$, respectively
\cite{ALICE5020CH}. We thus consider pions with the momentum
0.5682 GeV/c and kaons with the momentum 0.9177 GeV/$c$.

Denote by $\sqrt{s_{\pi \pi}}$ the Mandelstam variable which equals the square
of the sum of two pion four-momenta.
If two pions move in opposite directions, they give the maximum of 
$\sqrt{s_{\pi \pi}}$. If the two pions move in the same direction, they give
the minimum of $\sqrt{s_{\pi \pi}}$. The average of the maximum and the 
minimum is 0.723 GeV. From the cross-section formulas presented in Ref. 
\cite{XW}, we obtain
that the cross sections for elastic $\pi \pi$ scattering for $I=2$, $I=1$, and
$I=0$ at $\sqrt{s_{\pi \pi}}=0.723$ GeV are 8.5 mb, 260.3 mb, and 112.1 mb, 
respectively. The three values agree with the measured cross sections for I=2
\cite{CMSJS,CFSW,LCFMP} and the data derived from the measured phase shifts
for $I=1$ \cite{PABFF,Hyams,EM,Srinivasan,FP,BBMPS,GKPEY} and $I=0$ 
\cite{EM,Srinivasan,Rosselet,FP,BBMPS,AKMMP,GKPEY}. Consequently, the
isospin-averaged cross section for elastic $\pi \pi$ scattering is 
$\sigma_{\pi \pi}^{\rm un}=103.9$ mb.

If a pion and a kaon move in opposite (identical) directions, they give the
maximum (minimum) of the Mandelstam variable $\sqrt{s_{\pi K}}$ which is the 
square of the sum of the pion four-momentum and the kaon four-momentum. The
average of the maximum and the minimum is 1.127 GeV. We obtain that the cross
sections for elastic $\pi K$ scattering for $I=3/2$ and $I=1/2$ at
$\sqrt{s_{\pi K}}=1.127$ GeV are 2.59 mb and
21.7 mb, respectively. The two values agree with the measured cross sections 
for
$I=3/2$ \cite{JMTVH,Linglin} and the data derived from the experimental phase
shifts for $I=1/2$ \cite{Mercer,Estabrooks,Aston,OO}. Consequently, the
isospin-averaged cross section for elastic $\pi K$ scattering is 
$\sigma_{\pi K}^{\rm un}=8.96$ mb.

Denote the number density of pions in hadronic matter by $n_\pi$. The mean
free path of pions
due to elastic $\pi \pi$ scattering is $1/(n_\pi \sigma_{\pi \pi})$.
However, pions may also collide with kaons. To include $\pi K$ collisions, the
mean free path is taken as $1/[n_\pi (\sigma_{\pi \pi}^{\rm un} 
+\frac{n_K}{n_\pi}\sigma_{\pi K}^{\rm un})]$ where $n_K$ is the number density 
of kaons 
in hadronic matter. The contribution of kaons to the mean free path of pions
is weighted by $n_K/n_\pi$. When hadronic matter is produced at the
critical temperature, the $\pi$ and $K$ number densities are 0.243 
${\rm fm}^{-3}$ and 0.0915 ${\rm fm}^{-3}$, respectively, and hadronic matter
has a size of 20.1 fm along the Pb beam and a size
of 34.3 fm in the direction perpendicular to the beam. The mean free path 
of pions is 0.384 fm. When a pion moves from the center of hadronic matter
to matter surface along or perpendicular to the beam,
the collision number is 26.2 or 44.7. When hadronic matter expands, the
collision number decreases. At kinetic freeze-out the
$\pi$ and $K$ number densities are 0.0539 ${\rm fm}^{-3}$ and 0.0203 
${\rm fm}^{-3}$, respectively, and hadronic matter has a size of 38.45 fm along
the Pb beam and a size of 52.65 fm in the direction perpendicular to the beam.
The mean free path of pions is 1.73 fm. When a pion moves from the center of 
hadronic matter to matter surface along or perpendicular to the beam,
the collision number is 11.1 or 15.2. With the four collision numbers thermal
equilibrium can be established, and we suggest using hydrodynamics in hadronic
matter that is created in central Pb-Pb collisions at $\sqrt{s_{NN}}=5.02$ 
TeV.

The foundation of hydrodynamics has been related to the Klein-Gordon equation
and the Schr\"odinger equation in Refs. \cite{Wong1,Wong2}.
First- and second-order conformal viscous hydrodynamics was
derived from the exact solution of the Boltzmann equation in the
relaxation time approximation with Gubser symmetry in Ref. \cite{DHMNS}.
These works aid us in understanding the application of hydrodynamics to 
hadronic matter.

\vspace{0.5cm}
\leftline{\bf III. CROSS-SECTION FORMULAS}
\vspace{0.5cm}

For the twenty-seven reactions listed in Sect. II, we consider the meson-meson
scattering in which a meson  consists of the charm quark $c$ and the
light antiquark $\bar{q}_{2}$ and another meson consists of the light 
quark $q_1$ and the charm antiquark $\bar c$. When the two mesons collide,
interchange of
the $c$ quark and the $q_1$ quark leads to the reaction 
$c\bar{q}_2 + q_1\bar{c} \rightarrow q_1\bar{q}_2 + c\bar{c}$.
We denote the mass and the four-momentum of meson $i$ 
($i$ = $c\bar{q}_2$, $q_1\bar{c}$, $q_1\bar{q}_2$, $c\bar{c}$) by $m_i$ and 
$P_i=(E_i,\vec{P}_i)$, respectively. The Mandelstam variable is 
$s=(P_{c\bar{q}_2}+P_{q_1\bar{c}})^2$. 
The unpolarized cross section for 
$c\bar{q}_2 + q_1\bar{c} \rightarrow q_1\bar{q}_2 + c\bar{c}$ is 
\begin{eqnarray}
\sigma^{\rm unpol}(\sqrt {s},T) 
& = & \frac {1}{(2J_{c\bar{q}_2}+1)(2J_{q_1\bar{c}}+1)}
\frac{1}{64\pi s}\frac{|\vec{P}^{\prime }(\sqrt{s})|}{|\vec{P}(\sqrt{s})|}
              \nonumber    \\
& & \int_{0}^{\pi }d\theta 
\sum\limits_{J_{c\bar{q}_2z}J_{q_1\bar{c}z}J_{q_1\bar{q}_2z}J_{c\bar{c}z}}
(\mid {\cal M}_{\rm fi}^{\rm prior} \mid^2 
+ \mid {\cal M}_{\rm fi}^{\rm post} \mid^2) \sin \theta ,
\end{eqnarray}
where $J_{c\bar{q}_2z}$ ($J_{q_1\bar{c}z}$, $J_{q_1\bar{q}_2z}$, 
$J_{c\bar{c}z}$) denotes the magnetic projection
quantum number of the total angular momentum $J_{c\bar{q}_2}$ 
($J_{q_1\bar{c}}$, $J_{q_1\bar{q}_2}$, $J_{c\bar{c}}$) of 
meson $c\bar{q}_2$ ($q_1\bar{c}$, $q_1\bar{q}_2$, $c\bar{c}$);
$\vec{P}$ equals $\vec{P}_{c\bar{q}_2}$, and $\vec{P}^{'}$ equals
$\vec{P}_{q_1\bar{q}_2}$;
$\theta$ is the angle between $\vec{P}$ and $\vec{P}'$.

Quark interchange produces two forms in the Born-order meson-meson 
scattering, the prior form and the post form \cite{BS,Swanson}.
Scattering in the prior form means that gluon exchange
occurs prior to quark interchange. 
The transition amplitude for scattering in the prior form is
\begin{equation}
{\cal M}_{\rm{fi}}^{\rm{prior}} =4\sqrt{E_{c\bar{q}_2} E_{q_1\bar{c}} 
E_{q_1\bar{q}_2} E_{c\bar{c}}}
<\psi_{q_1\bar{q}_2} \mid <\psi_{c\bar{c}} \mid 
(V_{c\bar{c}}+V_{\bar{q}_2q_1}+V_{cq_1}+V_{\bar{q}_2\bar{c}} )
\mid \psi_{c\bar{q}_2}> \mid \psi_{q_1\bar{c}}>,
\end{equation}
where $\psi_{c\bar{q}_2}$ ($\psi_{q_1\bar{c}}$, $\psi_{q_1\bar{q}_2}$, 
$\psi_{c\bar{c}}$) is the wave function of meson $c\bar{q}_2$ ($q_1\bar{c}$, 
$q_1\bar{q}_2$, $c\bar{c}$), and is 
the product of the color wave function, the spin wave function, the flavor wave
function, and the mesonic quark-antiquark relative-motion wave function;
$V_{ab}$ is the potential between constituents $a$ and $b$.
Scattering in the post form means that quark interchange is followed by gluon 
exchange. The transition amplitude for scattering in the post form is
\begin{equation}
{\cal M}_{\rm{fi}}^{\rm{post}} =4\sqrt{E_{c\bar{q}_2} E_{q_1\bar{c}} 
E_{q_1\bar{q}_2} E_{c\bar{c}}}
<\psi_{q_1\bar{q}_2} \mid <\psi_{c\bar{c}} \mid 
(V_{q_1\bar{c}}+V_{\bar{q}_2c}+V_{cq_1}+V_{\bar{q}_2\bar{c}}) 
\mid \psi_{c\bar{q}_2}> \mid \psi_{q_1\bar{c}}>.
\end{equation}
Both ${\cal M}_{\rm{fi}}^{\rm{prior}}$ and ${\cal M}_{\rm{fi}}^{\rm{post}}$
contain $V_{cq_1}$ and $V_{\bar{q}_2\bar{c}}$. However, it is possible that
$<\psi_{q_1\bar{q}_2} \mid <\psi_{c\bar{c}} \mid 
(V_{c\bar{c}}+V_{\bar{q}_2q_1})
\mid \psi_{c\bar{q}_2}> \mid \psi_{q_1\bar{c}}>$ in 
${\cal M}_{\rm{fi}}^{\rm{prior}}$
is not the same as 
$<\psi_{q_1\bar{q}_2} \mid <\psi_{c\bar{c}} \mid 
(V_{q_1\bar{c}}+V_{\bar{q}_2c}) 
\mid \psi_{c\bar{q}_2}> \mid \psi_{q_1\bar{c}}>$ in
${\cal M}_{\rm{fi}}^{\rm{post}}$. Hence,
${\cal M}_{\rm{fi}}^{\rm{post}}$ may not equal 
${\cal M}_{\rm{fi}}^{\rm{prior}}$, which is the so-called post-prior 
discrepancy \cite{MM,BBS,WC}.

The transition amplitudes come from interactions between all constituent pairs
in different mesons. In this work, we consider a central
spin-independent potential, a spin-spin interaction, and a ternsor interaction.
Derived from perturbative QCD and lattice gauge calculations
\cite{BT,KLP,Xu2002}, the potential for $T<T_{\rm c}$ is,
\begin{eqnarray}
V_{ab}(\vec{r}_{ab}) & = &
- \frac {\vec{\lambda}_a}{2} \cdot \frac {\vec{\lambda}_b}{2}
\xi_1 \left[ 1.3- \left( \frac {T}{T_{\rm c}} \right)^4 \right] \tanh
(\xi_2 r_{ab}) + \frac {\vec{\lambda}_a}{2} \cdot \frac {\vec{\lambda}_b}{2}
\frac {6\pi}{25} \frac {v(\lambda r_{ab})}{r_{ab}} \exp (-\xi_3 r_{ab})
\nonumber  \\
& & -\frac {\vec{\lambda}_a}{2} \cdot \frac {\vec{\lambda}_b}{2}
\frac {16\pi^2}{25}\frac{d^3}{\pi^{3/2}}\exp(-d^2r^2_{ab})
\frac {\vec {s}_a \cdot \vec {s}_b} {m_am_b}
+\frac {\vec{\lambda}_a}{2} \cdot \frac {\vec{\lambda}_b}{2}\frac {4\pi}{25}
\frac {1} {r_{ab}} \frac {d^2v(\lambda r_{ab})}{dr_{ab}^2}
\frac {\vec {s}_a \cdot \vec {s}_b}{m_am_b}
\nonumber  \\
& & -\frac {\vec{\lambda}_a}{2} \cdot \frac {\vec{\lambda}_b}{2}
\frac {6\pi}{25m_am_b}\left[ v(\lambda r_{ab})
-r_{ab}\frac {dv(\lambda r_{ab})}{dr_{ab}} +\frac{r_{ab}^2}{3}
\frac {d^2v(\lambda r_{ab})}{dr_{ab}^2} \right]
\nonumber  \\
& & \left( \frac{3\vec {s}_a \cdot \vec{r}_{ab}\vec {s}_b \cdot \vec{r}_{ab}}
{r_{ab}^5} -\frac {\vec {s}_a \cdot \vec {s}_b}{r_{ab}^3} \right) ,
\end{eqnarray}
where $\vec{r}_{ab}$ is the relative coordinate of constituents $a$ and $b$;
$m_a$, $\vec{s}_a$, and $\vec{\lambda}_a$ are the mass, the spin, and the
Gell-Mann matrices for the color generators of constituent $a$,
respectively;
$\xi_1=0.525$ GeV, $\xi_2=1.5[0.75+0.25 (T/{T_{\rm c}})^{10}]^6$ GeV, 
$\xi_3=0.6$ GeV, $T_{\rm c}=0.175$ GeV, and
$\lambda=\sqrt{25/16\pi^2 \alpha'}$ with $\alpha'=1.04$ GeV$^{-2}$; the
function $v$ is given by Buchm\"uller and Tye \cite{BT}; the quantity 
$d$ is related to constituent masses \cite{lyLXW}.
The constituent masses are 0.32 GeV, 0.32 GeV, 0.5 GeV, and 1.51 GeV for
the up quark, the down quark, the strange quark, and the charm quark, 
respectively. 
The Schr\"odinger equation with $V_{ab} (\vec{r}_{ab})$
at zero temperature gives meson masses that are close to the experimental 
masses~\cite{PDG} of $\pi$, $\rho$, $K$, $K^*$, $\eta$, $J/\psi$, $\chi_{c}$, 
$\psi'$, $\psi (3770)$, $\psi (4040)$, $\psi (4160)$, $\psi (4415)$, $D$, 
$D^*$, $D_s$, and $D^*_s$ mesons. 
With the mesonic quark-antiquark relative-motion wave functions determined by
the Schr\"odinger equation, the experimental data 
\cite{CMSJS,DBGGT,LCFMP,Hoogland,PABFF,Hyams,EM,Srinivasan,Rosselet,FP,
BBMPS,AKMMP,GKPEY} of 
$S$- and $P$-wave elastic phase shifts for $\pi \pi$ scattering in
vacuum are reproduced in the Born approximation
\cite{ZX,SXW}.

By including color screening in medium, the lattice gauge calculations 
\cite{KLP}
provide a numerical spin-independent and temperature-dependent potential at
intermediate and large distances. The first and second terms on the right-hand
side of Eq. (10) fit the numerical quark potential at $T>0.55T_{\rm c}$ well
\cite{ZXG}. The expression
$\frac {\vec{\lambda}_a}{2} \cdot \frac {\vec{\lambda}_b}{2}
\frac {6\pi}{25} \frac {v(\lambda r_{ab})}{r_{ab}}$ in the second term
arises from one-gluon exchange plus perturbative one- and two-loop corrections
in a vacuum \cite{BT}, and the factor $\exp (-\xi_3 r_{ab})$ is a medium
modification factor. When the distance $r_{ab}$ increases from zero, the 
numerical 
potential increases and becomes a distance-independent value at
large distances at $T>0.55T_{\rm c}$. The value decreases with increasing
temperature, which means that confinement becomes weaker and weaker. The value
is parametrized as $- \frac {\vec{\lambda}_a}{2} \cdot 
\frac {\vec{\lambda}_b}{2}
\xi_1 \left[ 1.3- \left( \frac {T}{T_{\rm c}} \right)^4 \right]$ so that the
first term is obtained.
The function $\tanh (\xi_2 r_{ab})$ increases from 0 to 1 when $r_{ab}$
increases from 0 to $+\infty$.  $\xi_2$ increases when $T$ increases. The 
larger is $\xi_2$, the smaller is the distance at which $\tanh (\xi_2 r_{ab})$
is nearly 1, i.e., the stronger is the medium screening on the quark potential.
The first term is the 
confining potential that corresponds to the lattice results.
The third term is the smeared spin-spin interaction that
comes from one-gluon exchange between constituents $a$ and $b$. The
fourth term is the spin-spin interaction that originates from perturbative 
one- and two-loop corrections to one-gluon exchange. The fifth
term is the tensor interaction that arises from one-gluon exchange plus 
perturbative one- and two-loop corrections.

\vspace{0.5cm}
\centerline{\bf IV. NUMERICAL RESULTS AND DISCUSSIONS }
\vspace{0.5cm}

We solve the Schr\"odinger equation with the potential given in Eq. (10)
to obtain temperature-dependent meson masses
and mesonic quark-antiquark relative-motion wave functions in coordinate space.
The
transition amplitudes in the prior form and in the post form are calculated
from the Fourier transform of the potential and the wave functions.
Temperature-dependent unpolarized cross sections are obtained from Eq. (7). 
The cross sections are plotted in Figs. 1-27 for the twenty-seven 
reactions listed in Sect. II. The cross sections for 
$D_s^- D$, $D_s^- D^*$, $D_s^{*-} D$, $D_s^{*-} D^*$, and $D_s^- D_s^{*+}$
reactions equal those for $D_s^+ \bar{D}$, $D_s^+ \bar{D}^*$, 
$D_s^{*+} \bar{D}$, $D_s^{*+} \bar{D}^*$, and $D_s^+ D_s^{*-}$ reactions,
respectively. These unpolarized cross sections lead to the isospin-averaged
unpolarized cross sections in the source terms by a formula given in the
appendix of Ref. \cite{lyLXW}.

For the reaction $c\bar{q}_2 + q_1\bar{c} \to q_1\bar{q}_2 + c\bar{c}$, the 
absolute values of the three-dimensional momenta of mesons $c\bar{q}_2$ and 
$q_1\bar{q}_2$ in the center-of-mass frame are given by
\begin{displaymath}
\mid \vec{P} \mid =\frac {1}{2} 
\sqrt {\frac {(s-m_{c\bar{q}_2}^2-m_{q_1\bar c}^2)^2
-4m_{c\bar{q}_2}^2m_{q_1\bar c}^2}{s}},
\end{displaymath}
\begin{displaymath}
\mid \vec{P}^\prime \mid =\frac {1}{2} 
\sqrt {\frac {(s-m_{q_1\bar{q}_2}^2-m_{c\bar c}^2)^2
-4m_{q_1\bar{q}_2}^2m_{c\bar c}^2}{s}}.
\end{displaymath}
If the sum of the masses of the two initial mesons of a reaction is smaller 
than
that of the two final mesons, the reaction is endothermic. The threshold energy
equals the sum of the masses of the two final mesons. At the threshold, 
$\mid \vec{P} \mid \neq 0$, $\mid \vec{P}^\prime \mid = 0$, and the factor
$\mid \vec{P}^\prime \mid / \mid \vec{P} \mid$ in Eq. (7) gives
$\sigma^{\rm unpol}=0$. Given a temperature, every endothermic reaction has
at least a peak cross section. The initial mesons
need kinetic energies to satisfy energy conservation and to start the reaction,
and a certain amount of the kinetic energies are converted into the masses of 
the final mesons.
If the sum of the masses of the two initial mesons is larger than
that of the two final mesons, the reaction is exothermic. The threshold energy
equals the sum of the masses of the two initial mesons. At the threshold, 
$\mid \vec{P} \mid = 0$, $\mid \vec{P}^\prime \mid \neq 0$, and
$\mid \vec{P}^\prime \mid / \mid \vec{P} \mid$ in
Eq. (7) gives $\sigma^{\rm unpol}=+\infty$. Even
slowly-moving initial mesons may start the reaction, and a certain amount of 
the masses of the initial mesons are converted into the kinetic energies of the
final mesons. Since meson masses decrease with increasing temperature, the
sum of the masses of the two initial mesons may be smaller than
the one of the two final mesons at a temperature, but may be larger than
the one of the two final mesons at another temperature. Therefore, a reaction
may be endothermic at a temperature and exothermic at another temperature.
This phenomenon occurs to $D_s^+ \bar{D}^* \to K \psi(4040)$ in Fig. 4,
$D_s^{*+} \bar{D} \to K \psi(4040)$ in Fig. 10, 
$D_s^{*+} \bar{D} \to K \psi(4160)$ in Fig. 11, 
$D_s^{*+} \bar{D}^* \to K \psi(4040)$ in Fig. 16,
$D_s^{*+} \bar{D}^* \to K \psi(4160)$ in Fig. 17, 
$D_s^{*+} \bar{D}^* \to K \psi(4415)$ in Fig. 18,
$D_s^+ D_s^{*-} \to \eta \psi(4040)$ in Fig. 22,
$D_s^+ D_s^{*-} \to \eta \psi(4160)$ in Fig. 23,
$D_s^+ D_s^{*-} \to \eta \psi(4415)$ in Fig. 24,
$D_s^{*+} D_s^{*-} \to \eta \psi(4040)$ in Fig. 25,
$D_s^{*+} D_s^{*-} \to \eta \psi(4160)$ in Fig. 26, and
$D_s^{*+} D_s^{*-} \to \eta \psi(4415)$ in Fig. 27.

The Schr\"odinger equation with the potential given in Eq. (10) yields energy
eigenvalues and quark-antiquark relative-motion wave functions in coordinate
space. Through the Schr\"odinger equation, a meson mass is given as
a sum of the quark mass, the antiquark mass, and an eigenvalue. 
Since the potential decreases with increasing temperature, 
eigenvalues and meson masses decrease \cite{wxLXW}. Threshold energies, which
are the sum
of the masses of the two final (initial) mesons for endothermic (exothermic)
reactions, decrease with increasing temperature as seen in Figs. 1-27. The
reduced amounts of meson masses are different for different mesons. For 
example, from 
$T=0$ to $T=0.85T_{\rm c}$ the $K^\ast$ and $\psi(4040)$ masses are reduced by
0.399 GeV and 0.859 GeV, respectively, and the threshold energy of
$D_s^+ \bar{D} \to K^\ast \psi(4040)$ is reduced by 1.258 GeV.

When $\sqrt s$ increases from the threshold energy, $\mid \vec{P} \mid$ of
any endothermic reaction increases from a nonzero value, 
$\mid \vec{P}^\prime \mid$ increases from zero, and 
$\mid \vec{P}^\prime \mid / \mid \vec{P} \mid$ causes a rapid increase in the 
cross section close to the threshold energy. Since every mesonic 
quark-antiquark relative momentum is a linear combination of 
$\vec{P}$ and $\vec{P}^\prime$, its absolute value 
increases with increasing $\sqrt s$. The radial wave functions of the
quark-antiquark relative motion of $D$, $\bar D$, $D^*$, $\bar{D}^*$, $D_s^+$,
$D_s^-$, $D_s^{*+}$, $D_s^{*-}$, $K$, $\bar K$, $K^*$, $\bar{K}^*$, and $\eta$
mesons are decreasing functions of relative momenta. The radial wave functions
of $\psi(4040)$, $\psi(4160)$, and $\psi(4415)$ mesons have nodes and are
decreasing functions of large relative momenta. The transition amplitudes may 
increase and then decrease rapidly with increasing $\sqrt s$.
The rising $\mid \vec{P}^\prime \mid / \mid \vec{P} \mid$ and the falling
transition amplitudes produce a narrow peak in the cross-section curve near the
threshold energy.

Meson masses at zero temperature give that the sum of the masses of the two
initial mesons of any reaction listed in Sect. II is smaller than that of the
two final mesons, and the reaction is endothermic. Concerning three 
reactions such as $D_s^+ \bar{D} \to K^\ast \psi(4040)$, 
$D_s^+ \bar{D} \to K^\ast \psi(4160)$, and 
$D_s^+ \bar{D} \to K^\ast \psi(4415)$, which have the same initial mesons and
one identical final meson, a characteristic observed in Figs. 1-27 is that the 
peak cross
section of producing $\psi(4040)$ is larger than those of producing 
$\psi(4160)$ and $\psi(4415)$ and the peak cross section of producing 
$\psi(4160)$ is similar to the one of producing $\psi(4415)$.
The $\mid \vec{P}^\prime \mid / s\mid \vec{P} \mid$ value corresponding to
the peak cross section of producing $\psi(4040)$ is larger than that of
producing $\psi(4160)$, and the latter is larger than that of producing 
$\psi(4415)$. 
$\mid \vec{P}^\prime \mid / s\mid \vec{P} \mid$ in Eq. (7) reveals that the
peak cross section of producing $\psi(4040)$ is larger than those of
producing $\psi(4160)$ and $\psi(4415)$. In Eqs. (8) and (9), the wave function
$\psi_{c\bar c}$ contains the $c\bar c$ relative-motion wave
function which is the product of the radial wave
function of the relative motion and the spherical harmonics 
$Y_{L_{c\bar c}M_{c\bar c}}$ where $L_{c\bar c}$ is the 
orbital-angular-momentum quantum number
and $M_{c\bar c}$ is the magnetic projection quantum number.
According to the quantum numbers of $\psi(4160)$ and $\psi(4415)$ mesons, the
relative-motion wave functions of $\psi(4160)$ and $\psi(4415)$ contain
$Y_{2M_{c\bar c}}$ ($M_{c\bar c}$=-2, -1, 0, 1, 2) and $Y_{00}$, respectively.
Since the $\psi(4160)$ mass is smaller than the $\psi(4415)$ mass, it is
expected that producing $\psi(4160)$ is easier than producing $\psi(4415)$.
However,
the cross section for producing $\psi(4160)$ is more reduced by 
$Y_{2M_{c\bar c}}$ than the cross section for producing $\psi(4415)$ by 
$Y_{00}$. Consequently,
the peak cross section of producing $\psi(4160)$ is similar to
the one of producing $\psi(4415)$.

Figs. 1-21 show that, below the critical temperature, the following
reactions are endothermic:
\begin{displaymath}
D_s^+ \bar{D} \to K^\ast \psi(4040),~D_s^+ \bar{D} \to K^\ast \psi(4160),
~D_s^+ \bar{D} \to K^\ast \psi(4415),
\end{displaymath}
\begin{displaymath}
D_s^+ \bar{D}^\ast \to K \psi(4160),~D_s^+ \bar{D}^\ast \to K \psi(4415),
~D_s^+ \bar{D}^\ast \to K^\ast \psi(4040),
\end{displaymath}
\begin{displaymath}
~D_s^+ \bar{D}^\ast \to K^\ast \psi(4160),
~D_s^+ \bar{D}^\ast \to K^\ast \psi(4415),
~D_s^{\ast +} \bar{D} \to K \psi(4415),
\end{displaymath}
\begin{displaymath}
D_s^{\ast +} \bar{D} \to K^\ast \psi(4040),
~D_s^{\ast +} \bar{D} \to K^\ast \psi(4160),
~D_s^{\ast +} \bar{D} \to K^\ast \psi(4415),
\end{displaymath}
\begin{displaymath}
D_s^{\ast +} \bar{D}^\ast \to K^\ast \psi(4040),
~D_s^{\ast +} \bar{D}^\ast \to K^\ast \psi(4160),
~D_s^{\ast +} \bar{D}^\ast \to K^\ast \psi(4415).
\end{displaymath}
As the temperature increases from zero, 
confinement shown by the potential in Eq. (10) becomes weaker and weaker,
the Schr\"odinger equation gives increasing meson radii, and mesonic
quark-antiquark states become looser and looser. On one hand, the weakening
confinement with increasing temperature makes combining final 
quarks and antiquarks into final mesons more difficult, and thus reduces
cross sections; On the other hand, the increasing radii of initial
mesons cause increasing cross sections as the temperature goes up. 
The two factors determine the change in the peak cross section with respect to
the temperature, which is shown in Figs. 1-21.

The $K^\ast$ meson is a loose bound state and is more 
affected by confinement. Regarding the six reactions,
\begin{displaymath}
D_s^+ \bar{D} \to K^\ast \psi(4040),~D_s^+ \bar{D} \to K^\ast \psi(4160),
~D_s^+ \bar{D} \to K^\ast \psi(4415),
\end{displaymath}
\begin{displaymath}
D_s^+ \bar{D}^\ast \to K^\ast \psi(4040),
~D_s^+ \bar{D}^\ast \to K^\ast \psi(4160),
~D_s^+ \bar{D}^\ast \to K^\ast \psi(4415),
\end{displaymath}
at $T/T_{\rm c}=0.65$ and 0.75, the reduced amount of the cross section due to 
the weakening confinement exceeds the increased amount of the cross section due
to the increasing radii of the initial mesons, and the peak cross sections thus
go down as $T/T_{\rm c}$ increases from 0.65 to 0.75. The six reactions have 
the characteristic that the peak cross sections initially decrease and then 
generally increase as the
temperature goes up from $0.65T_{\rm c}$. Note that the six reactions have the
initial meson $D_s^+$. If we replace the $D_s^+$ meson by the $D_s^{*+}$ 
meson, which has a radius larger than the $D_s^+$ meson, the reduced amount of 
the cross section due to the weakening confinement is smaller than the 
increased amount of the cross section due to the increasing radii of the 
initial mesons. Thus,
the peak cross sections of the following reactions,
\begin{displaymath}
D_s^{\ast +} \bar{D} \to K^\ast \psi(4040),
~D_s^{\ast +} \bar{D} \to K^\ast \psi(4160),
~D_s^{\ast +} \bar{D} \to K^\ast \psi(4415),
\end{displaymath}
\begin{displaymath}
D_s^{\ast +} \bar{D}^\ast \to K^\ast \psi(4040),
~D_s^{\ast +} \bar{D}^\ast \to K^\ast \psi(4160),
~D_s^{\ast +} \bar{D}^\ast \to K^\ast \psi(4415),
\end{displaymath}
go up as $T/T_{\rm c}$ increases from 0.65 to 0.75. These reactions have 
the characteristic
that the peak cross sections initially increase and then decrease as the
temperature goes up from $0.65T_{\rm c}$.

The $K$ meson is a tight bound state and is less affected by confinement.
As to the three reactions, 
\begin{displaymath}
D_s^+ \bar{D}^\ast \to K \psi(4160),~D_s^+ \bar{D}^\ast \to K \psi(4415),
~D_s^{\ast +} \bar{D} \to K \psi(4415),
\end{displaymath}
at $T/T_{\rm c}=0.65$ and 0.75, the decrease in cross sections due to the
weakening confinement can not balance the increase in cross sections due to
the increasing radii of the initial mesons, and the peak cross sections thus 
increase as
$T/T_{\rm c}$ increases from 0.65 to 0.75. The three reactions have the 
characteristic that the peak cross sections initially go up and then go down as
the temperature increases from $0.65T_{\rm c}$.

Now, we examine Fig. 4, Fig. 10, Fig. 11, Figs. 16-18, and Figs. 22-27 in which
reactions are endothermic at some temperatures and exothermic at other 
temperatures. The final light-quark mesons of the reactions are the $K$ meson
and the $\eta$ meson. The $\eta$ meson is also a tight bound
state and is less affected by confinement. Since cross sections for 
exothermic reactions are infinite at the threshold energy, 
$m_{c\bar{q}_2}+m_{q_1\bar{c}}$, we start calculations of the cross sections
at $\sqrt{s}=m_{c\bar{q}_2}+m_{q_1\bar{c}}
+\Delta \sqrt{s}$ with $\Delta \sqrt{s} = 10^{-4}$ GeV. 
At $\sqrt{s}=m_{c\bar{q}_2}+m_{q_1\bar{c}}+\Delta \sqrt{s}$,
\begin{displaymath}
\frac{\vec{P}^{~\prime 2}}{\vec{P}^2} \approx
\frac{m_{c\bar{q}_2}+m_{q_1\bar{c}}-m_{q_1\bar{q}_2}-m_{c\bar{c}}}
{\Delta \sqrt{s}} 
\frac{m_{q_1\bar{q}_2}m_{c\bar{c}}}{m_{c\bar{q}_2}m_{q_1\bar{c}}}
\frac{m_{c\bar{q}_2}+m_{q_1\bar{c}}}{m_{q_1\bar{q}_2}+m_{c\bar{c}}} ,
\end{displaymath}
which depends on the difference between the sum of the masses of the two 
initial
mesons and the sum of the masses of the two final mesons. If the difference is
not close to zero, $\mid \vec{P}^{~\prime} \mid / \mid \vec{P} \mid$ is not
small. When $\sqrt s$ increases from the threshold energy, 
$\mid \vec{P}^{~\prime} \mid / \mid \vec{P} \mid$ decreases rapidly first and
then slowly to a minimum value, and further increases slowly. The transition 
amplitudes
may increase and then decrease with increasing $\sqrt s$. From these changes,
Eq. (7) may yield a peak in the cross-section curve of an exothermic reaction
such as
$D_s^+ \bar{D}^* \to K \psi (4040)$ in Fig. 4 in the region
$\sqrt{s} > m_{c\bar{q}_2}+m_{q_1\bar{c}} + 10^{-4}$ GeV.
Therefore, every exothermic reaction has the characteristic that the cross 
section decreases rapidly
and then may increase to form a peak when $\sqrt s$ increases from the 
threshold energy. The cross sections shown in Fig. 4, Fig. 10, Fig. 11, 
Figs. 16-18, and Figs. 22-27 have the second characteristic that the peak
cross sections initially increase and then generally decrease as the 
temperature goes up from zero.

\newpage
\begin{figure}[htbp]
  \centering
    \includegraphics[scale=0.6]{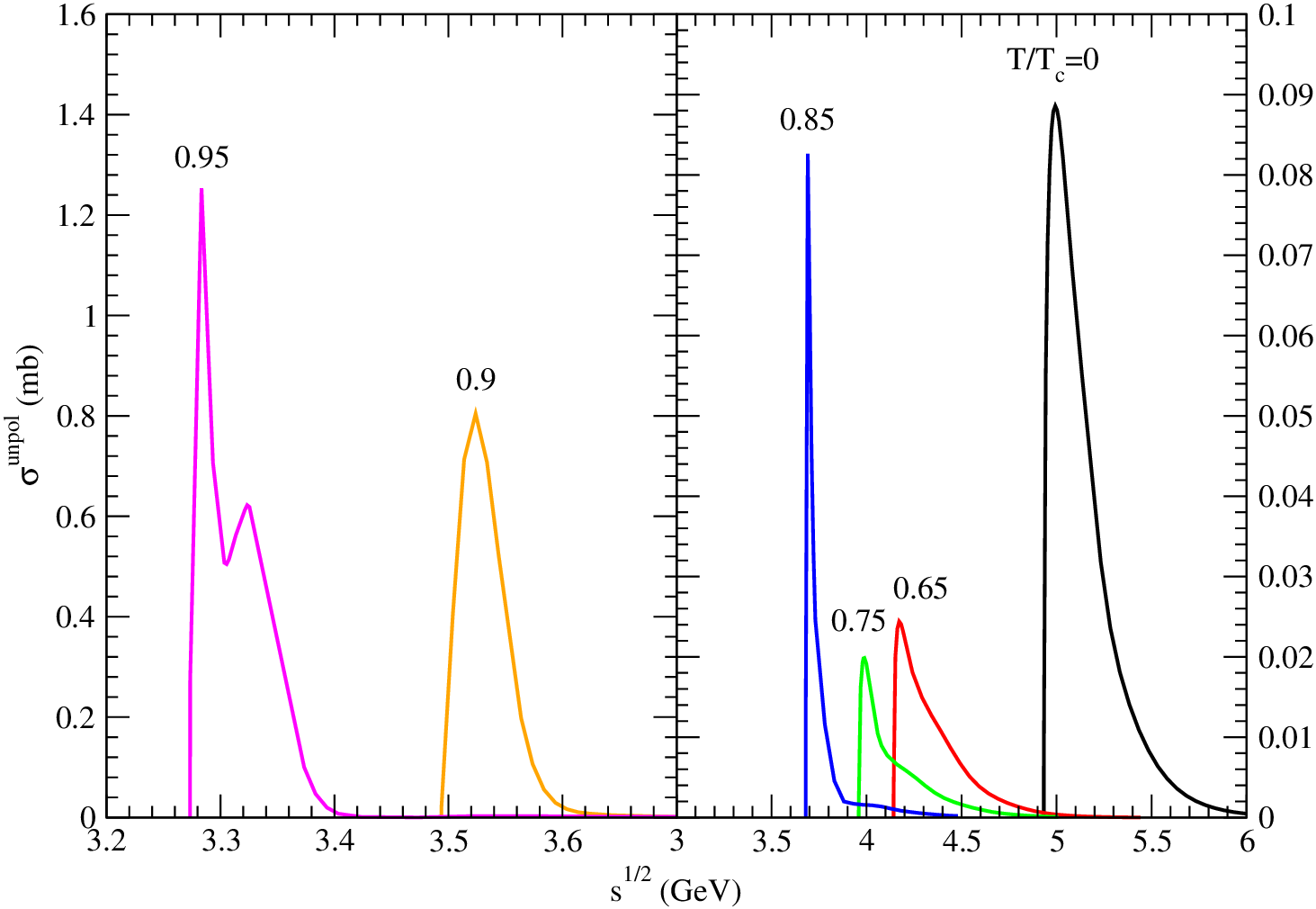}
\caption{Cross sections for $D_s^+ \bar{D} \rightarrow K^\ast \psi(4040)$
at various temperatures.}
\label{fig1}
\end{figure}

\newpage
\begin{figure}[htbp]
  \centering
    \includegraphics[scale=0.6]{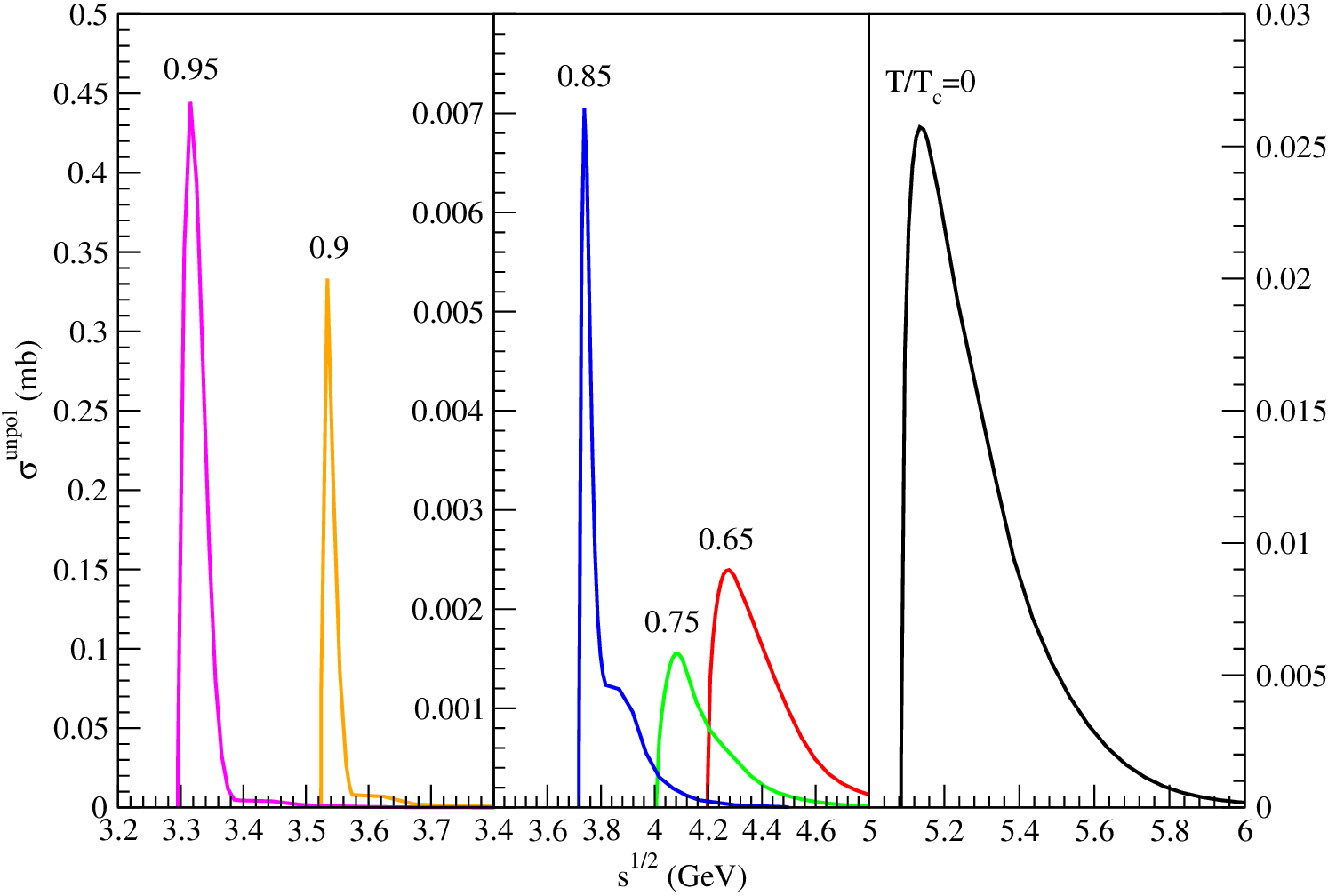}
\caption{Cross sections for $D_s^+ \bar{D} \rightarrow K^\ast \psi(4160)$
at various temperatures.}
\label{fig2}
\end{figure}

\newpage
\begin{figure}[htbp]
  \centering
    \includegraphics[scale=0.6]{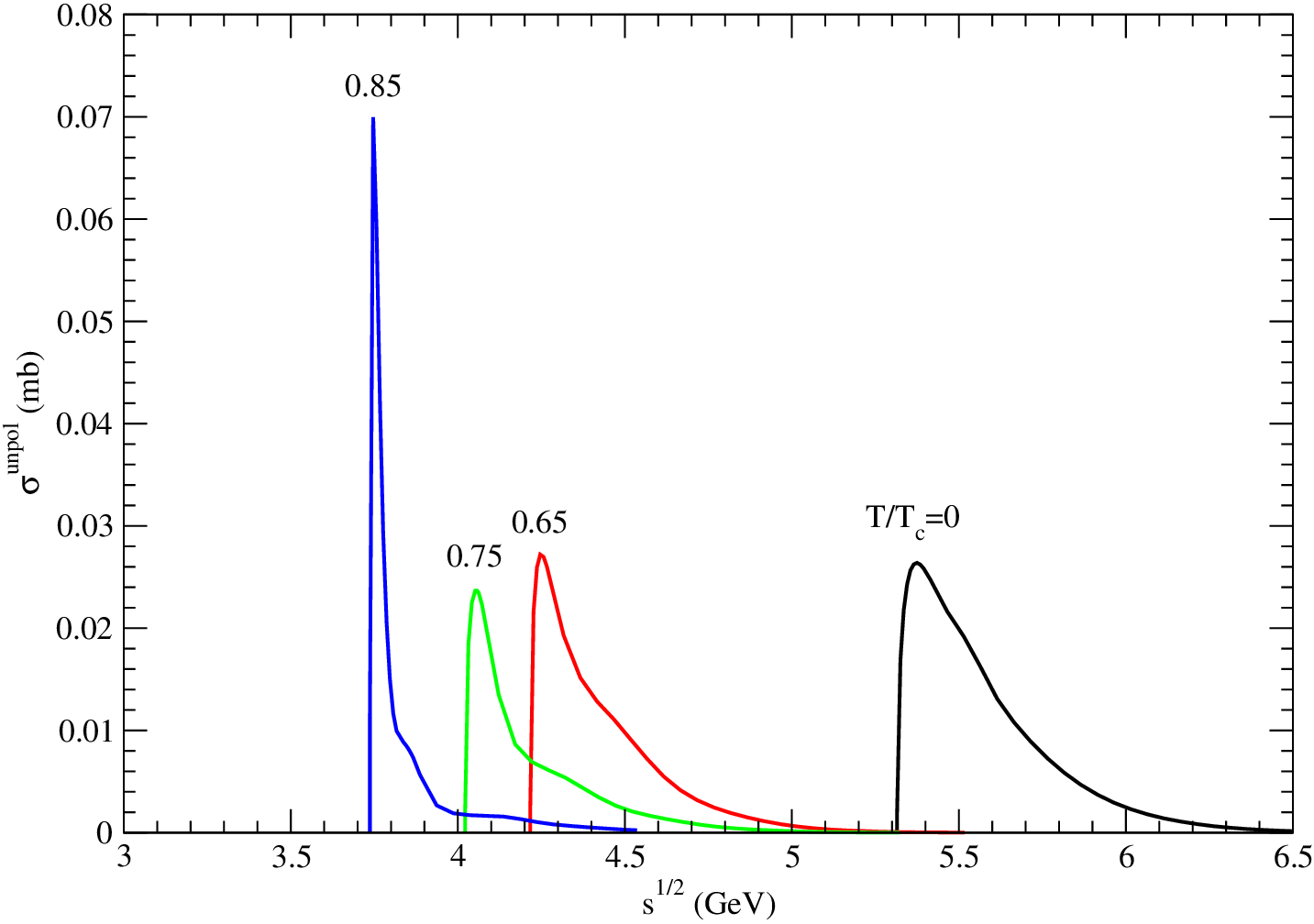}
\caption{Cross sections for $D_s^+ \bar{D} \rightarrow K^\ast \psi(4415)$
at various temperatures.}
\label{fig3}
\end{figure}

\newpage
\begin{figure}[htbp]
  \centering
    \includegraphics[scale=0.6]{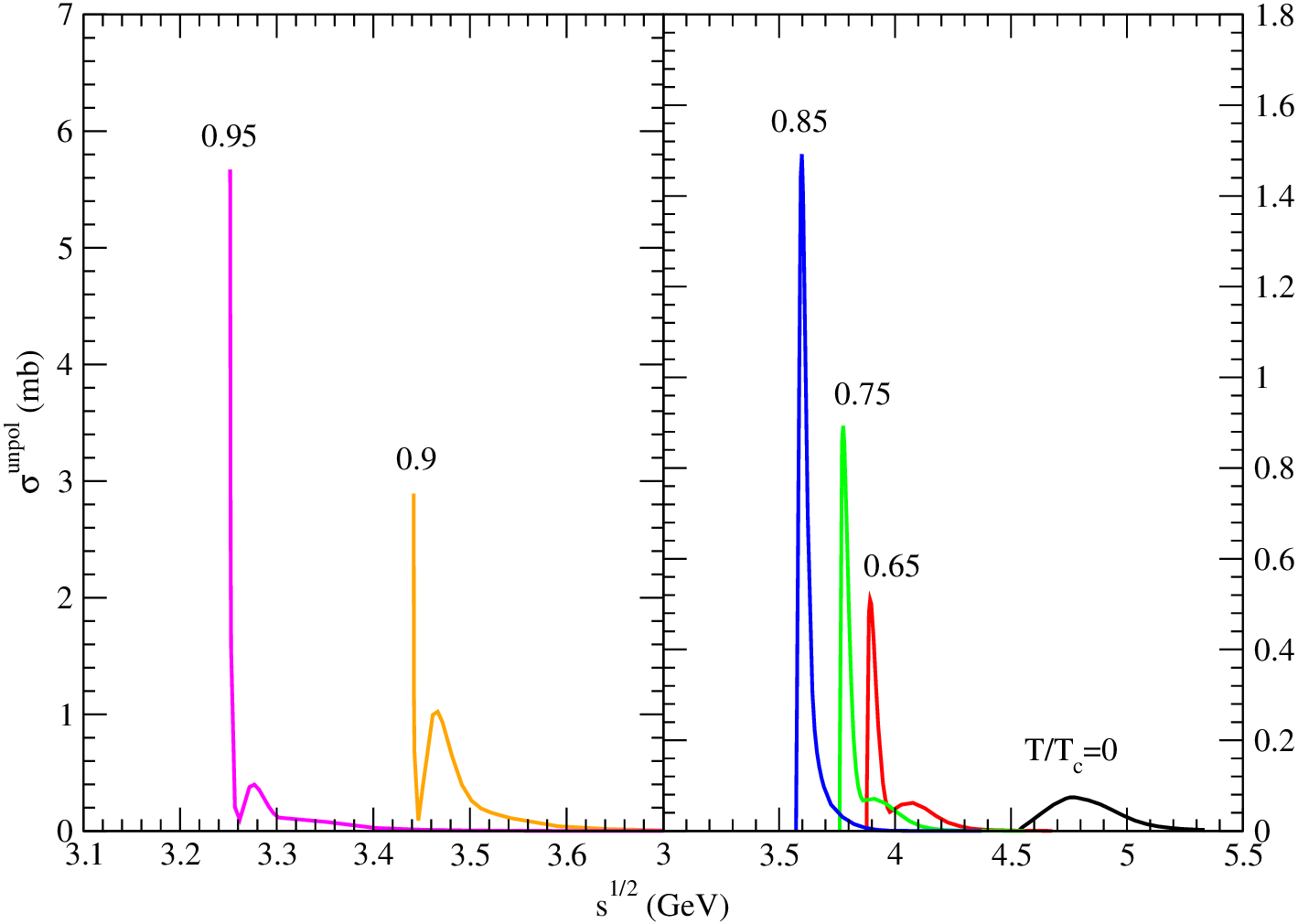}
\caption{Cross sections for $D_s^+ \bar{D}^{\ast} \rightarrow K \psi(4040)$ 
at various temperatures.}
\label{fig4}
\end{figure}

\newpage
\begin{figure}[htbp]
  \centering
    \includegraphics[scale=0.6]{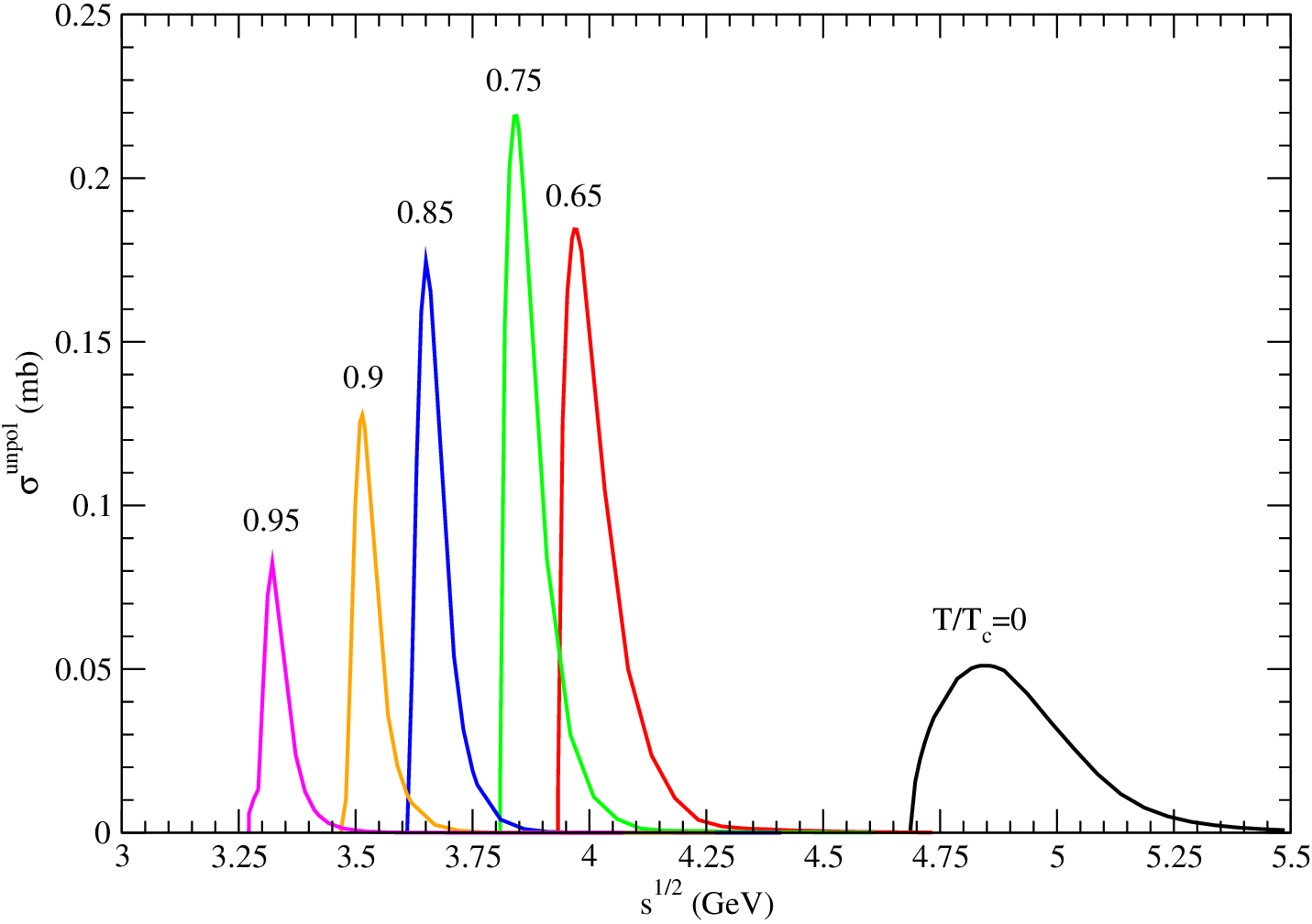}
\caption{Cross sections for $D_s^+ \bar{D}^{\ast} \rightarrow K \psi(4160)$ 
at various temperatures.}
\label{fig5}
\end{figure}

\newpage
\begin{figure}[htbp]
  \centering
    \includegraphics[scale=0.6]{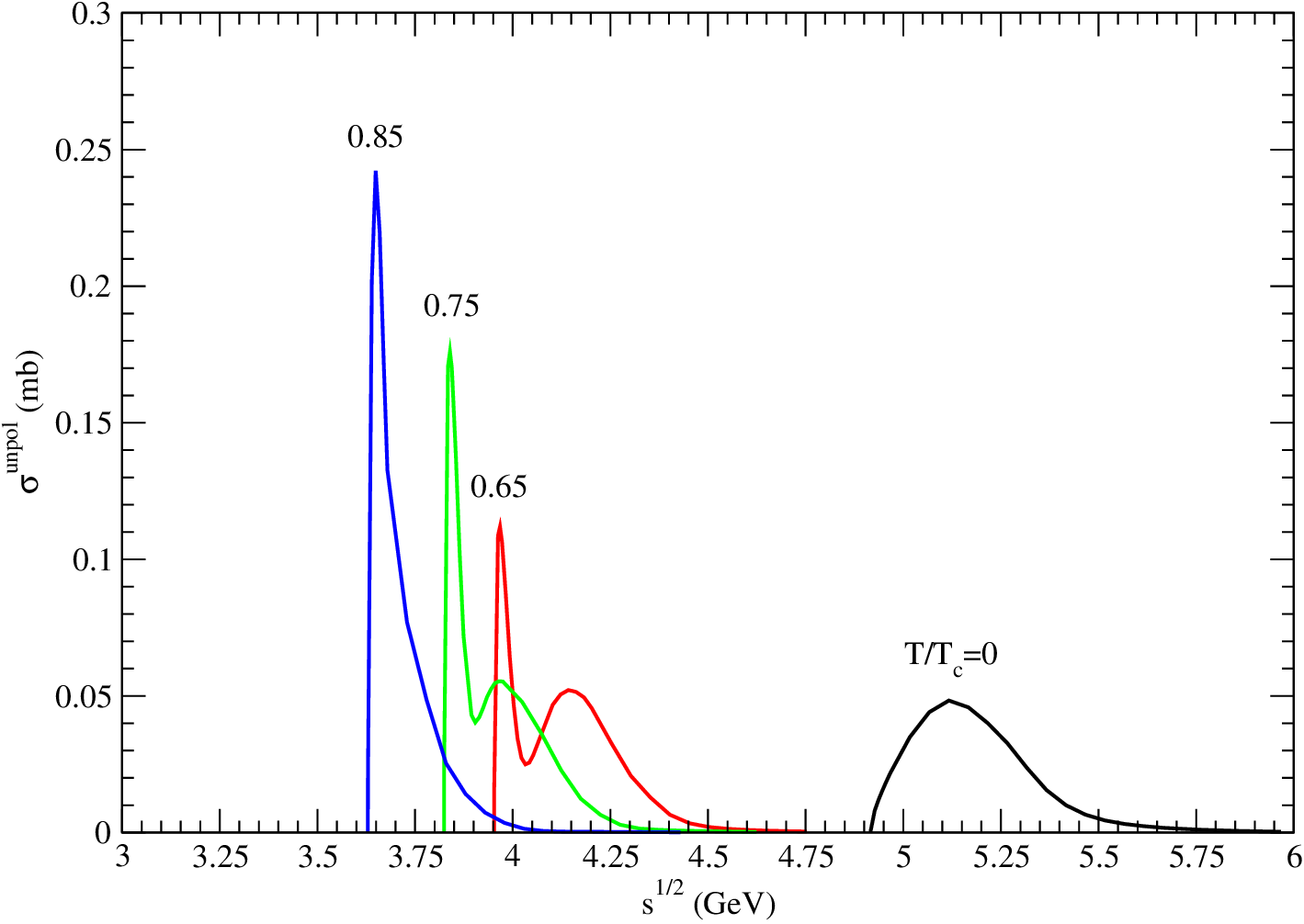}
\caption{Cross sections for $D_s^+ \bar{D}^{\ast} \rightarrow K \psi(4415)$ 
at various temperatures.}
\label{fig6}
\end{figure}

\newpage
\begin{figure}[htbp]
  \centering
    \includegraphics[scale=0.6]{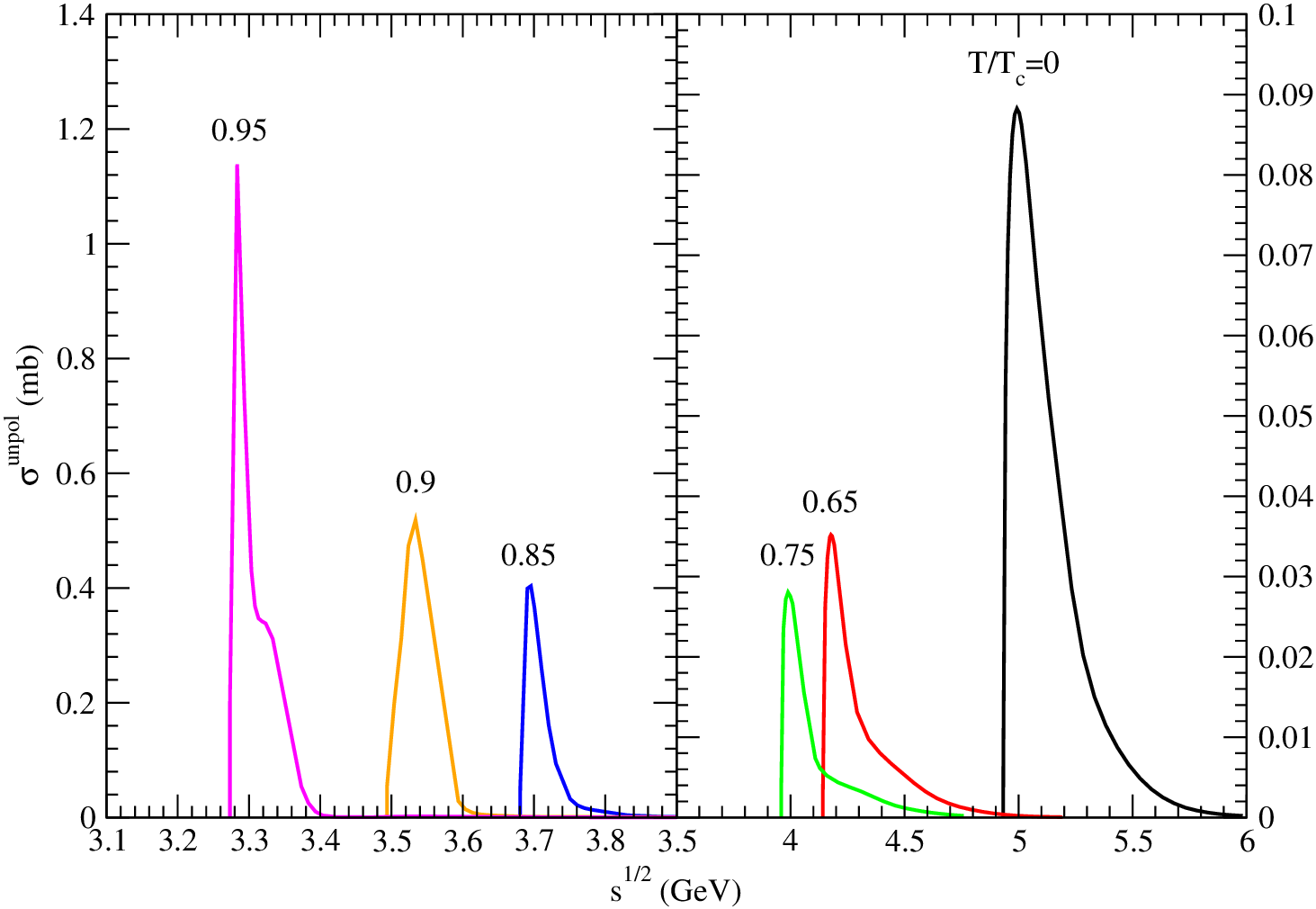}
\caption{Cross sections for $D_s^+ \bar{D}^{\ast} \rightarrow K^\ast 
\psi(4040)$ at various temperatures.}
\label{fig7}
\end{figure}

\newpage
\begin{figure}[htbp]
  \centering
    \includegraphics[scale=0.6]{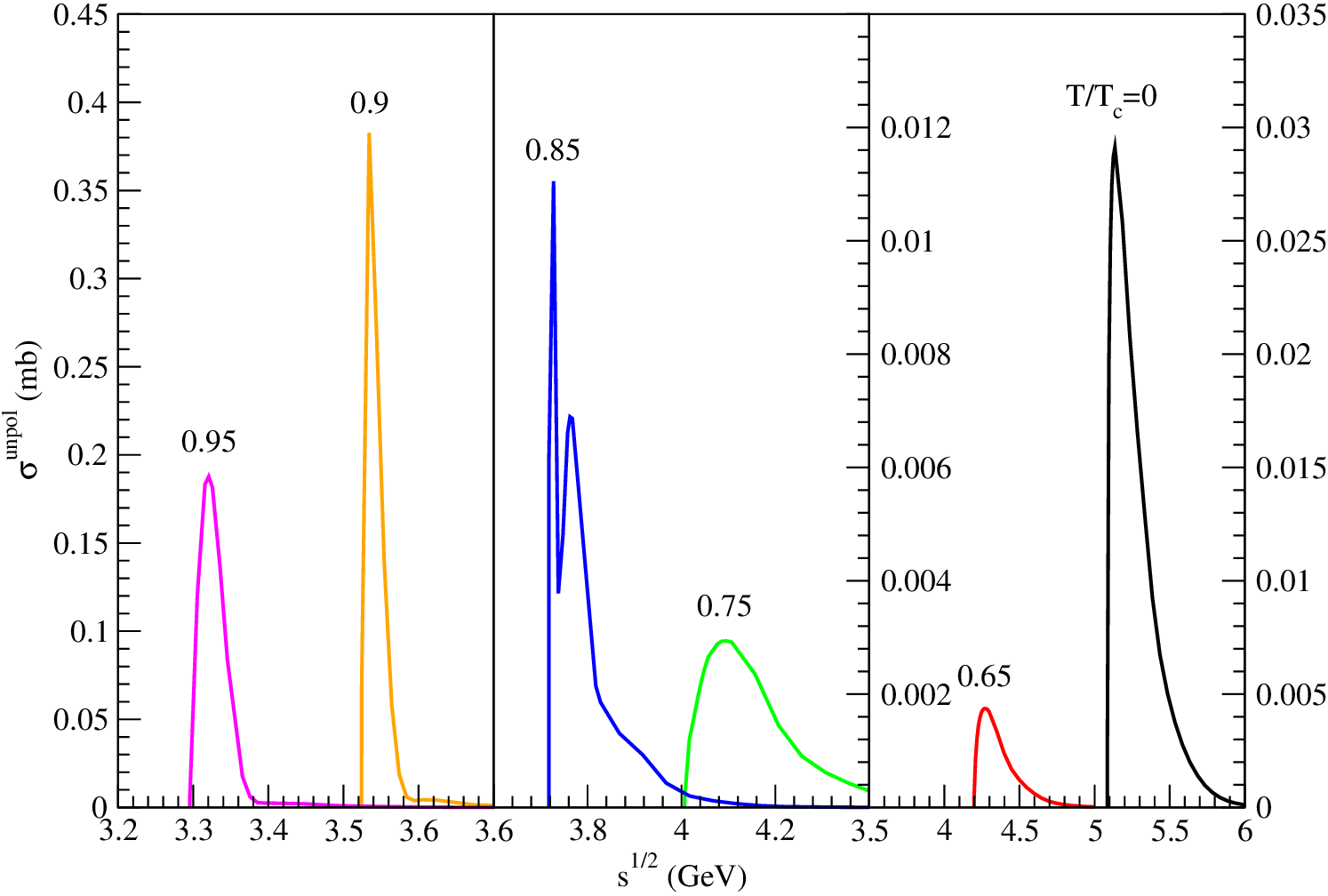}
\caption{Cross sections for $D_s^+ \bar{D}^{\ast} \rightarrow K^\ast 
\psi(4160)$ at various temperatures.}
\label{fig8}
\end{figure}

\newpage
\begin{figure}[htbp]
  \centering
    \includegraphics[scale=0.6]{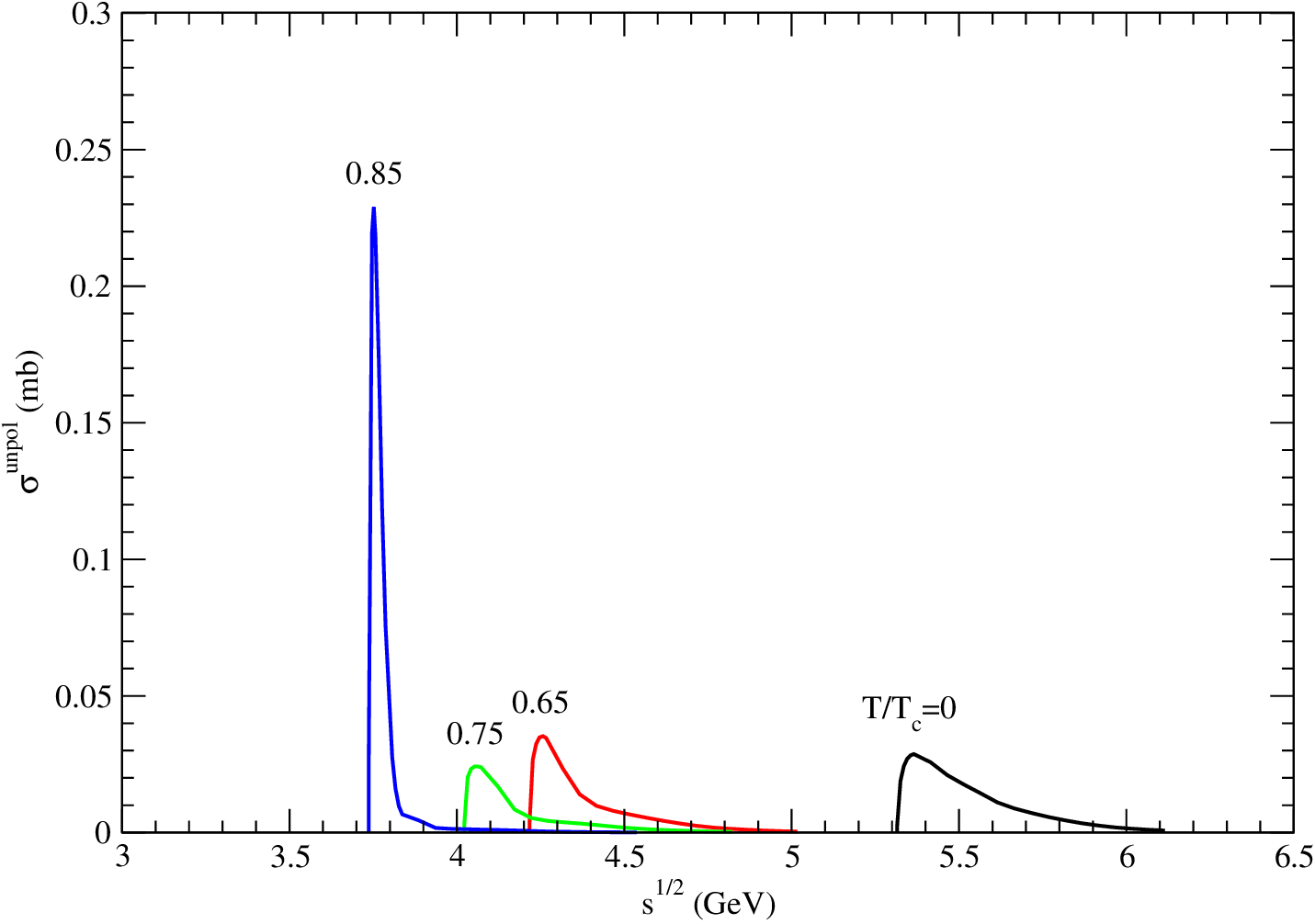}
\caption{Cross sections for $D_s^+ \bar{D}^{\ast} \rightarrow K^\ast 
\psi(4415)$ at various temperatures.}
\label{fig9}
\end{figure}

\newpage
\begin{figure}[htbp]
  \centering
    \includegraphics[scale=0.6]{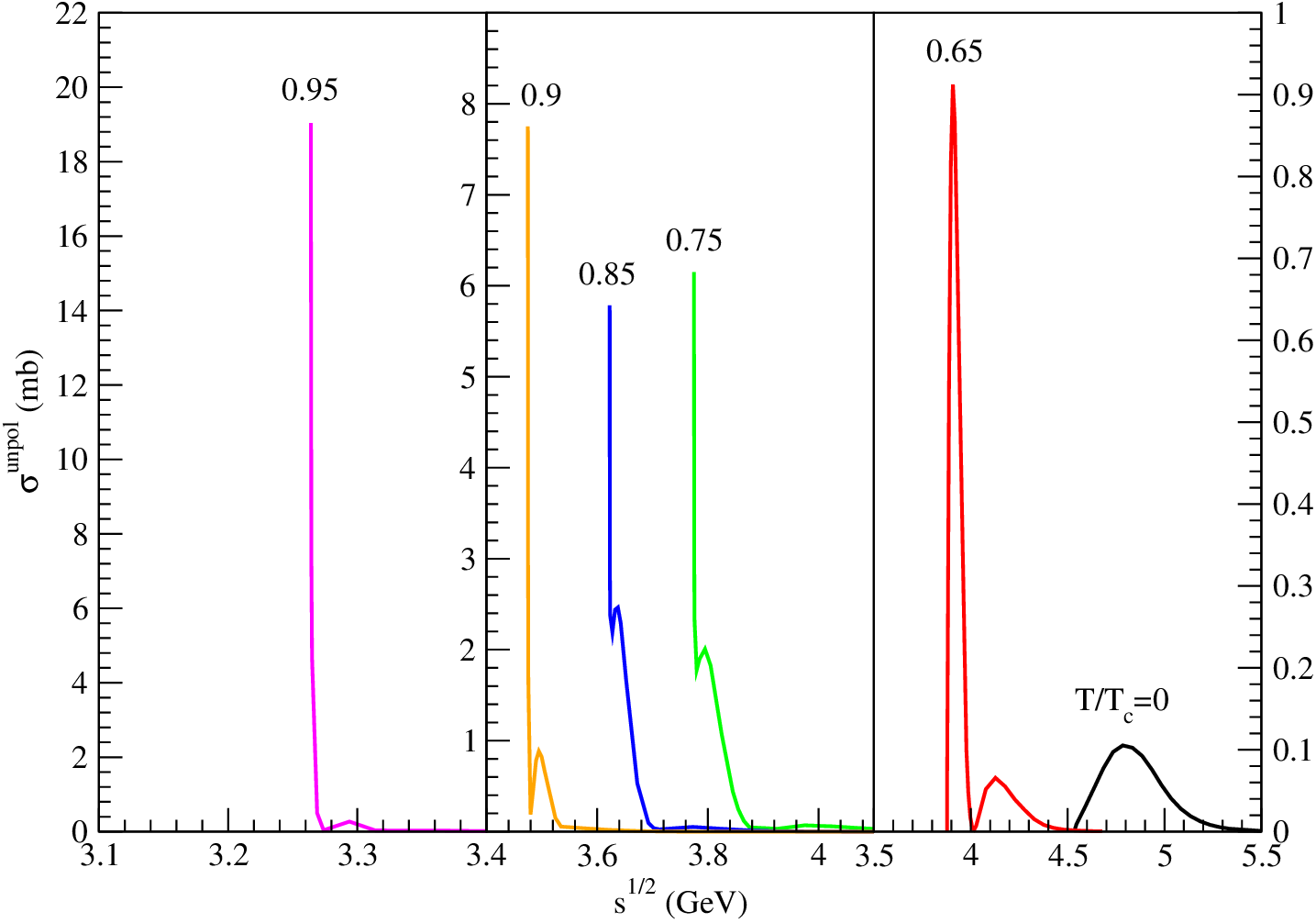}
\caption{Cross sections for $D_s^{\ast +} \bar{D} \rightarrow K \psi(4040)$
at various temperatures.}
\label{fig10}
\end{figure}

\newpage
\begin{figure}[htbp]
  \centering
    \includegraphics[scale=0.6]{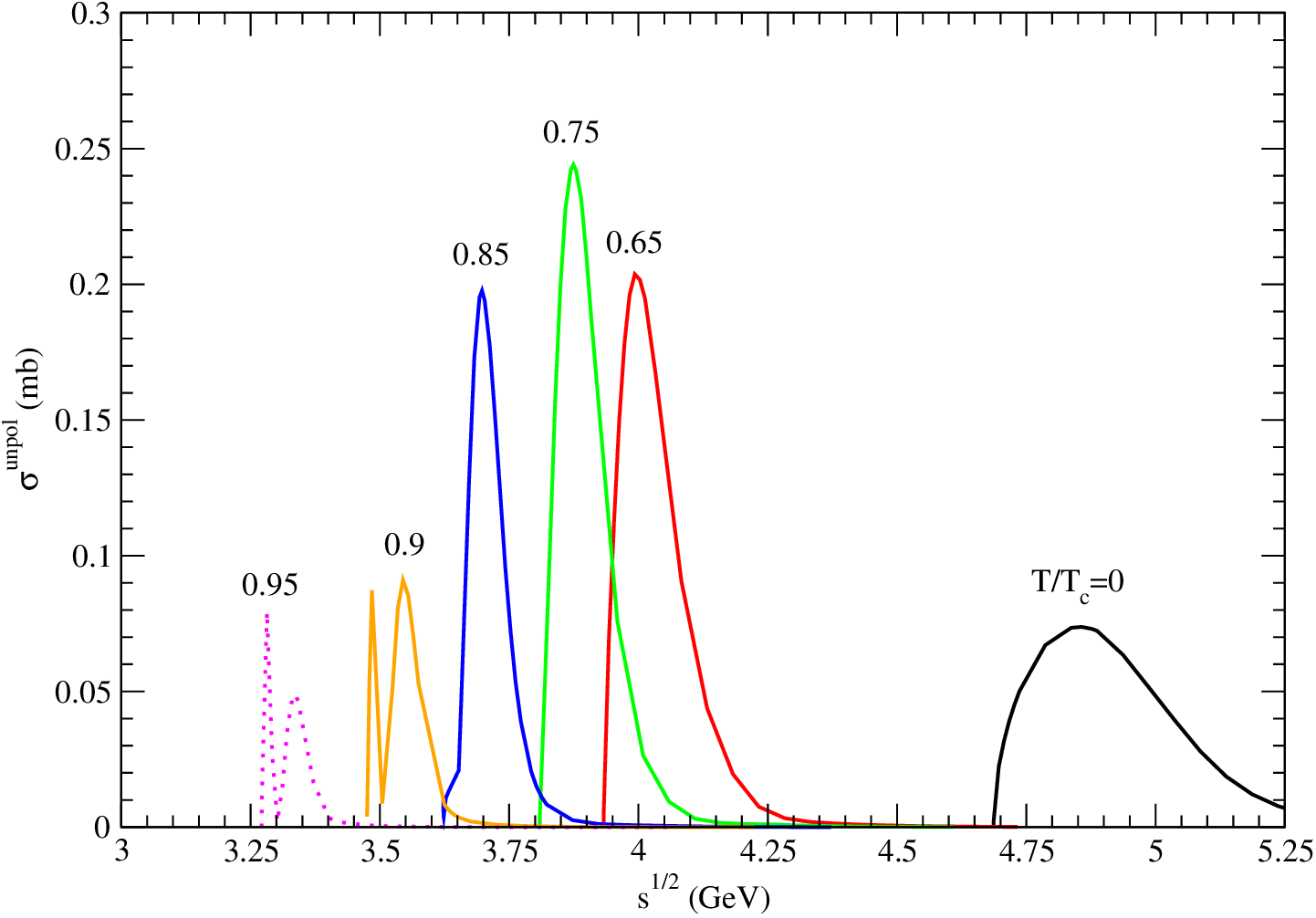}
\caption{Cross sections for $D_s^{\ast +} \bar{D} \rightarrow K \psi(4160)$
at various temperatures.}
\label{fig11}
\end{figure}

\newpage
\begin{figure}[htbp]
  \centering
    \includegraphics[scale=0.6]{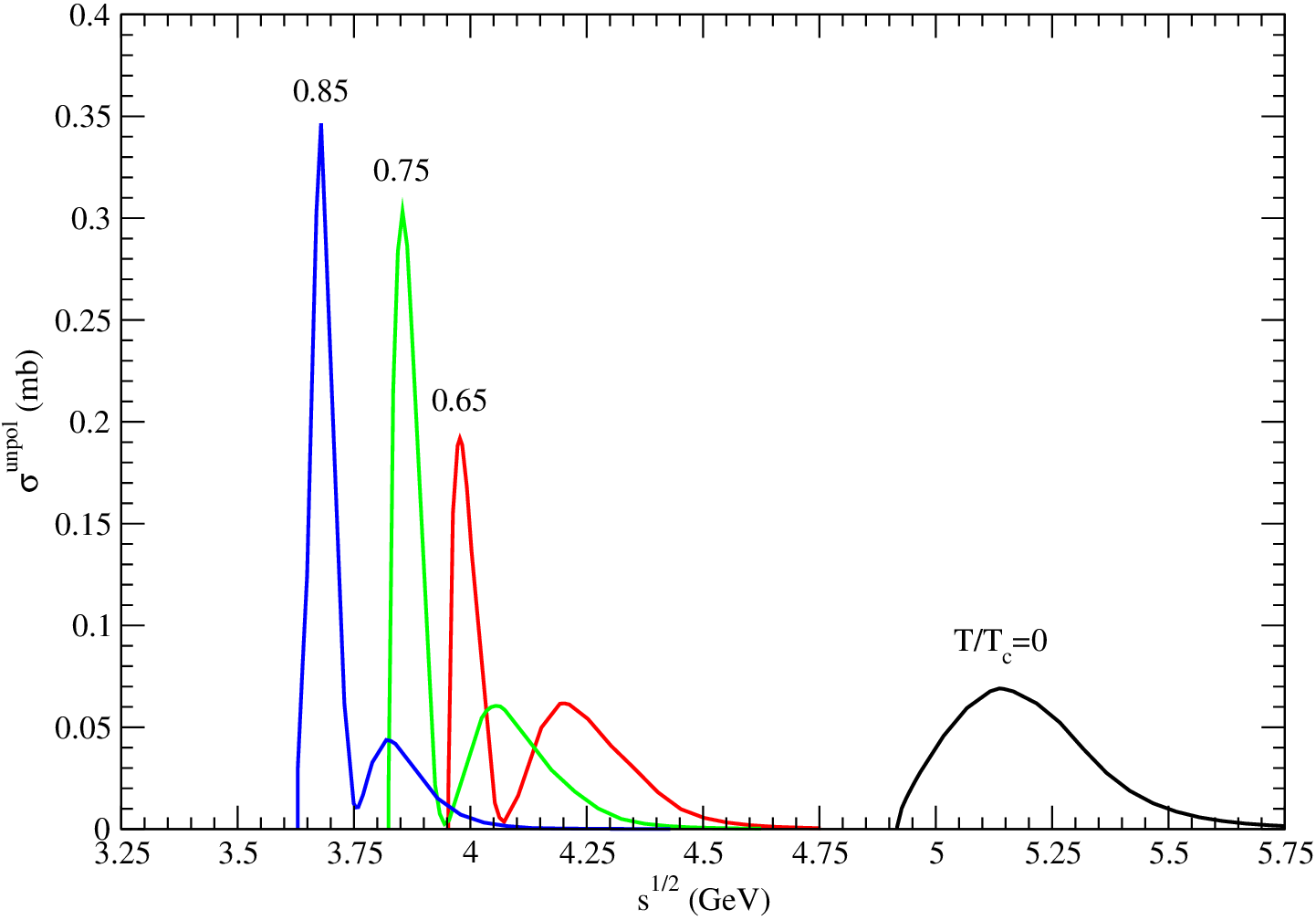}
\caption{Cross sections for $D_s^{\ast +} \bar{D} \rightarrow K \psi(4415)$
at various temperatures.}
\label{fig12}
\end{figure}

\newpage
\begin{figure}[htbp]
  \centering
    \includegraphics[scale=0.6]{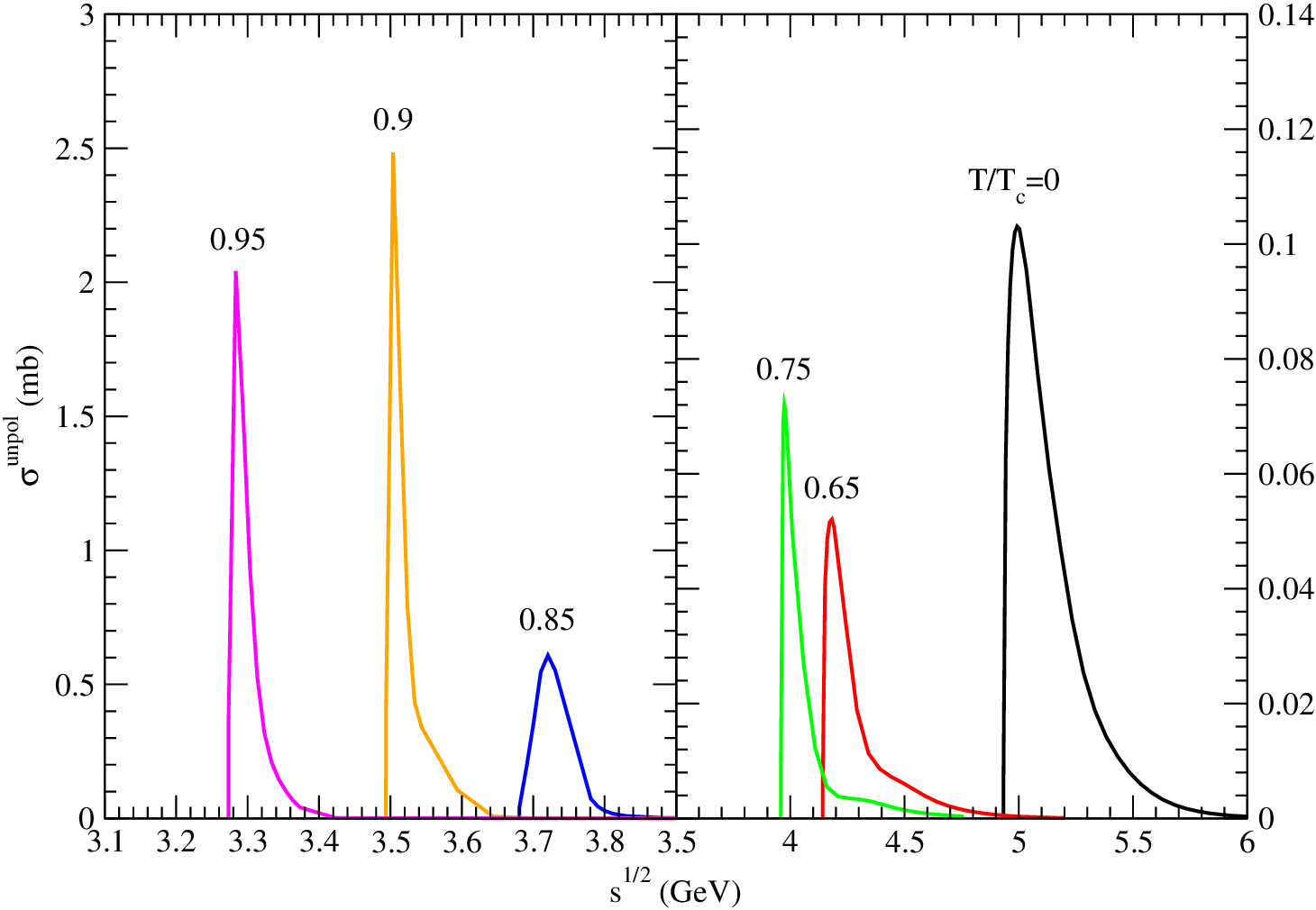}
\caption{Cross sections for $D_s^{\ast +} \bar{D} \rightarrow K^\ast 
\psi(4040)$ at various temperatures.}
\label{fig13}
\end{figure}

\newpage
\begin{figure}[htbp]
  \centering
    \includegraphics[scale=0.6]{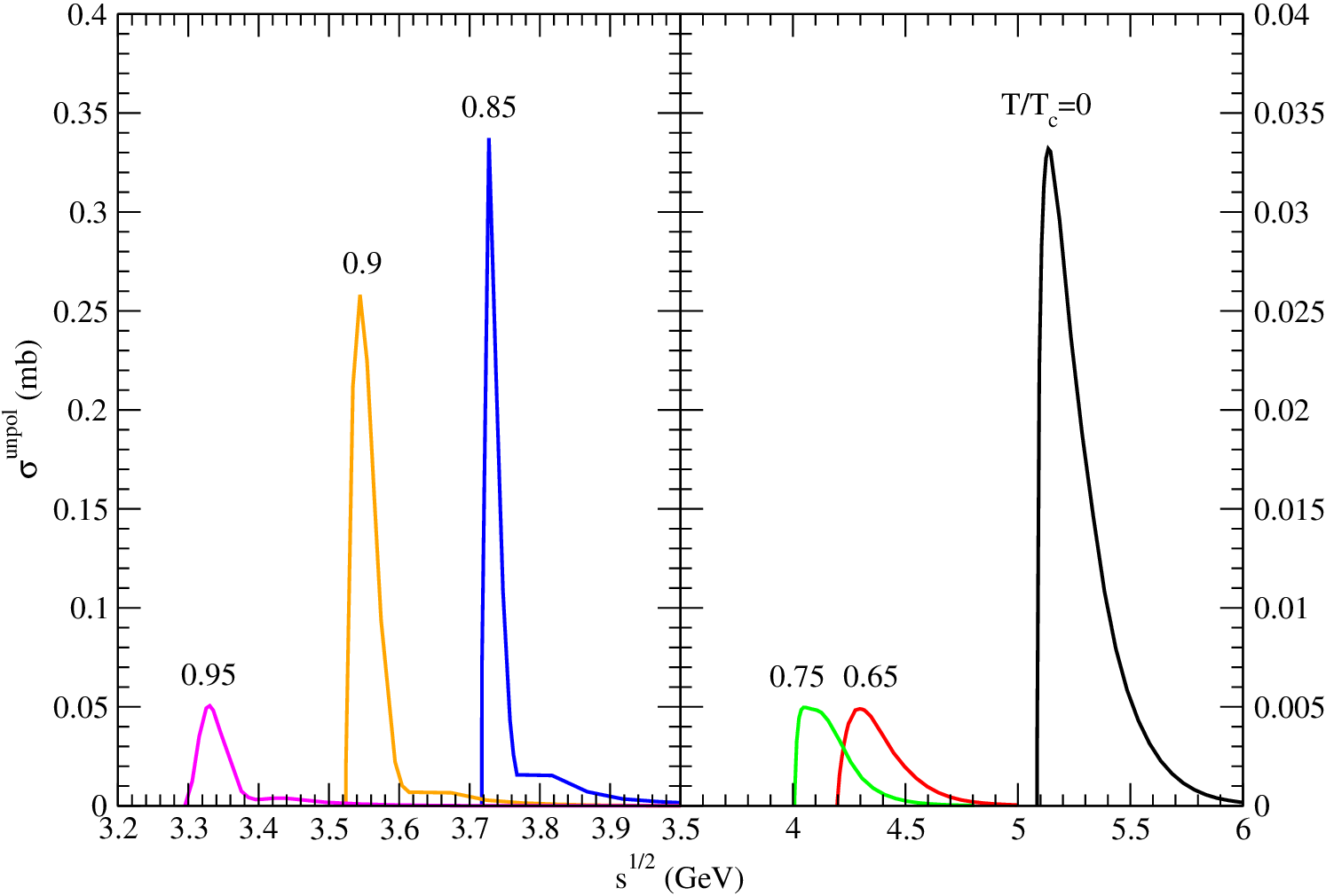}
\caption{Cross sections for $D_s^{\ast +} \bar{D} \rightarrow K^\ast
\psi(4160)$ at various temperatures.}
\label{fig14}
\end{figure}

\newpage
\begin{figure}[htbp]
  \centering
    \includegraphics[scale=0.6]{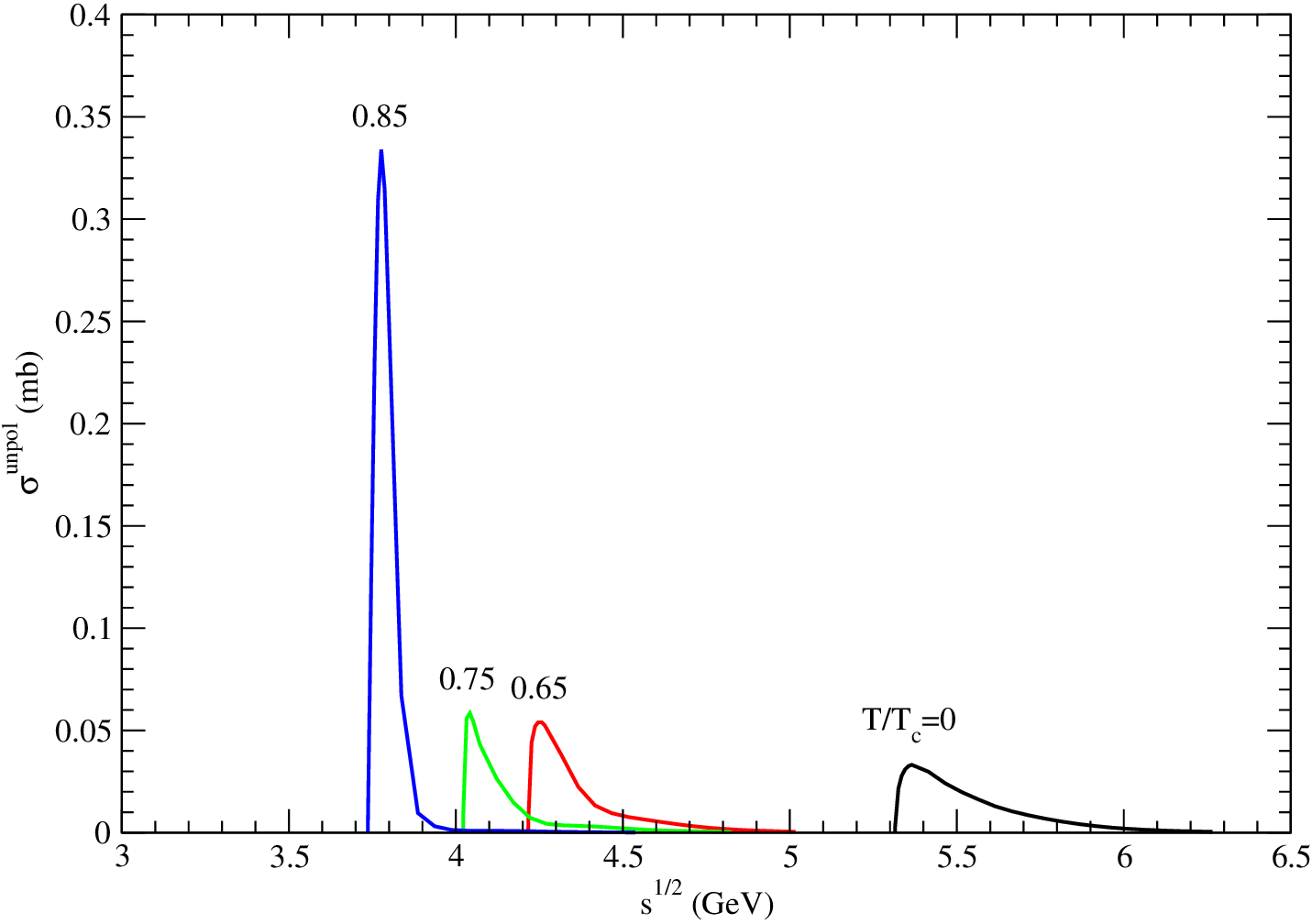}
\caption{Cross sections for $D_s^{\ast +}\bar{D} \rightarrow K^\ast
\psi(4415)$ at various temperatures.}
\label{fig15}
\end{figure}

\newpage
\begin{figure}[htbp]
  \centering
    \includegraphics[scale=0.6]{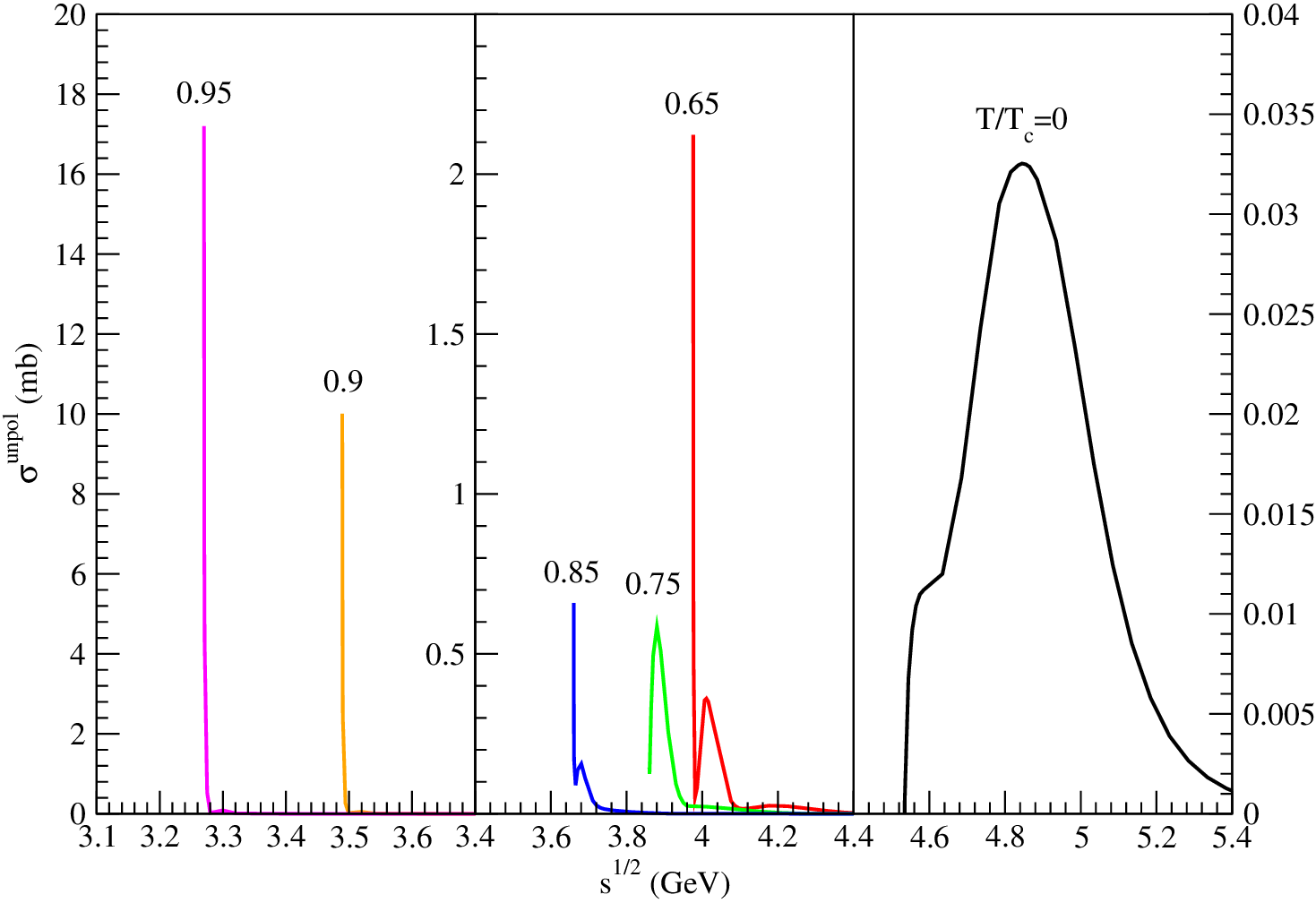}
\caption{Cross sections for $D_s^{\ast +} \bar{D}^{\ast} \rightarrow K 
\psi(4040)$ at various temperatures.}
\label{fig16}
\end{figure}

\newpage
\begin{figure}[htbp]
  \centering
    \includegraphics[scale=0.6]{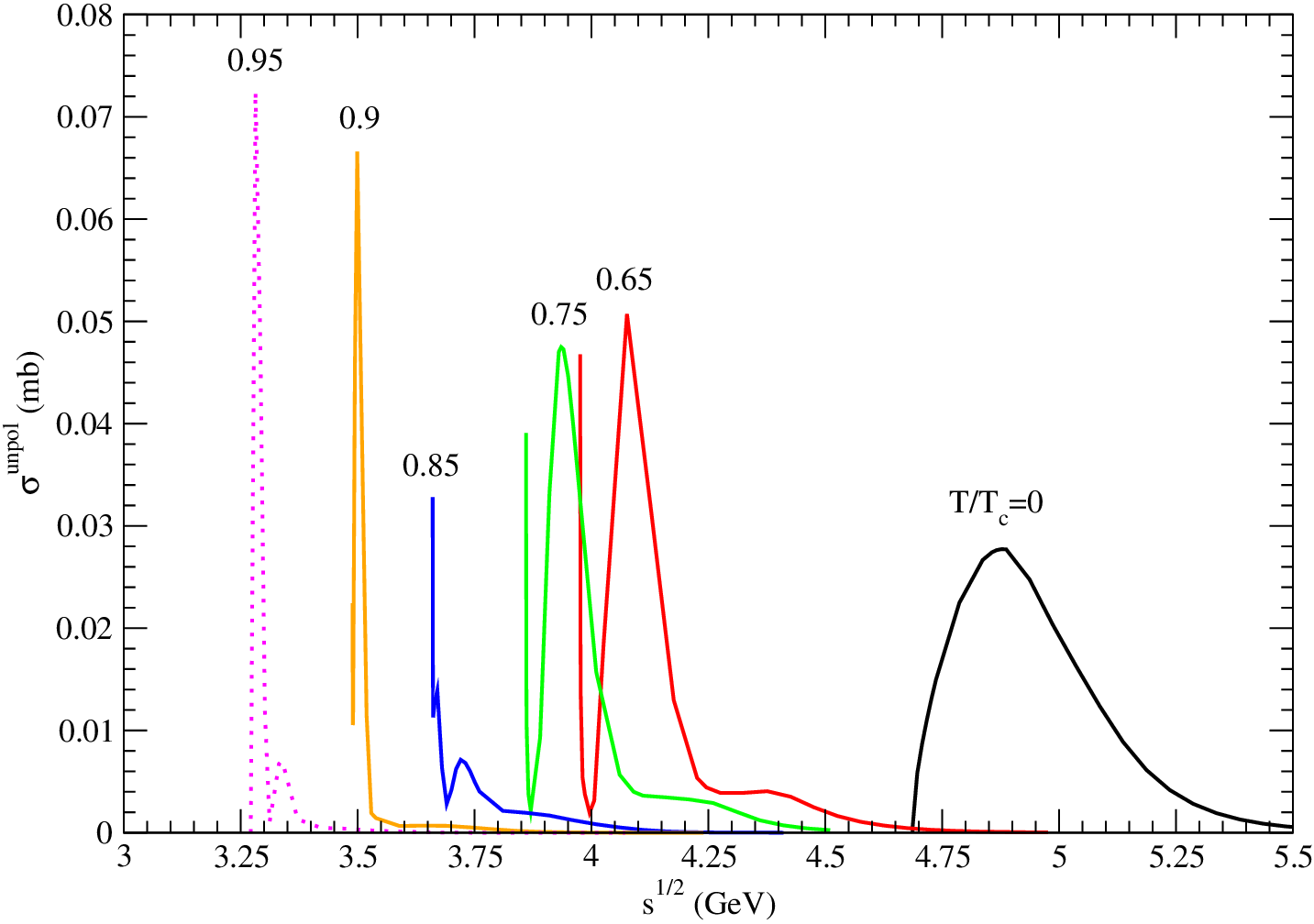}
\caption{Cross sections for $D_s^{\ast +} \bar{D}^{\ast} \rightarrow K
\psi(4160)$ at various temperatures.}
\label{fig17}
\end{figure}

\newpage
\begin{figure}[htbp]
  \centering
    \includegraphics[scale=0.6]{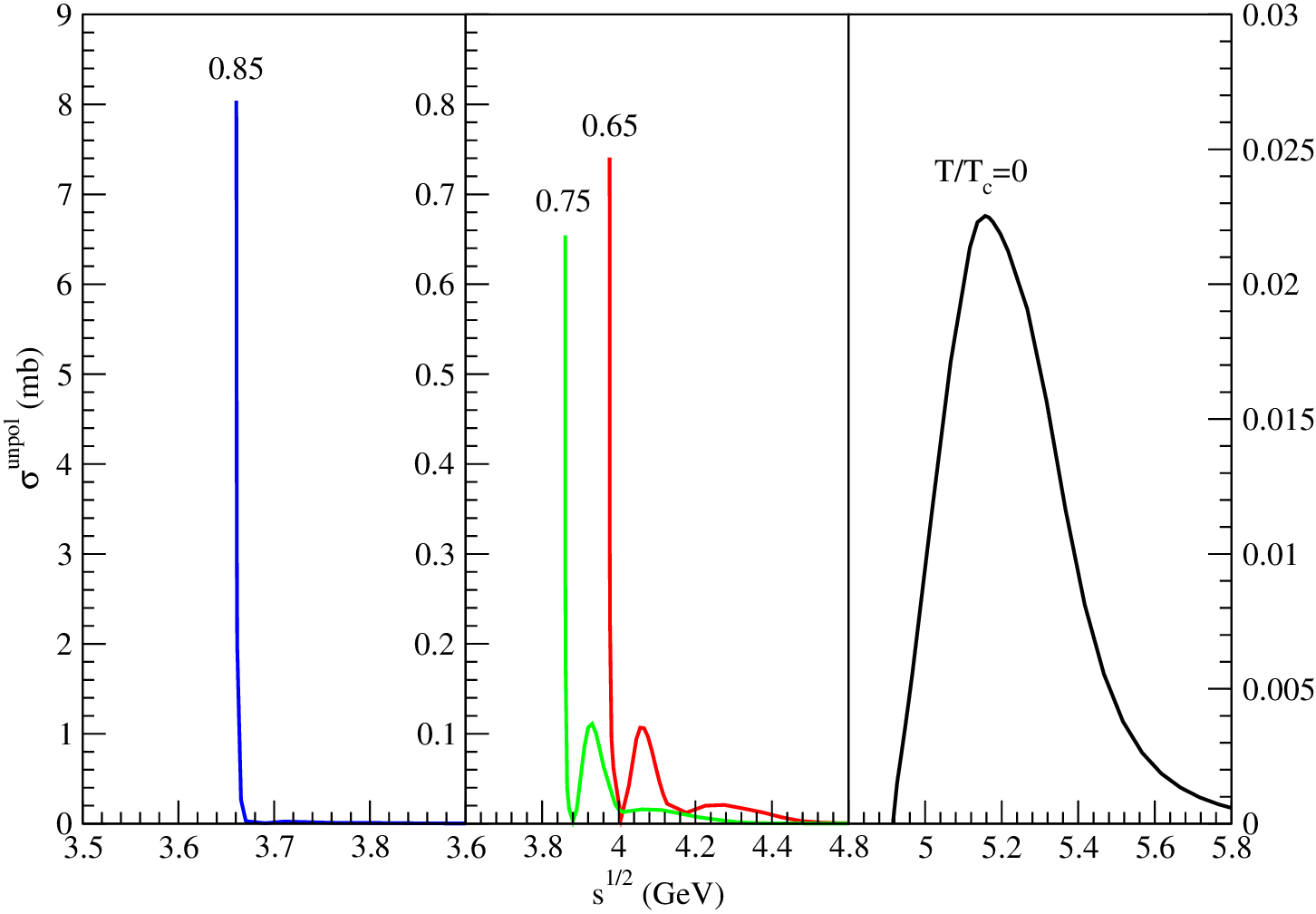}
\caption{Cross sections for $D_s^{\ast +} \bar{D}^{\ast} \rightarrow K
\psi(4415)$ at various temperatures.}
\label{fig18}
\end{figure}

\newpage
\begin{figure}[htbp]
  \centering
    \includegraphics[scale=0.6]{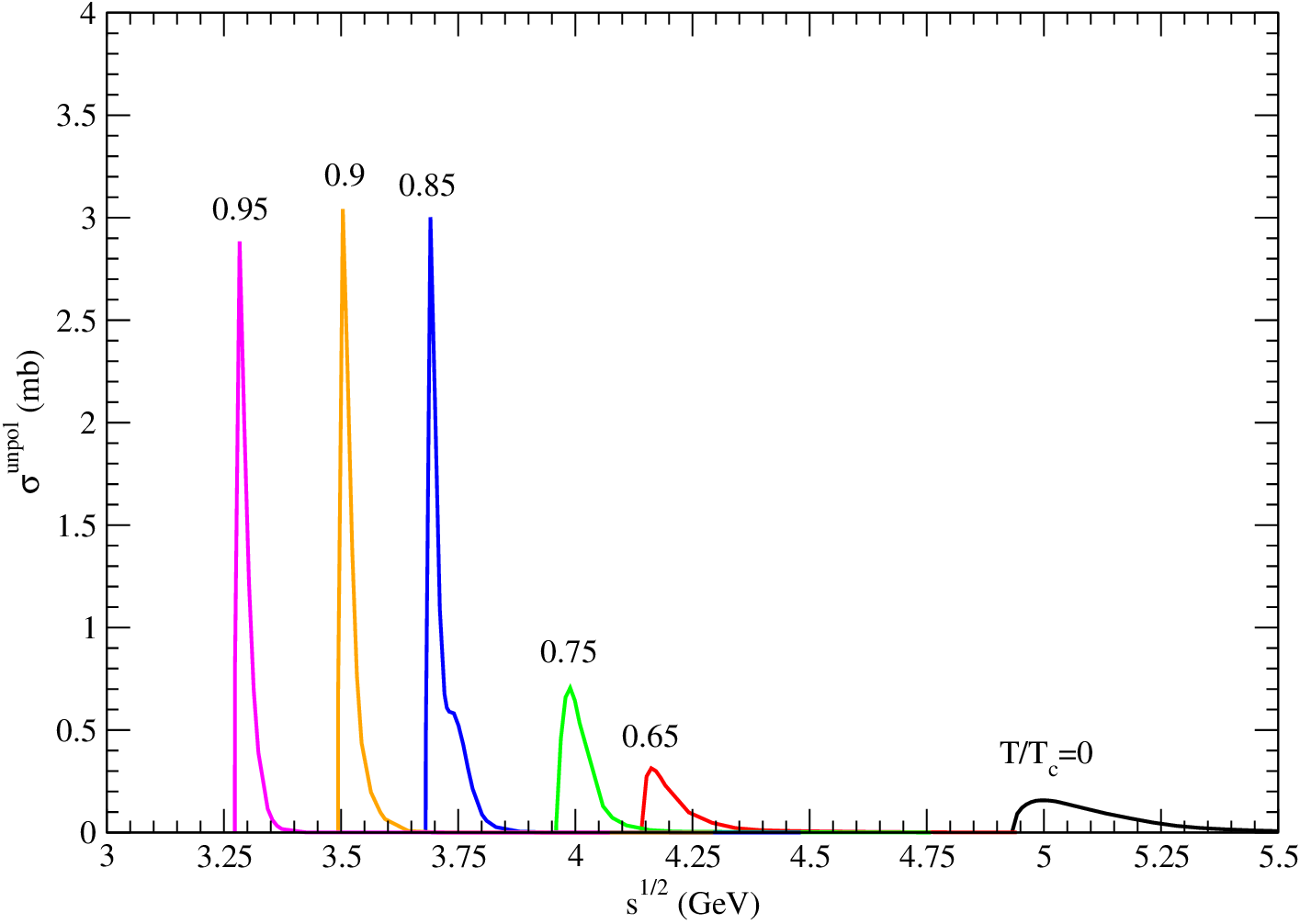}
\caption{Cross sections for $D_s^{\ast +} \bar{D}^{\ast} \rightarrow K^\ast
\psi(4040)$ at various temperatures.}
\label{fig19}
\end{figure}

\newpage
\begin{figure}[htbp]
  \centering
    \includegraphics[scale=0.6]{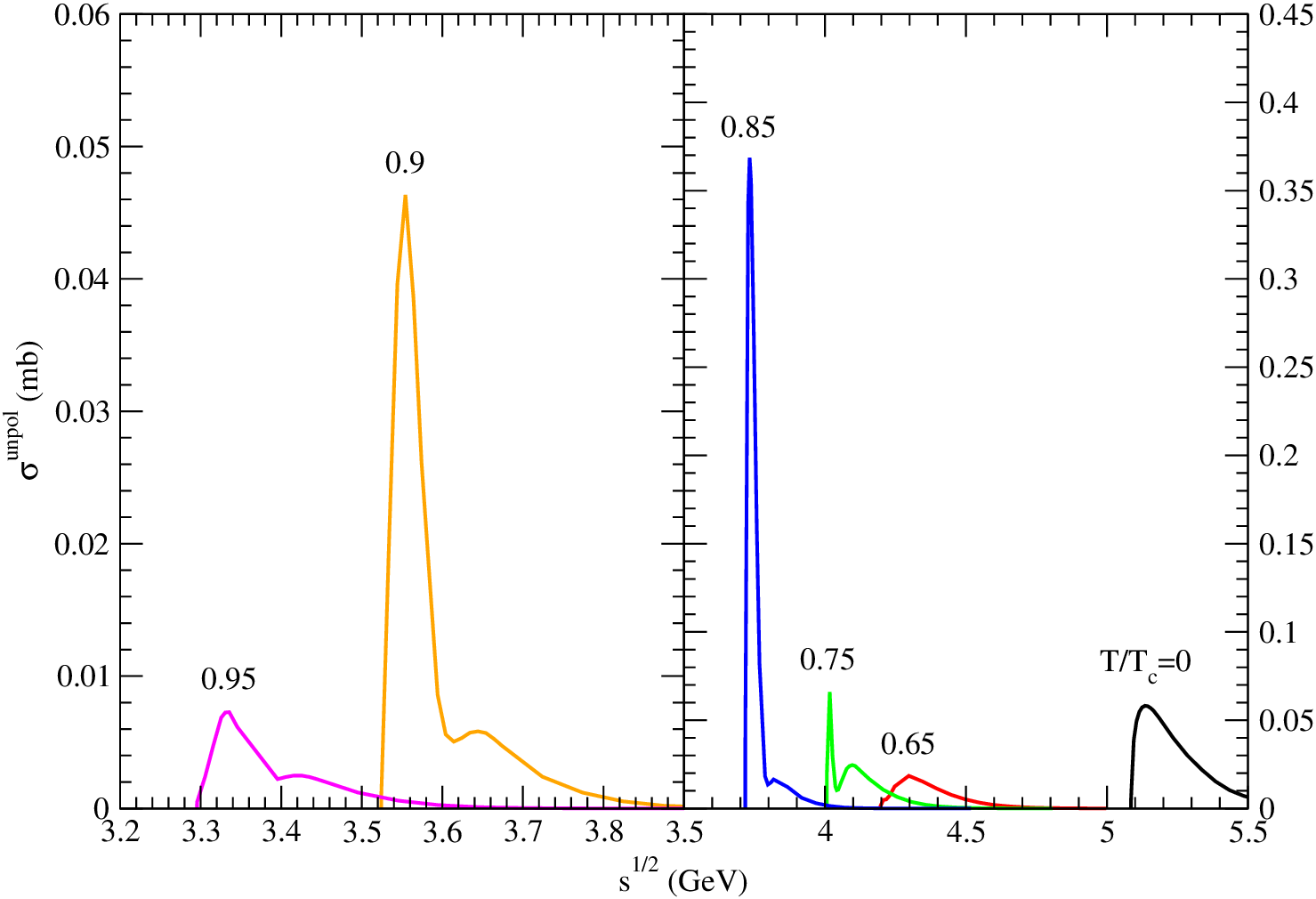}
\caption{Cross sections for $D_s^{\ast +} \bar{D}^{\ast} \rightarrow K^\ast
\psi(4160)$ at various temperatures.}
\label{fig20}
\end{figure}

\newpage
\begin{figure}[htbp]
  \centering
    \includegraphics[scale=0.6]{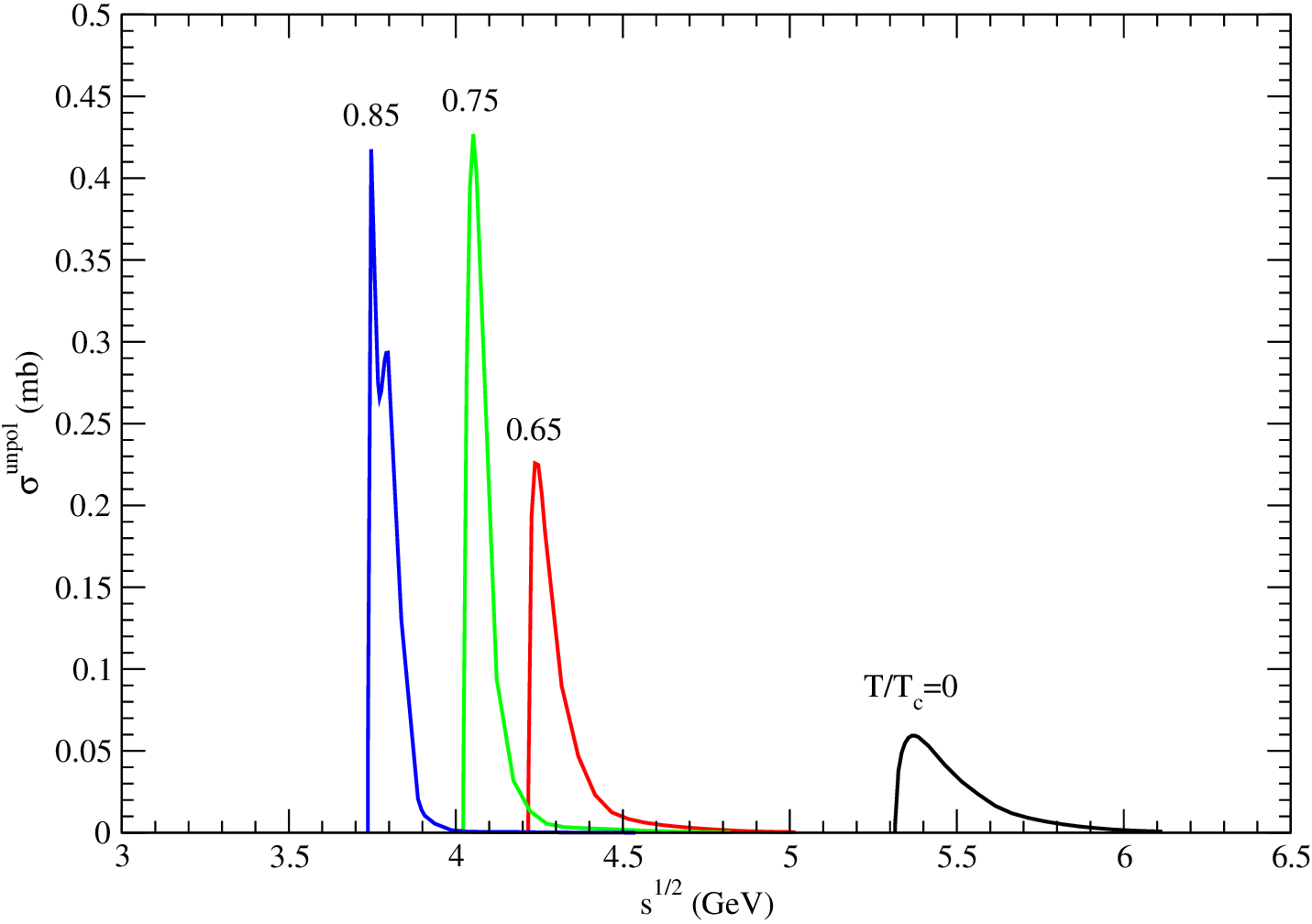}
\caption{Cross sections for $D_s^{\ast +} \bar{D}^{\ast} \rightarrow K^\ast
\psi(4415)$ at various temperatures.}
\label{fig21}
\end{figure}

\newpage
\begin{figure}[htbp]
  \centering
    \includegraphics[scale=0.6]{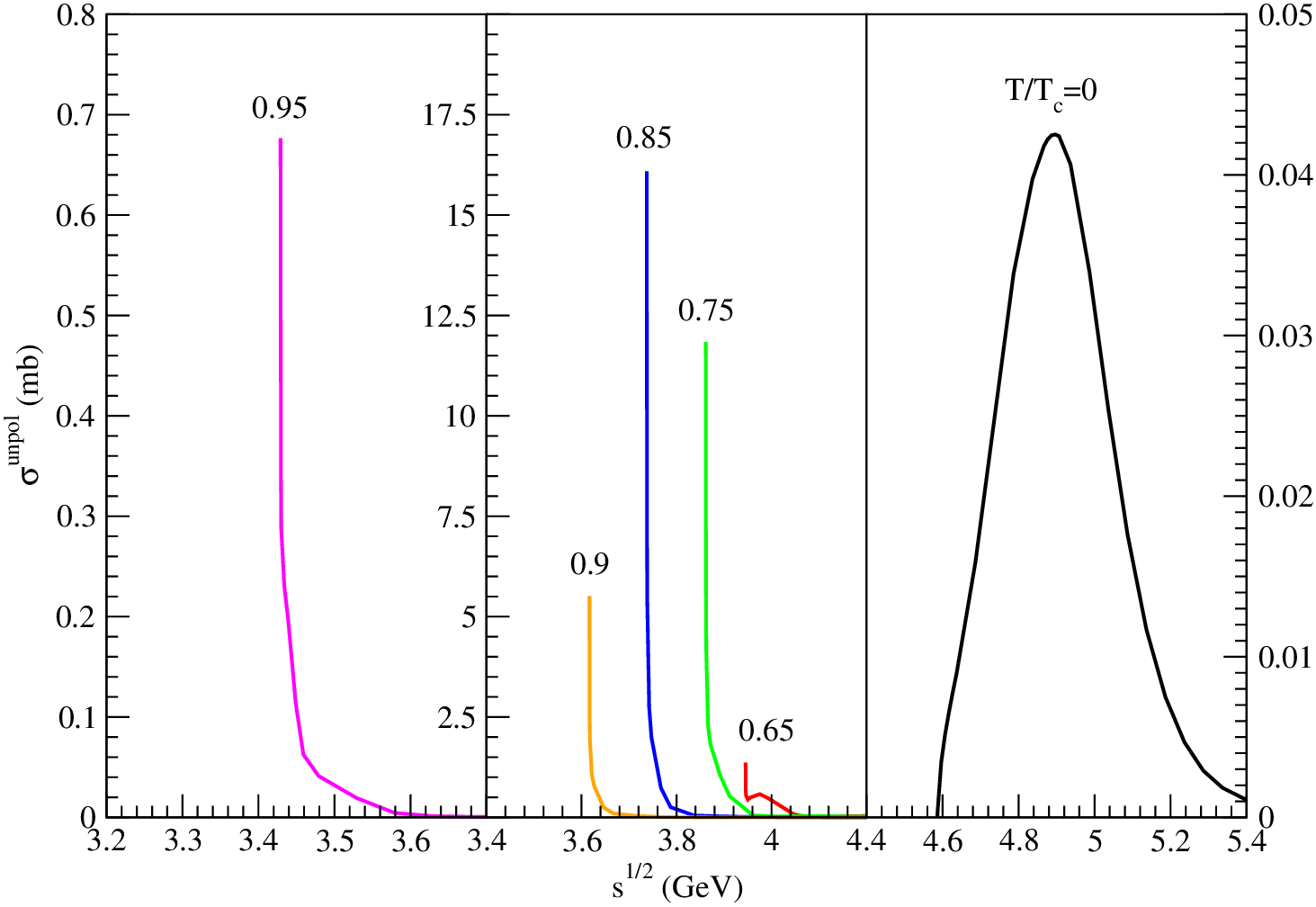}
\caption{Cross sections for $D_s^+ D_s^{\ast -} \rightarrow \eta 
\psi(4040)$ at various temperatures.}
\label{fig22}
\end{figure}

\newpage
\begin{figure}[htbp]
  \centering
    \includegraphics[scale=0.6]{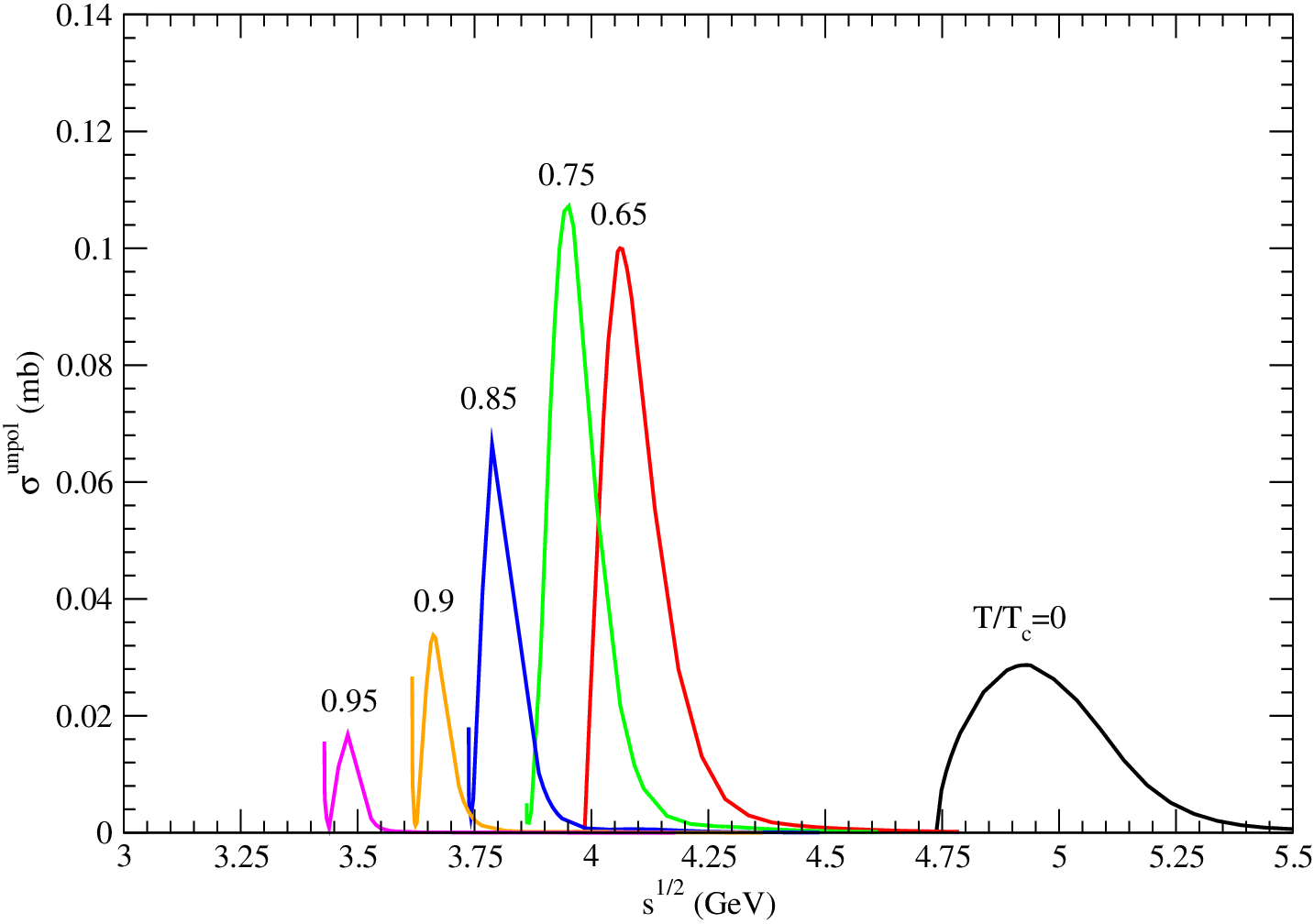}
\caption{Cross sections for $D_s^+ D_s^{\ast -} \rightarrow \eta
\psi(4160)$ at various temperatures.}
\label{fig23}
\end{figure}

\newpage
\begin{figure}[htbp]
  \centering
    \includegraphics[scale=0.6]{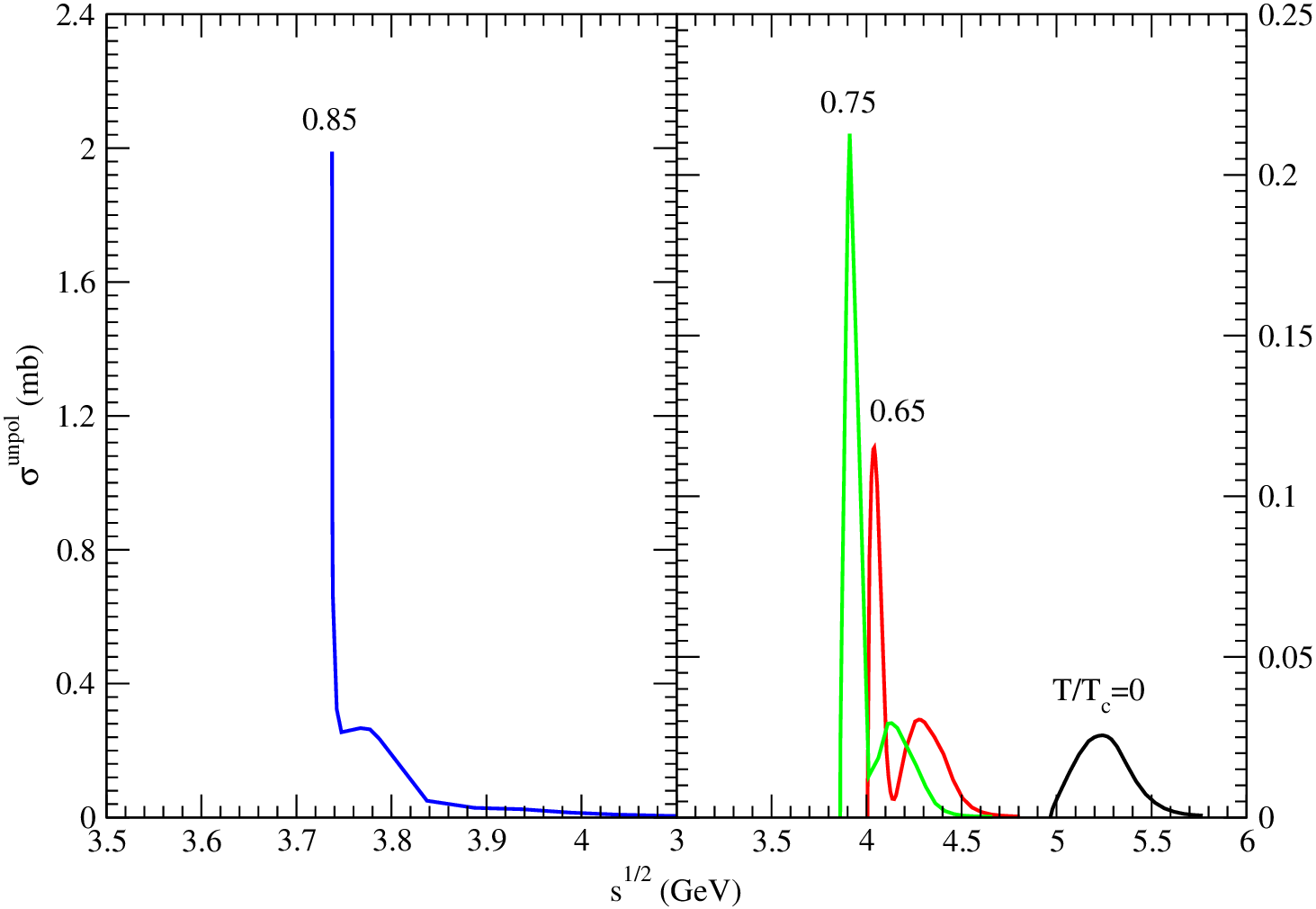}
\caption{Cross sections for $D_s^+ D_s^{\ast -} \rightarrow \eta
\psi(4415)$ at various temperatures.}
\label{fig24}
\end{figure}

\newpage
\begin{figure}[htbp]
  \centering
    \includegraphics[scale=0.6]{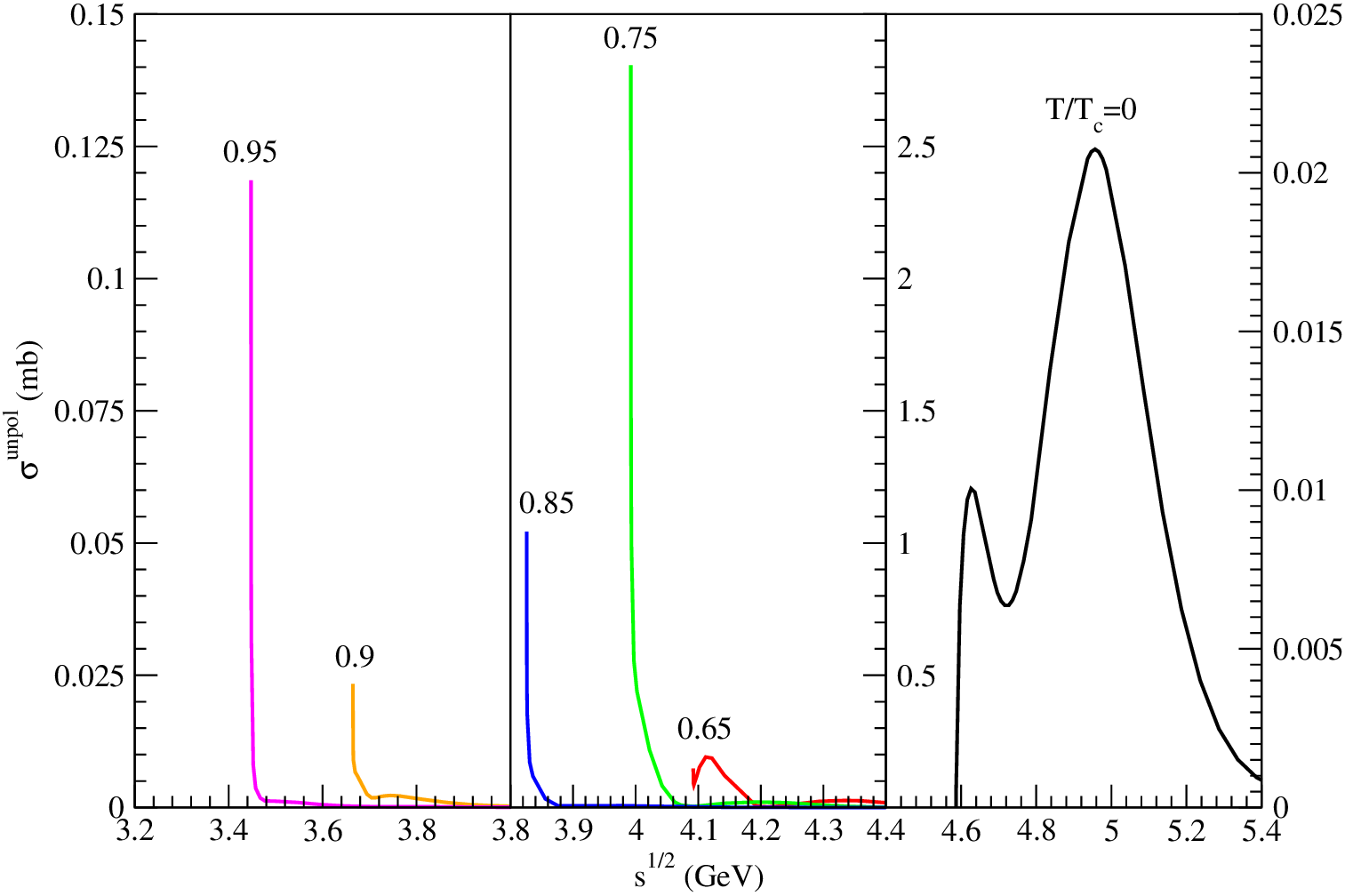}
\caption{Cross sections for $D_s^{\ast +} D_s^{\ast -} \rightarrow \eta 
\psi(4040)$ at various temperatures.}
\label{fig25}
\end{figure}

\newpage
\begin{figure}[htbp]
  \centering
    \includegraphics[scale=0.6]{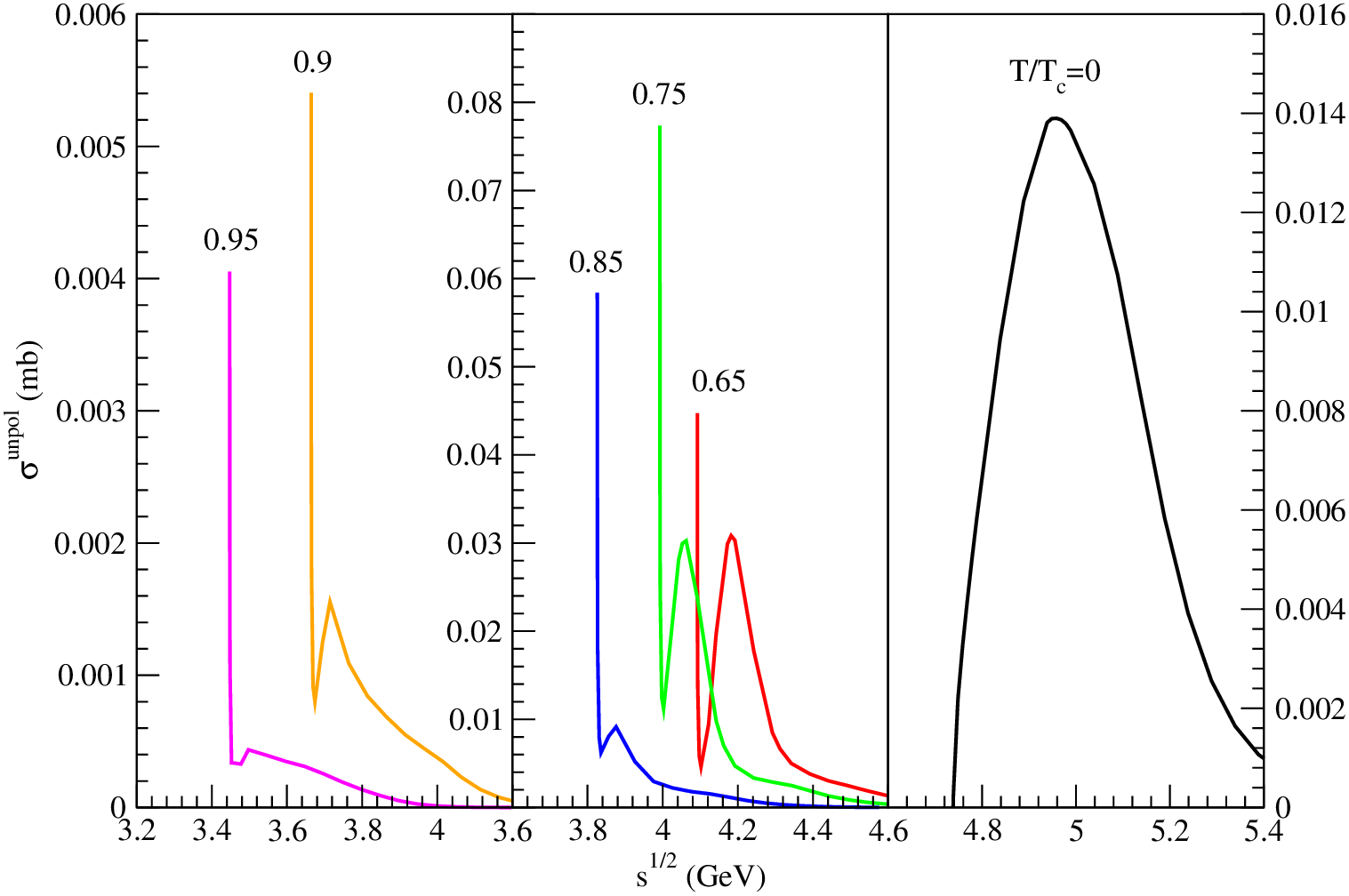}
\caption{Cross sections for $D_s^{\ast +} D_s^{\ast -} \rightarrow \eta
\psi(4160)$ at various temperatures.}
\label{fig26}
\end{figure}

\newpage
\begin{figure}[htbp]
  \centering
    \includegraphics[scale=0.6]{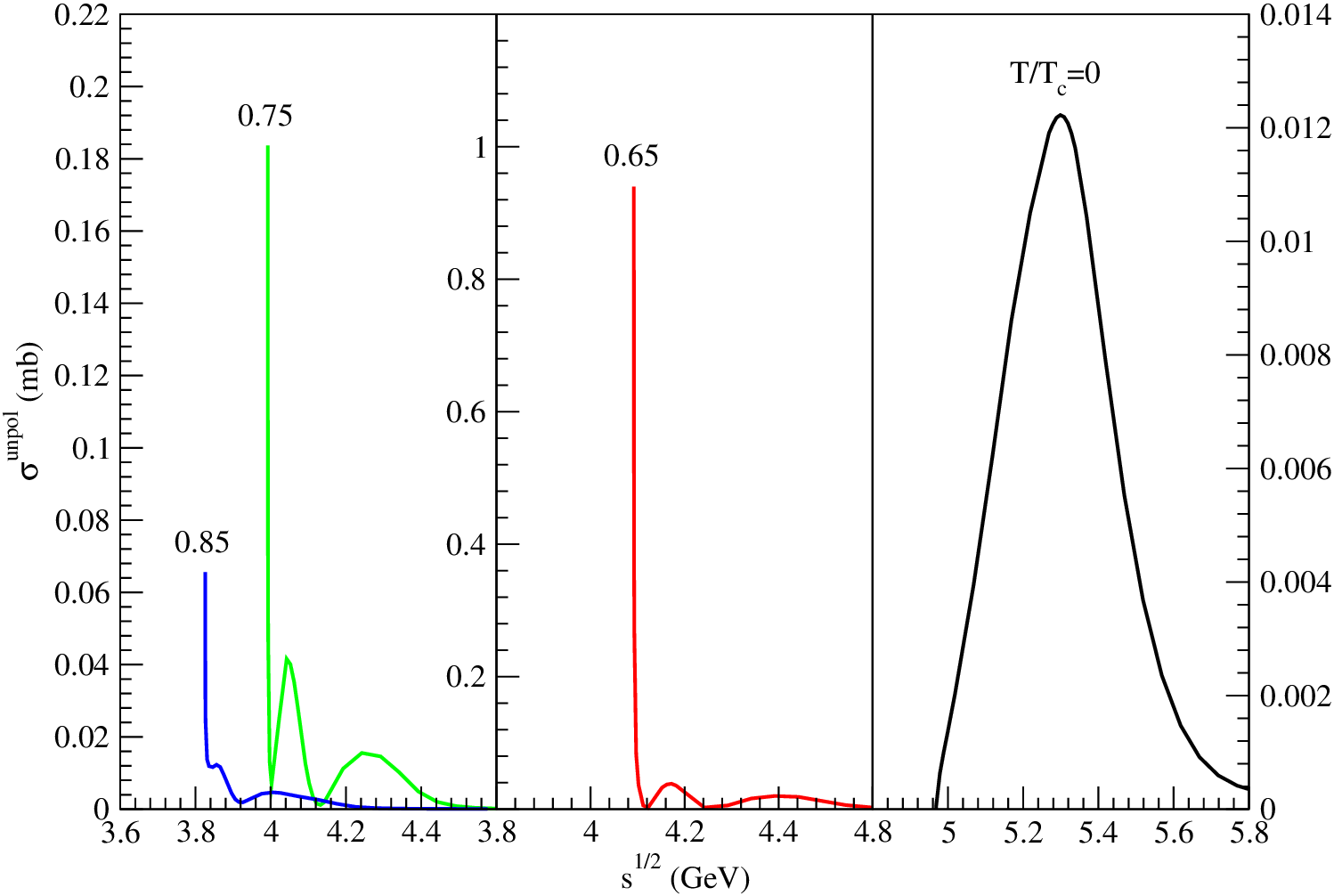}
\caption{Cross sections for $D_s^{\ast +} D_s^{\ast -} \rightarrow \eta
\psi(4415)$ at various temperatures.}
\label{fig27}
\end{figure}

Since the naive quark model was proposed by Gell-Mann and Zweig in 1964, cross 
sections for meson-meson scattering were first thought to be four times the 
cross
section for quark-quark scattering. However, this additive picture of cross
sections has been deemed to be approximate since QCD was
established. One reason is that the cross section for elastic quark-quark
scattering is not identical to the one for elastic quark-antiquark scattering
which involves quark-antiquark annihilation and creation \cite{XMCW}. 
Assuming that wave functions of quarks and antiquarks are
plane waves, the cross sections for quark-quark scattering and 
quark-antiquark scattering were obtained in perturbative QCD.
In low-energy meson-meson scattering, confinement of quarks and antiquarks in
mesons needs to be taken into account. Wave functions of
quarks and antiquarks are no longer plane waves, and cross sections for
meson-meson scattering look like 
those in Figs. 1-27. In the present work
low-energy meson-meson scattering produces two mesons. When the total 
center-of-mass energy ($\sqrt s$) increases, three, four, and more mesons are 
produced. Two-to-three meson-meson scattering, two-to-four meson-meson 
scattering, and so on lead to finite cross sections for meson-meson scattering.

At zero temperature, all reactions shown in Figs. 1-27 are endothermic.
However, at $T=0.65T_{\rm c}$, $0.75T_{\rm c}$, $0.85T_{\rm c}$, 
$0.9T_{\rm c}$, or $0.95T_{\rm c}$, a reaction may be endothermic or 
exothermic. Hence, we use the following two expressions to parametrize the
numerical cross sections shown in Figs. 1-27:
\begin{eqnarray}
\sigma^{\rm unpol}(\sqrt {s},T) & = &
a_1 \left( \frac {\sqrt {s} -\sqrt {s_0}}{b_1} \right)^{c_1}
\exp \left[ c_1 \left( 1-\frac {\sqrt {s} -\sqrt {s_0}}{b_1} \right) \right]
                   \notag   \\
& & + a_2 \left( \frac {\sqrt {s} -\sqrt {s_0}}{b_2} \right)^{c_2}
\exp \left[ c_2 \left( 1-\frac {\sqrt {s} -\sqrt {s_0}}{b_2} \right) \right] ,
\end{eqnarray}
for endothermic reactions and
\begin{eqnarray}
\sigma^{\rm unpol}(\sqrt {s},T) & = &
\frac{{\vec {P'}}^2}{{\vec P}^2}
\left\{a_1 \left( \frac {\sqrt {s} -\sqrt {s_0}}{b_1} \right)^{c_1}
\exp \left[ c_1 \left( 1-\frac {\sqrt {s} -\sqrt {s_0}}{b_1}
\right) \right]\right.
                   \notag   \\
& & \left. + a_2 \left( \frac {\sqrt {s} -\sqrt {s_0}}{b_2} \right)^{c_2}
\exp \left[ c_2 \left( 1-\frac {\sqrt {s} -\sqrt {s_0}}{b_2}
\right) \right]\right\} ,
\end{eqnarray}
for exothermic reactions. $\sqrt{s}_0$ is the threshold energy. 
In order to use the two parametrizations in the 
master rate equations, we require the separation ($d_0$) between the peak's
location on the $\sqrt s$ axis and the threshold energy and the square root 
($\sqrt{s_z}$) of the Mandelstam variable at which the cross section is 1/100 
of the peak cross section. Values of $a_1$, $b_1$, $c_1$, $a_2$, $b_2$, $c_2$,
$d_0$, and $\sqrt{s_z}$ are listed in Tables 1-9.

The expression on the right-hand side of Eq. (11) equals 0 at 
$\sqrt{s} = \sqrt{s_0}$, and thus can be used to parametrize numerical cross
sections for endothermic reactions. Denote the spins of mesons $c\bar{q}_2$,
$q_1\bar{c}$, $q_1\bar{q}_2$, and $c\bar{c}$ by $S_{c\bar{q}_2}$,
$S_{q_1\bar{c}}$, $S_{q_1\bar{q}_2}$, and $S_{c\bar{c}}$, respectively. If
the reaction $c\bar{q}_2 + q_1\bar{c} \to q_1\bar{q}_2 + c\bar{c}$ is
exothermic, its cross section can be related to the endothermic reaction
$q_1\bar{q}_2 + c\bar{c} \to c\bar{q}_2 + q_1\bar{c}$ using the detailed 
balance
\begin{equation}
\sigma_{c\bar{q}_2 + q_1\bar{c} \to q_1\bar{q}_2 + c\bar{c}}^{\rm unpol} =
\frac{(2S_{q_1\bar{q}_2}+1)(2S_{c\bar{c}}+1)}
{(2S_{c\bar{q}_2}+1)(2S_{q_1\bar c}+1)}
\frac{\vec{P}^{\prime~2}}{\vec{P}^2}
\sigma_{q_1\bar{q}_2 + c\bar{c} \to c\bar{q}_2 + q_1\bar{c}}^{\rm unpol} .
\end{equation}
Hence, the expression on the right-hand side of Eq. (12) has the factor 
$\vec{P}^{\prime~2} / \vec{P}^2$.

We study the production of $\psi(4040)$, $\psi(4160)$, and $\psi(4415)$ mesons
in central Pb-Pb collisions at the LHC. Hadronic matter produced in the
collisions exhibits cylindrical
symmetry, and the hydrodynamic equation is solved in terms of
the cylindrical polar
coordinates ($r$, $\phi$, $z$) \cite{GMKR}, where the $z$-axis in the rest 
frame of hadronic matter is set along the moving direction of a nucleus and 
passes through the nuclear center, $r$ is the distance from the fluid-element
center to the $z$-axis, and $\phi$ is the azimuth. 
With the shear viscosity given in Ref. \cite{NEP},
the hydrodynamic equation provides the temperature and the transverse
velocity of hadronic matter that expands.

In the source terms of the master rate equations, $n_D$, $n_{\bar D}$,
$n_{D^*}$, $n_{\bar{D}^*}$, $n_{D_s^+}$, $n_{D_s^-}$, $n_{D_s^{*+}}$, and
$n_{D_s^{*-}}$ are obtained from momentum distribution functions of charmed
mesons and charmed strange mesons. Unlike pion-pion scattering, cross sections
for pion scattering by open-charm mesons are small. Thermal states of 
open-charm mesons may not be established by such small cross sections.
We give a Lorentz-invariant form of the
momentum distribution functions of open-charm mesons,
\begin{equation}\label{distribution}
f_i(k_i) =\frac{1+\sum_{l=1}^{\infty} c_l 
(k_i\cdot u)^l}{e^{k_i\cdot u/T_{\rm dec}}-1},
\end{equation}
where $T_{\rm dec}$ is the inverse slope parameter. If 
$\sum_{l=1}^{\infty} c_l (k_i\cdot u)^l =0$, $f_i(k_i)$ becomes the 
Bose-Einstein distribution function. The term $\sum_{l=1}^{\infty} c_l 
(k_i\cdot u)^l$ indicates deviation from thermal equilibrium. After
fits to the experimental data \cite{ALICE5020D} of $dN/dp_T$ of prompt $D^+$, 
$D^0$, $D^{*+}$, and $D_s^+$ mesons at $p_T<8$ GeV/$c$ in central Pb-Pb 
collisions at $\sqrt{s_{NN}}=5.02$ TeV, the values of $l$ and $c_l$ for
$D^+$, $D^0$, and $D^{*+}$ mesons are listed in Ref. \cite{lyLXW}, and those 
for $D_s^+$ mesons here:
\begin{displaymath}
l=15,~~~~~c_l=6\times 10^{-17};
\end{displaymath}
\begin{displaymath}
l \ne 15,~~~~~c_l=0.
\end{displaymath}
$T_{\rm dec}$ determined from the experimental data is 0.1686 GeV, and the 
value is close to the critical temperature. This means that open-charm
mesons decouple early from hadronic matter. We thus use the momentum 
distribution functions (Eq. (14)) to obtain the average cross section defined
in Eq. (3) for the first twenty-four terms on the right-hand side of Eq. (2). 
Temperature dependence of the average cross section weighted by 
the relative velocity arises from temperature dependence of 
$\sigma_{ij \to i'j'}$ and of $v_{ij}$.

In central Pb-Pb collisions at $\sqrt{s_{NN}}$=5.02~\rm{TeV}, hadronic matter
is produced at the proper time 10.05 fm/$c$ \cite{lyLXW}. We start solving
the hydrodynamic equation and the master rate equations at the time and get
number densities at kinetic freeze-out. Using the momentum distribution
functions $1/(e^{k_i\cdot u/T}-1)$ for pions, kaons, and vector kaons in the
Cooper-Frye formula \cite{CF}, fits to the experimental data of momentum
spectra \cite{ALICE5020CH,ALICE5020KA} to obtain the freeze-out time 21.07 
fm/$c$ and the freeze-out temperature 0.126 GeV.
The average cross sections in the last twenty-four terms on the right-hand side
of Eq. (2) involve momentum distribution functions of $\psi(4040)$, 
$\psi(4160)$, and $\psi(4415)$. Currently, we assume the distribution
functions have the form $\lambda_i/(e^{k_i\cdot u/T_i}-1)$, where 
$\lambda_i$
are constants and the inverse slope parameters $T_i$ equal the dissociation
temperatures of $\psi(4040)$, $\psi(4160)$, and $\psi(4415)$. We need not know
values of $\lambda_i$ because $\lambda_i$ in the numerator and in the 
denominator in Eq. (3) cancel each other out.
Variation of number densities with respect to the proper time ($\tau$)
at $r=0$ fm is drawn as upper solid, upper dashed, and upper dotted curves
in Fig. 28, and $r$ dependence of number densities at kinetic freeze-out is 
plotted as upper solid, upper dashed, and upper dotted curves
in Fig. 29. Number densities of $\psi(4040)$, $\psi(4160)$, and $\psi(4415)$
mesons were obtained with an early version of FORTRAN code which numerically
solves the master rate equations in Ref. \cite{lyLXW}. After several errors
are removed, a new version is used to calculate number densities, which are
smaller than those shown in  Ref. \cite{lyLXW}. When the proper time increases 
from 10.83 fm/$c$, 11.38 fm/$c$, and 13.95 fm/$c$, respectively,
the number densities of $\psi(4040)$, $\psi(4160)$, and $\psi(4415)$ 
increase. However, the three mesons produced at $r=0$ fm spread out, and this
reduces the number densities. When the reduced amount exceeds the increased
amount, the number densities decrease as seen in Fig. 28.

The upper solid, upper dashed, and upper dotted curves shown in Figs. 28 and 29
result from reactions between two charmed mesons, between a charmed meson and
a charmed strange meson, and between two charmed strange mesons as well as
their reverse reactions. To show 
contributions of charmed strange mesons in producing $\psi(4040)$, 
$\psi(4160)$, and $\psi(4415)$ mesons, we plot lower solid, lower dashed, and
lower dotted curves that only result from reactions between two charmed mesons.
From the lower curves to the upper curves, changes of the number densities are 
obvious.
For example, the $\psi(4040)$, $\psi(4160)$, and $\psi(4415)$ number densities
at kinetic freeze-out at $r=0$ increase by 19.2\%, 14.5\%, and 18.4\%, 
respectively, owing to the reactions of charmed strange mesons.

\begin{table*}[htbp]
\caption{\label{table1}Values of the parameters in Eq. (11) for 
$D_s^+ \bar{D} \to K^\ast \psi(4040)$, $K^\ast \psi(4160)$, and 
$K^\ast \psi(4415)$. $a_1$ and $a_2$ are
in units of millibarns; $b_1$, $b_2$, $d_0$, and $\sqrt{s_{\rm z}}$ are
in units of GeV; $c_1$ and $c_2$ are dimensionless.}
\tabcolsep=5pt
\begin{tabular}{cccccccccc}
  \hline
  \hline
final state & $T/T_{\rm c} $ & $a_1$ & $b_1$ & $c_1$ & $a_2$ & $b_2$ & $c_2$ &
$d_0$ & $\sqrt{s_{\rm z}} $\\
\hline
 $K^\ast \psi(4040)$
   & 0     & 0.01  & 0.03  & 0.42  & 0.08  & 0.07   & 0.51 & 0.06 & 5.88\\
   & 0.65  & 0.011 & 0.183 & 1.91  & 0.023 & 0.0243 & 0.49 & 0.03 & 5.06\\
   & 0.75  & 0.008 & 0.07  & 0.26  & 0.013 & 0.024  & 0.9  & 0.03 & 4.87\\
   & 0.85  & 0.014 & 0.048 & 0.25  & 0.072 & 0.008  & 0.56 & 0.01 & 4.19\\
   & 0.9   & 0.21  & 0.007 & 0.71  & 0.76  & 0.031  & 3    & 0.03 & 3.62\\
   & 0.95  & 0.81  & 0.012 & 10.7  & 0.88  & 3.37  & 0.132 & 0.01 & 3.4\\
  \hline
 $K^\ast \psi(4160)$
   & 0     & 0.004 & 0.02 & 0.55  & 0.023 & 0.07  & 0.48  & 0.05  & 5.95\\
   & 0.65  & 0.001 & 0.019 & 0.44 & 0.002 & 0.113 & 1.08  & 0.08  & 5.02\\
   & 0.75  & 0.0001 & 0.0003 & 1.22 & 0.00142 & 0.0727 & 0.7  & 0.08 & 4.77\\
   & 0.85  & 0.0016 & 0.024  & 0.29 & 0.0056  & 0.02   & 1.66 & 0.02 & 4.17\\
   & 0.9   & 0.08   & 0.0003 & 0.43 & 0.34    & 0.0086 & 1.11 & 0.01 & 3.66\\
   & 0.95  & 0.07   & 0.003  & 0.35 & 0.44    & 0.021  & 1.86 & 0.02 & 3.4\\
  \hline
 $K^\ast \psi(4415)$
   & 0     & 0.005 & 0.025 & 0.59  & 0.023 & 0.088 & 0.47  & 0.06  & 6.41\\
   & 0.65  & 0.018 & 0.068 & 0.37  & 0.011 & 0.022 & 0.91  & 0.03  & 5.14\\
   & 0.75  & 0.006 & 0.123 & 0.19  & 0.019 & 0.03  & 0.8   & 0.035 & 4.93\\
   & 0.85  & 0.007 & 0.03  & 0.07  & 0.063 & 0.01  & 0.64  & 0.01  & 4.33\\
  \hline
  \hline
\end{tabular}
\end{table*}
\begin{table*}[htbp]
\caption{\label{table2}Values of the parameters in Eqs. (11) and (12) for 
$D_s^+ \bar{D}^* \to K \psi(4040)$, $K \psi(4160)$, and 
$K \psi(4415)$. $a_1$ and $a_2$ are
in units of millibarns; $b_1$, $b_2$, $d_0$, and $\sqrt{s_{\rm z}}$ are
in units of GeV; $c_1$ and $c_2$ are dimensionless.}
\tabcolsep=5pt
\begin{tabular}{cccccccccc}
  \hline
  \hline
final state & $T/T_{\rm c} $ & $a_1$ & $b_1$ & $c_1$ & $a_2$ & $b_2$ & $c_2$ &
$d_0$ & $\sqrt{s_{\rm z}} $\\
\hline
 $K \psi(4040)$
   & 0     & 0.014 & 0.081 & 0.585 & 0.066 & 0.236  & 3.05  & 0.22  & 5.42\\
   & 0.65  & 0.059 & 0.001 & 0.12  & 0.502 & 0.017  & 0.91  & 0.015 & 4.34\\
   & 0.75  & 0.053 & 0.0001 & 0.03 & 0.905 & 0.0134 & 0.731 & 0.015 & 4.12\\
   & 0.85  & 0.042 & 0.0306 & 0.276 & 1.38 & 0.0244 & 2.07  & 0.025 & 3.82\\
   & 0.9   & 0.46  & 0.00058 & 0.57 & 2.38 & 0.0266 & 3.2   & 0.025 & 3.67\\
   & 0.95  & 1.55  & 0.0009  & 0.48 & 1.14 & 0.0257 & 4.6   & 0.001 & 3.48\\
  \hline
  $K \psi(4160)$
   & 0     & 0.022 & 0.04  & 0.54  & 0.044 & 0.18  & 2.24 & 0.15  & 5.53\\
   & 0.65  & 0.04  & 0.051 & 2.16  & 0.15  & 0.03  & 0.52 & 0.035 & 4.29\\
   & 0.75  & 0.11  & 0.014 & 0.49  & 0.14  & 0.044 & 1.53 & 0.035 & 4.09\\
   & 0.85  & 0.01  & 0.01  & 0.65  & 0.17  & 0.042 & 2.23 & 0.04  & 3.85\\
   & 0.9   & 0.005 & 0.001 & 1.02  & 0.129 & 0.047 & 3.33 & 0.045 & 3.7\\
   & 0.95  & 0.01  & 0.09  & 0.09  & 0.083 & 0.053 & 7.8  & 0.05  & 3.5\\
  \hline
  $K \psi(4415)$
   & 0     & 0.0151 & 0.0658 & 0.561 & 0.0418 & 0.213 & 2.99 & 0.2   & 5.9\\
   & 0.65  & 0.114  & 0.012  & 0.54  & 0.053  & 0.197 & 5.19 & 0.015 & 4.58 \\
   & 0.75  & 0.179  & 0.0119 & 0.56  & 0.056  & 0.164 & 4.8  & 0.015 & 4.31\\
   & 0.85  & 0.105  & 0.032  & 0.35  & 0.145  & 0.018 & 1.84 & 0.02  & 4.01\\
  \hline
  \hline
\end{tabular}
\end{table*}
\begin{table*}[htbp]
\caption{\label{table3}The same as Table 1 except for 
$D_s^+ \bar{D}^* \to K^\ast \psi(4040)$, $K^\ast \psi(4160)$, and 
$K^\ast \psi(4415)$.}
\tabcolsep=5pt
\begin{tabular}{cccccccccc}
  \hline
  \hline
final state & $T/T_{\rm c} $ & $a_1$ & $b_1$ & $c_1$ & $a_2$ & $b_2$ & $c_2$ &
$d_0$ & $\sqrt{s_{\rm z}} $\\
\hline
 $K^\ast \psi(4040)$
   & 0     & 0.04  & 0.02  & 0.46  & 0.068 & 0.1   & 0.96 & 0.06  & 5.78\\
   & 0.65  & 0.008 & 0.1   & 0.19  & 0.028 & 0.032 & 0.76 & 0.035 & 4.94\\
   & 0.75  & 0.002 & 0.041 & 0.1   & 0.027 & 0.029 & 0.6  & 0.03  & 4.73\\
   & 0.85  & 0.06  & 0.002 & 0.12  & 0.37  & 0.014 & 1.09 & 0.015 & 3.83\\
   & 0.9   & 0.18  & 0.003 & 0.47  & 0.5   & 0.04  & 2.79 & 0.04  & 3.63\\
   & 0.95  & 0.32  & 0.035 & 0.1   & 0.88  & 0.008 & 1.39 & 0.01  & 3.39\\
  \hline
 $K^\ast \psi(4160)$
   & 0     & 0.009  & 0.016 & 0.41  & 0.024  & 0.08  & 0.66  & 0.05 & 5.89\\
   & 0.65  & 0.0002 & 0.028 & 0.04  & 0.0041 & 0.07  & 0.73  & 0.07 & 4.93\\
   & 0.75  & 0.001  & 0.0024 & 0.82 & 0.003  & 0.076  & 1.04 & 0.09 & 4.67\\
   & 0.85  & 0.007  & 0.0495 & 3.76 & 0.0275 & 0.0027 & 0.65 & 0.01 & 4.07\\
   & 0.9   & 0.015  & 0.0001 & 0.025 & 0.407 & 0.0073 & 0.609 & 0.01  & 3.63\\
   & 0.95  & 0.06   & 0.005  & 0.75  & 0.18  & 0.026  & 2.77  & 0.025 & 3.46\\
  \hline
 $K^\ast \psi(4415)$
   & 0     & 0.01  & 0.04  & 0.71  & 0.02  & 0.08  & 0.43  & 0.05 & 6.2\\
   & 0.65  & 0.006 & 0.148 & 0.25  & 0.03  & 0.031 & 0.56  & 0.04 & 5.02\\
   & 0.75  & 0.0024 & 0.076 & 19.8 & 0.0235 & 0.0316 & 0.443 & 0.04  & 4.82\\
   & 0.85  & 0.052  & 0.003 & 0.3  & 0.207  & 0.017  & 1.17  & 0.015 & 3.93\\
  \hline
  \hline
\end{tabular}
\end{table*}
\begin{table*}[htbp]
\caption{\label{table4}The same as Table 2 except for 
$D_s^{\ast +}\bar{D} \to K \psi(4040)$, $K \psi(4160)$, and $K \psi(4415)$.}
\tabcolsep=5pt
\begin{tabular}{cccccccccc}
  \hline
  \hline
final state & $T/T_{\rm c} $ & $a_1$ & $b_1$ & $c_1$ & $a_2$ & $b_2$ & $c_2$ &
$d_0$ & $\sqrt{s_{\rm z}} $\\
\hline
 $K \psi(4040)$
   & 0     & 0.018 & 0.074 & 0.589 & 0.097 & 0.252  & 3.05 & 0.25 & 5.49\\
   & 0.65  & 0.08  & 0.003 & 0.34  & 0.9   & 0.031  & 1.38 & 0.03 & 4\\
   & 0.75  & 0.32  & 0.002 & 0.52  & 3.07  & 0.0249 & 1.59 & 0.03 & 4.16\\
   & 0.85  & 0.19  & 0.006 & 0.61  & 1.57  & 0.023  & 2.21 & 0.02 & 3.9\\
   & 0.9   & 0.118 & 0.0009 & 0.559 & 0.89 & 0.0242 & 4    & 0.025 & 3.69\\
   & 0.95  & 0.2   & 0.003  & 1.42  & 0.88 & 0.001  & 0.52 & 0.001 & 3.43\\
  \hline
 $K \psi(4160)$
   & 0     & 0.033 & 0.044 & 0.54  & 0.062 & 0.189 & 2.35  & 0.17  & 5.48\\
   & 0.65  & 0.08  & 0.04  & 0.54  & 0.13  & 0.07  & 2.23  & 0.06  & 4.33\\
   & 0.75  & 0.04  & 0.017 & 0.64  & 0.23  & 0.067 & 2.55  & 0.065 & 4.13\\
   & 0.85  & 0.02  & 0.06  & 0.95  & 0.23  & 0.08  & 5.2   & 0.075 & 3.89\\
   & 0.9   & 0.15  & 0.073 & 18    & 0.069 & 0.0129 & 1.7  & 0.07  & 3.72\\
   & 0.95  & 0.062 & 0.066 & 18.1  & 0.189 & 0.0044 & 1.18 & 0.01  & 3.47\\
  \hline
 $K \psi(4415)$
   & 0     & 0.02  & 0.072 & 0.554 & 0.059 & 0.228  & 2.96  & 0.22  & 5.86\\
   & 0.65  & 0.073 & 0.253 & 18    & 0.224 & 0.0142 & 0.78  & 0.025 & 4.6\\
   & 0.75  & 0.07  & 0.003 & 0.49  & 0.3   & 0.03   & 1.26  & 0.03  & 3.94\\
   & 0.85  & 0.078 & 0.009 & 0.27  & 0.301 & 0.05   & 5.6   & 0.05  & 4.03\\
  \hline
  \hline
\end{tabular}
\end{table*}
\begin{table*}[htbp]
\caption{\label{table5}The same as Table 1 except for 
$D_s^{\ast +} \bar{D} \to K^\ast \psi(4040)$, $K^\ast \psi(4160)$, and 
$K^\ast \psi(4415)$.}
\tabcolsep=5pt
\begin{tabular}{cccccccccc}
  \hline
  \hline
final state & $T/T_{\rm c} $ & $a_1$ & $b_1$ & $c_1$ & $a_2$ & $b_2$ & $c_2$ &
$d_0$ & $\sqrt{s_{\rm z}} $\\
\hline
 $K^\ast \psi(4040)$
   & 0     & 0.005 & 0.17  & 0.47  & 0.099 & 0.06  & 0.5   & 0.06  & 5.81\\
   & 0.65  & 0.003 & 0.026 & 0.03  & 0.05  & 0.034 & 0.57  & 0.04  & 4.93\\
   & 0.75  & 0.034 & 0.01  & 0.54  & 0.043 & 0.03  & 0.46  & 0.015 & 4.63\\
   & 0.85  & 0.2   & 0.03  & 0.33  & 0.41  & 0.04  & 4.79  & 0.04  & 3.85\\
   & 0.9   & 0.145 & 0.0001 & 0.05  & 2.63 & 0.0074 & 0.71 & 0.01 & 3.63\\
   & 0.95  & 0.128 & 0.0001 & 0.041 & 2.08 & 0.0085 & 0.63 & 0.01 & 3.4\\
  \hline
 $K^\ast \psi(4160)$
   & 0     & 0.01  & 0.19  & 2.03  & 0.03  & 0.04  & 0.5   & 0.05  & 5.91\\
   & 0.65  & 0.001 & 0.044 & 0.33  & 0.004 & 0.108 & 1.12  & 0.1   & 4.98\\
   & 0.75  & 0.0049 & 0.0424 & 0.573 & 0.00193 & 0.175 & 4.08 & 0.04 & 4.68\\
   & 0.85  & 0.07  & 0.001 & 0.03  & 0.28  & 0.01  & 1.44  & 0.01  & 3.92\\
   & 0.9   & 0.03  & 0.003 & 0.19  & 0.25  & 0.02  & 1.67  & 0.02  & 3.73\\
   & 0.95  & 0.004 & 0.006 & 0.94  & 0.05  & 0.034 & 3.36  & 0.035 & 3.58\\
  \hline
 $K^\ast \psi(4415)$
   & 0     & 0.018  & 0.13  & 0.8  & 0.021 & 0.03  & 0.48  & 0.05 & 6.3\\
   & 0.65  & 0.004  & 0.008 & 0.35 & 0.053 & 0.037 & 0.51  & 0.04 & 5.01\\
   & 0.75  & 0.016  & 0.011 & 0.84 & 0.045 & 0.027 & 0.39  & 0.02 & 4.74\\
   & 0.85  & 0.07   & 0.007 & 0.79 & 0.33  & 0.04  & 2.71  & 0.04 & 3.94\\
  \hline
  \hline
\end{tabular}
\end{table*}
\begin{table*}[htbp]
\caption{\label{table6}The same as Table 2 except for 
$D_s^{\ast +}\bar{D}^\ast \to K \psi(4040)$, $K \psi(4160)$, and 
$K \psi(4415)$.}
\tabcolsep=5pt
\begin{tabular}{cccccccccc}
  \hline
  \hline
final state & $T/T_{\rm c} $ & $a_1$ & $b_1$ & $c_1$ & $a_2$ & $b_2$ & $c_2$ &
$d_0$ & $\sqrt{s_{\rm z}} $\\
\hline
 $K \psi(4040)$
   & 0     & 0.011 & 0.053   & 0.56  & 0.031 & 0.321  & 5.24  & 0.31  & 5.51\\
   & 0.65  & 0.011 & 0.00112 & 0.54  & 0.233 & 0.0439 & 3.73  & 0.04  & 4.43\\
   & 0.75  & 0.01  & 0.014   & 0.87  & 0.26  & 0.03   & 2.26  & 0.03  & 4.23\\
   & 0.85  & 0.016 & 0.059   & 0.4   & 0.063 & 0.024  & 3.6   & 0.02  & 3.94\\
   & 0.9   & 0.037 & 0.003   & 1     & 0.099 & 0.001  & 0.499 & 0.001 & 3.67\\
   & 0.95  & 0.128 & 0.03    & 15    & 0.79  & 0.00135 & 0.75 & 0.001 & 3.36\\
  \hline
 $K \psi(4160)$
   & 0     & 0.002  & 0.006 & 0.395 & 0.027  & 0.148 & 1.19  & 0.19 & 5.59\\
   & 0.65  & 0.0015 & 0.008 & 0.55  & 0.0849 & 0.109 & 6.1   & 0.1  & 4.67\\
   & 0.75  & 0.0008 & 0.0054 & 0.496 & 0.0679 & 0.0852 & 5.3   & 0.08 & 4.47\\
   & 0.85  & 0.0062 & 0.097  & 1.66  & 0.0027 & 0.0089 & 0.962 & 0.07 & 4.21\\
   & 0.9   & 0.0024 & 0.025  & 0.46  & 0.063  & 0.013  & 2.92  & 0.01 & 3.8\\
   & 0.95  & 0.002  & 0.0008 & 0.142 & 0.078  & 0.0125 & 3.55  & 0.01 & 3.41\\
  \hline
 $K \psi(4415)$
   & 0     & 0.0027 & 0.055  & 0.59  & 0.0218 & 0.24  & 2.8  & 0.24 & 5.94\\
   & 0.65  & 0.044  & 0.0052 & 0.625 & 0.223  & 0.088 & 10.5 & 0.09 & 4.54\\
   & 0.75  & 0.195 & 0.0727 & 8.79 & 0.0156 & 0.00282 & 0.603 & 0.07  & 4.34\\
   & 0.85  & 0.052 & 0.001  & 1.27 & 0.111  & 0.002   & 0.32  & 0.001 & 3.98\\
  \hline
  \hline
\end{tabular}
\end{table*}
\begin{table*}[htbp]
\caption{\label{table7}The same as Table 1 except for 
$D_s^{\ast +} \bar{D}^\ast \to K^\ast \psi(4040)$, $K^\ast \psi(4160)$, and 
$K^\ast \psi(4415)$.}
\tabcolsep=5pt
\begin{tabular}{cccccccccc}
  \hline
  \hline
final state & $T/T_{\rm c} $ & $a_1$ & $b_1$ & $c_1$ & $a_2$ & $b_2$ & $c_2$ &
$d_0$ & $\sqrt{s_{\rm z}} $\\
\hline
 $K^\ast \psi(4040)$
   & 0     & 0.051  & 0.015 & 0.43  & 0.138 & 0.09  & 0.93 & 0.07 & 5.71\\
   & 0.65  & 0.0001 & 0.002 & 0.002 & 0.305 & 0.021 & 0.5  & 0.02 & 4.63\\
   & 0.75  & 0.05   & 0.002 & 0.29  & 0.7   & 0.028 & 1.27 & 0.03 & 4.19\\
   & 0.85  & 0.49   & 0.03  & 0.04  & 2.53  & 0.01  & 1.57 & 0.01 & 3.83\\
   & 0.9   & 0.01   & 0.001 & 0.01  & 3.24  & 0.007 & 0.48 & 0.01 & 3.62\\
   & 0.95  & 0.16   & 0.02  & 0.95  & 2.98  & 0.006 & 0.43 & 0.01 & 3.37\\
  \hline
 $K^\ast \psi(4160)$
   & 0     & 0.01   & 0.03  & 0.47  & 0.05   & 0.06   & 0.5  & 0.05  & 5.81\\
   & 0.65  & 0.0056 & 0.003 & 0.51  & 0.0178 & 0.108  & 1.7  & 0.1   & 4.84\\
   & 0.75  & 0.025  & 0.096 & 3.74  & 0.094  & 0.0042 & 0.57 & 0.01  & 4.46\\
   & 0.85  & 0.041  & 0.001 & 0.05  & 0.361  & 0.016  & 1.51 & 0.015 & 3.96\\
   & 0.9   & 0.014  & 0.012 & 0.71  & 0.037  & 0.031  & 3.72 & 0.03  & 3.84\\
   & 0.95  & 0.00234 & 0.22 & 0.23  & 0.00522 & 0.0336 & 3.79 & 0.04 & 3.67\\
  \hline
 $K^\ast \psi(4415)$
   & 0     & 0.01  & 0.11  & 0.63  & 0.05  & 0.05  & 0.49  & 0.05 & 6.13\\
   & 0.65  & 0.04  & 0.04  & 0.2   & 0.19  & 0.023 & 0.67  & 0.02 & 4.75\\
   & 0.75  & 0.12  & 0.008 & 0.48  & 0.37  & 0.034 & 1.61  & 0.03 & 4.3\\
   & 0.85  & 0.25  & 0.065 & 6.2   & 0.42  & 0.009 & 0.59  & 0.01 & 3.95\\
  \hline
  \hline
\end{tabular}
\end{table*}
\begin{table*}[htbp]
\caption{\label{table8}The same as Table 2 except for 
$D_s^+ D_s^{\ast -} \to \eta \psi(4040)$, $\eta \psi(4160)$, and 
$\eta \psi(4415)$.}
\tabcolsep=5pt
\begin{tabular}{cccccccccc}
  \hline
  \hline
final state & $T/T_{\rm c} $ & $a_1$ & $b_1$ & $c_1$ & $a_2$ & $b_2$ & $c_2$ &
$d_0$ & $\sqrt{s_{\rm z}} $\\
\hline
 $\eta \psi(4040)$
   & 0     & 0.009 & 0.113 & 0.59  & 0.037 & 0.285 & 4.7  & 0.31 & 5.46\\
   & 0.65  & 0.14  & 0.009 & 0.55  & 0.8   & 0.037 & 1.66 & 0.03 & 4.44\\
   & 0.75  & 0.3   & 0.011 & 0.53  & 0.54  & 0.03  & 1.73 & 0.03 & 4.29\\
   & 0.85  & 0.19  & 0.009 & 0.53  & 0.23  & 0.023 & 1.61 & 0.01 & 4.07\\
   & 0.9   & 0.032 & 0.099 & 0.25  & 0.109 & 0.014 & 0.93 & 0.01 & 3.9\\
   & 0.95  & 0.03  & 0.04  & 0.6   & 0.022 & 0.012 & 2.82 & 0.02 & 3.75\\
  \hline
 $\eta \psi(4160)$
   & 0     & 0.012  & 0.053 & 0.54  & 0.024 & 0.205 & 2.71 & 0.19  & 5.62\\
   & 0.65  & 0.03   & 0.057 & 0.58  & 0.07  & 0.078 & 2.69 & 0.075 & 4.48\\
   & 0.75  & 0.0014 & 0.057 & 0.55  & 0.19  & 0.088 & 3.33 & 0.09  & 4.26\\
   & 0.85  & 0.001  & 0.23  & 0.44  & 0.064 & 0.068 & 3.9  & 0.05  & 4.19\\
   & 0.9   & 0.00023 & 0.0024 & 0.57  & 0.0278 & 0.0551 & 5.04 & 0.05 & 3.86\\
   & 0.95  & 0.00028 & 0.0044 & 0.695 & 0.0162 & 0.0529 & 7.1  & 0.05 & 3.62\\
  \hline
 $\eta \psi(4415)$
   & 0     & 0.0053 & 0.065 & 0.58  & 0.0242 & 0.24   & 3.28  & 0.27  & 5.86\\
   & 0.65  & 0.038  & 0.271 & 21.3  & 0.117  & 0.0188 & 0.7   & 0.035 & 4.64\\
   & 0.75  & 0.07   & 0.026 & 0.26  & 0.15   & 0.055  & 3.68  & 0.05  & 4.41\\
   & 0.85  & 0.08   & 0.022 & 0.55  & 0.17   & 0.047  & 3.69  & 0.05  & 4.17\\
  \hline
  \hline
\end{tabular}
\end{table*}
\begin{table*}[htbp]
\caption{\label{table9}The same as Table 2 except for 
$D_s^{\ast +} D_s^{\ast -} \to \eta \psi(4040)$, $\eta \psi(4160)$, and 
$\eta \psi(4415)$.}
\tabcolsep=5pt
\begin{tabular}{cccccccccc}
  \hline
  \hline
final state & $T/T_{\rm c} $ & $a_1$ & $b_1$ & $c_1$ & $a_2$ & $b_2$ & $c_2$ &
$d_0$ & $\sqrt{s_{\rm z}} $\\
\hline
 $\eta \psi(4040)$
   & 0     & 0.0097 & 0.042  & 0.58  & 0.0197 & 0.365 & 7.7  & 0.37 & 5.5\\
   & 0.65  & 0.016  & 0.245  & 16.1  & 0.0343 & 0.026 & 1.33 & 0.03 & 4.63\\
   & 0.75  & 0.055  & 0.0154 & 0.81  & 0.025  & 0.201 & 27   & 0.03 & 4.45\\
   & 0.85  & 0.0118  & 0.011  & 0.69  & 0.0064  & 0.16  & 6.9  & 0.01 & 4.42\\
   & 0.9   & 0.00076 & 0.026  & 0.75  & 0.00145 & 0.139 & 3.9  & 0.1  & 4.42\\
   & 0.95  & 0.0007  & 0.004  & 0.39  & 0.0009  & 0.083 & 1.58 & 0.1  & 4.21\\
  \hline
 $\eta \psi(4160)$
   & 0     & 0.005  & 0.1   & 0.55  & 0.0105 & 0.229  & 3.25 & 0.22 & 5.64\\
   & 0.65  & 0.0046 & 0.842 & 0.496 & 0.0259 & 0.103  & 4.15 & 0.1  & 4.86\\
   & 0.75  & 0.003  & 0.147 & 0.52  & 0.018  & 0.079  & 3.46 & 0.07 & 4.71\\
   & 0.85  & 0.002  & 0.128 & 0.56  & 0.003  & 0.075  & 3.24 & 0.05 & 4.52\\
   & 0.9   & 0.0003  & 0.4   & 0.5  & 0.0009   & 0.106 & 1.9 & 0.1  & 4.33\\
   & 0.95  & 0.00002 & 0.002 & 0.33 & 0.000475 & 0.132 & 1.3 & 0.15 & 4.08\\
  \hline
 $\eta \psi(4415)$
   & 0     & 0.002 & 0.14  & 0.61  & 0.0101 & 0.29  & 4.26  & 0.33 & 5.92\\
   & 0.65  & 0.042 & 0.08  & 11.7  & 0.03   & 0.31  & 15.9  & 0.08 & 4.77\\
   & 0.75  & 0.025 & 0.275 & 18.9  & 0.032  & 0.059 & 7.4   & 0.06 & 4.6\\
   & 0.85  & 0.0052 & 0.227 & 13.3 & 0.0033 & 0.036 & 0.82  & 0.04 & 4.64\\
  \hline
  \hline
\end{tabular}
\end{table*}

\newpage
\begin{figure}[htbp]
  \centering
    \includegraphics[scale=0.6]{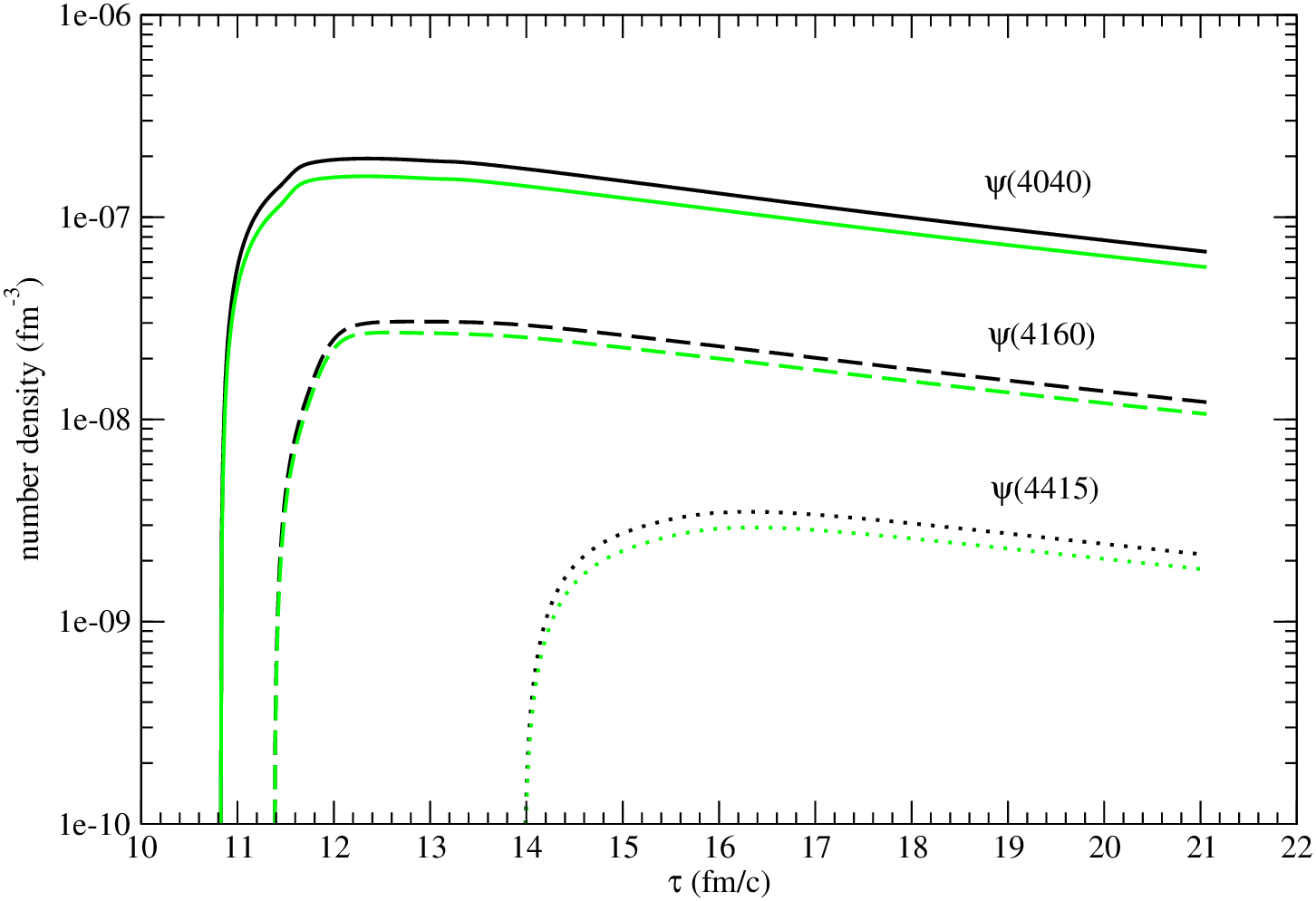}
\caption{Number densities as functions of $\tau$ at $r=0$ fm. The upper solid, 
upper dashed, and upper dotted curves result from reactions between two
open-charm mesons and their reverse reactions, and the lower solid, lower 
dashed, and lower dotted curves from reactions between two charmed mesons and
their reverse reactions.}
\label{fig28}
\end{figure}

\newpage
\begin{figure}[htbp]
  \centering
    \includegraphics[scale=0.6]{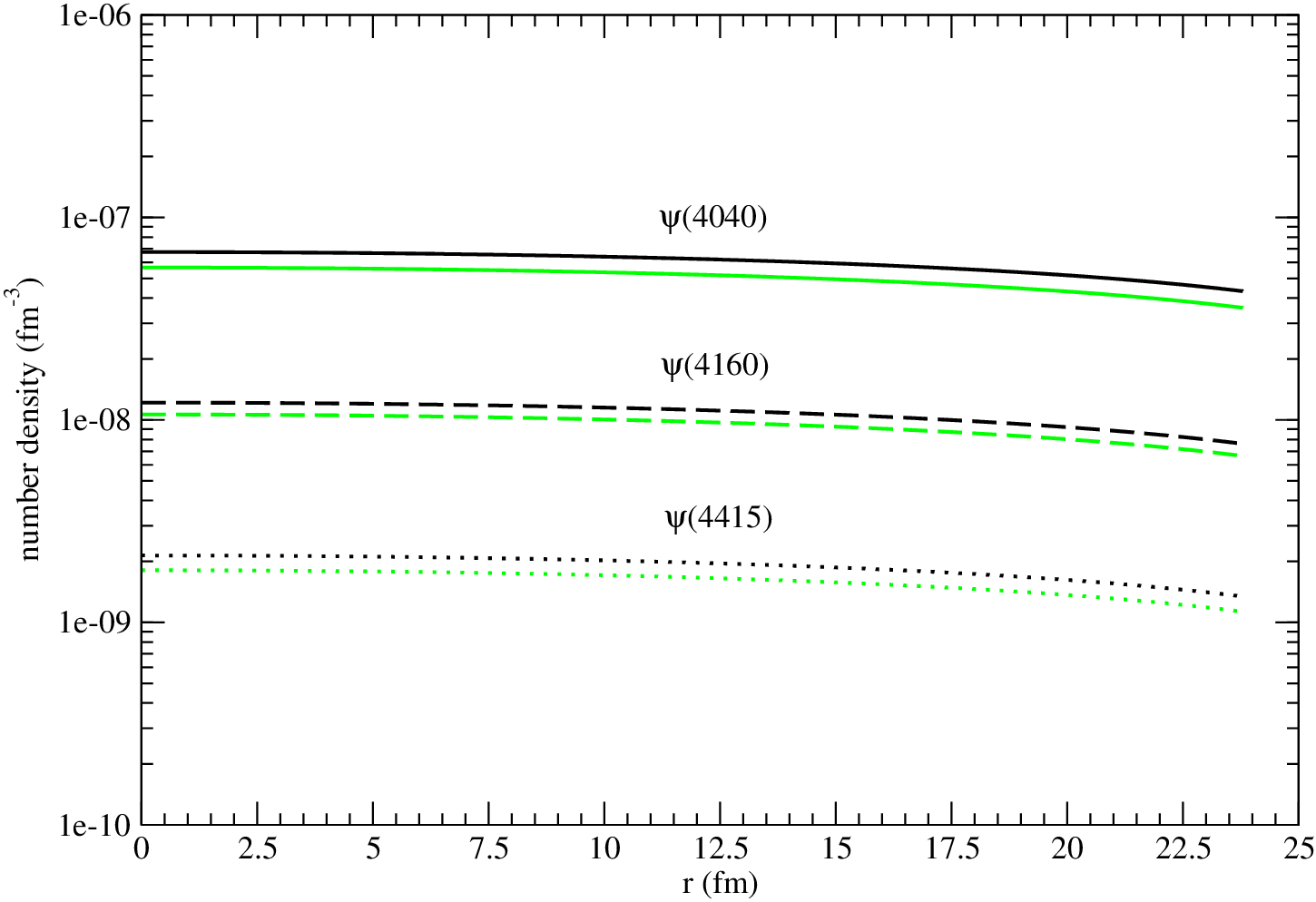}
\caption{The same as Fig. 28, but for $r$ dependence at kinetic freeze-out.}
\label{fig29}
\end{figure}

The potential given in Eq. (10) shows explicit dependence on temperature.
The Schr\"odinger equation with the potential gives temperature dependence
of meson masses and mesonic quark-antiquark relative-motion wave functions.
Since $\sqrt{s_0}$, $\sqrt s$, $\mid \vec{P} \mid$, and 
$\mid \vec{P}^\prime \mid$ relate to the meson masses, they
depend on temperature. Consequently, the cross sections for the production
of $\psi (4040)$, $\psi (4160)$, and $\psi (4415)$ mesons depend on 
temperature as seen in Figs. 1-27. When hadronic matter expands,
contributions of the twenty-seven reactions to the charmonium production differ
at different temperatures. This may also be understood
from the maximum of the five peak cross sections of an endothermic reaction,
which correspond to the
temperatures $0.65T_{\rm c}$, $0.75T_{\rm c}$, $0.85T_{\rm c}$, $0.9T_{\rm c}$,
and $0.95T_{\rm c}$. For example, the largest peak cross section of 
$D_s^+ \bar{D} \to K^\ast \psi (4160)$ 
($D_s^+ \bar{D}^\ast \to K \psi (4160)$, 
$D_s^+ \bar{D}^\ast \to K^\ast \psi (4160)$, 
$D_s^{*+} \bar{D} \to K^\ast \psi (4160)$,
$D_s^{*+} \bar{D}^\ast \to K^\ast \psi (4160)$)
appears at the temperature $0.95T_{\rm c}$ ($0.75T_{\rm c}$, $0.9T_{\rm c}$,
$0.85T_{\rm c}$, $0.85T_{\rm c}$). The endothermic reaction
$D_s^+ \bar{D} \to K^\ast \psi (4160)$ 
($D_s^+ \bar{D}^\ast \to K \psi (4160)$, 
$D_s^+ \bar{D}^\ast \to K^\ast \psi (4160)$, 
$D_s^{*+} \bar{D} \to K^\ast \psi (4160)$,
$D_s^{*+} \bar{D}^\ast \to K^\ast \psi (4160)$)
may thus produce the largest amount of $\psi (4160)$ mesons at
$0.95T_{\rm c}$ ($0.75T_{\rm c}$, $0.9T_{\rm c}$,
$0.85T_{\rm c}$, $0.85T_{\rm c}$) during evolution of hadronic matter.

Temperature dependence of the interquark potential has been obtained
in the lattice gauge calculations. It is shown from the potential that the
interaction range drops sharply around the critical temperature \cite{Satz}.
Consequently,
the $c\bar c$ relative-motion wave functions obtained from the Schr\"odinger
equation with the potential exhibit that the spatial size of each wave 
function increases rapidly when the temperature increases
from a certain value. This value is near the critical
temperature and is different for different quantum numbers of
$c\bar c$ states. Thus, we take this value as the dissociation temperature of 
the $c\bar c$ state. This method is valid not only for bound states
such as $J/\psi$, $\chi_c$, and $\psi^\prime$ but also for resonances such as
$\psi(4040)$, $\psi(4160)$, and $\psi(4415)$.

Number densities obtained from the master rate equations depend on the
average cross sections weighted by the relative velocity and the
dissociation temperatures of $\psi(4040)$, $\psi(4160)$, and $\psi(4415)$
mesons. The averages 
$\langle \sigma_{ij\to i^\prime \psi(4040)} v_{ij}\rangle$ and
$\langle \sigma_{ij\to i^\prime \psi(4415)} v_{ij}\rangle$ are typically 5
and 3.3 times 
$\langle \sigma_{ij\to i^\prime \psi(4160)} v_{ij}\rangle$, respectively.
In addition, the $\psi(4040)$ dissociation temperature is higher
than the $\psi(4160)$ dissociation temperature. Hadronic matter takes a long
time to produce $\psi(4040)$ than to produce $\psi(4160)$.
We thus see that the 
$\psi(4040)$ number density is larger than the $\psi(4160)$ number density
in Figs. 28 and 29. However,
because the $\psi(4415)$ dissociation temperature is lower than the 
$\psi(4160)$ dissociation temperature, hadronic matter takes a short
time to produce $\psi(4415)$ than to produce $\psi(4160)$. This factor causes
the $\psi(4415)$ number density to be
smaller than the $\psi(4160)$ number density. The number
densities and the volume of hadronic matter at kinetic freeze-out
give 0.0034, 0.0006, and 0.00011
as the numbers of $\psi(4040)$, $\psi(4160)$, and $\psi(4415)$ mesons produced
in a central Pb-Pb collision at $\sqrt{s_{NN}}=5.02$ TeV, respectively.

The first twenty-four terms on the right-hand side of Eq. (2) are gain terms
of producing $\psi(4040)$, $\psi(4160)$, and $\psi(4415)$ mesons, and the other
terms are loss terms of breaking the three charmonia. The cross sections for
$q_1\bar{q}_2 + c\bar{c} \to c\bar{q}_2 + q_1\bar{c}$ in the loss terms are
obtained from those for $c\bar{q}_2 + q_1\bar{c} \to q_1\bar{q}_2 + c\bar{c}$
using detailed balance. The inclusion of the loss terms in the master rate 
equations reduces the number densities of the charmonia, but
the difference between the number densities obtained
from Eq. (1) with the loss terms and without the loss terms is small. For
example, the $\psi(4040)$ ($\psi(4160)$, $\psi(4415)$) number density
at $r=0$ fm at kinetic freeze-out is $6.75 \times 10^{-8}$ 
${\rm fm}^{-3}$ ($1.22 \times 10^{-8}$ ${\rm fm}^{-3}$, 
$2.14 \times 10^{-9}$ ${\rm fm}^{-3}$) with the
loss terms and $7.35 \times 10^{-8}$ ${\rm fm}^{-3}$
($1.25 \times 10^{-8}$ ${\rm fm}^{-3}$, $2.21 \times 10^{-9}$ ${\rm fm}^{-3}$)
without the loss terms. The reason for the small difference is that the number 
densities of the charmonia are small. The maximum number densities appear
at $r=0$ fm. It is shown by the upper solid, upper dashed, and upper dotted
curves in Fig. 28 that the maximum number densities of $\psi(4040)$, 
$\psi(4160)$, and $\psi(4415)$ are $1.95 \times 10^{-7}$ $\rm{fm}^{-3}$,
$3.05 \times 10^{-8}$ $\rm{fm}^{-3}$, and $3.49 \times 10^{-9}$ $\rm{fm}^{-3}$,
respectively. The number densities of pions, kaons, $\eta$ mesons, $\rho$ 
mesons, and vector kaons are obtained from the 
Bose-Einstein distribution. The product of the number densities of the 
charmonia and the light-quark mesons gives small loss terms that cause the 
small difference.

\vspace{0.5cm}
\leftline{\bf V. SUMMARY }
\vspace{0.5cm}

We have studied the production of $\psi(4040)$, $\psi(4160)$, and $\psi(4415)$
mesons in ultrarelativistic heavy-ion collisions at the LHC. This research
includes two parts. In one part we have studied the charmonium production
from the reactions between charmed strange mesons and open-charm mesons.
These reactions arise from quark interchange in association with color
interactions between all constituent pairs in different mesons.
Fifty-one reactions are considered, and we have presented numerical
unpolarized cross sections and their parametrizations for the twenty-seven
reactions: $D_s^+ \bar{D} \to K^\ast R$, $D_s^+ \bar{D}^\ast \to K R$,
$D_s^+ \bar{D}^\ast \to K^\ast R$, $D_s^{\ast +} \bar{D} \to K R$,
$D_s^{\ast +} \bar{D} \to K^\ast R$, $D_s^{\ast +} \bar{D}^\ast \to K R$,
$D_s^{\ast +} \bar{D}^\ast \to K^\ast R$, $D_s^+ D_s^{\ast -} \to \eta R$,
and $D_s^{\ast +} D_s^{\ast -} \to \eta R$, where $R$ indicates 
$\psi(4040)$, $\psi(4160)$, or $\psi(4415)$. 
We have presented characteristics of the endothermic reactions below the 
critical
temperature and of the reactions which are endothermic at some temperatures
and exothermic at other temperatures. The characteristics are related to 
confinement, mesonic quark-antiquark relative-motion wave functions, and $\mid 
\vec{P}^{~\prime} \mid / \mid \vec{P} \mid$. In another part we have studied
the production of $\psi(4040)$, $\psi(4160)$, and $\psi(4415)$ in 
hadronic matter that results from the quark-gluon plasma created in Pb-Pb
collisions at the LHC. We have established the master rate equations with the
new source terms that include the reactions between charmed strange mesons
and open-charm mesons and their reverse reactions. 
The temperature dependence of the cross sections reflects
different contributions of the fifty-one reactions to the charmonium production
at different temperatures.
The master rate equations in association with the
hydrodynamic equation are solved to obtain number densities of
$\psi(4040)$, $\psi(4160)$, and $\psi(4415)$. In central Pb-Pb 
collisions at $\sqrt{s_{NN}}=5.02$ TeV, the $\psi(4040)$ number density is 
larger than the $\psi(4160)$ number density, and the latter is larger than the
$\psi(4415)$ number density. The reactions between charmed strange mesons and
open-charm mesons increase the number densities. The small number densities
lead the loss terms to be small.

\vspace{0.5cm}
\leftline{\bf ACKNOWLEDGEMENTS}
\vspace{0.5cm}

We thank Prof. H. J. Weber for careful readings of the manuscript.
This work was supported by the project STRONG-2020 of the European Center for
Theoretical Studies in Nuclear Physics and Related Areas.


\begin{thebibliography}{00}
\bibitem{Augustin} J.-E. Augustin {\it et al.}, Phys. Rev. Lett. 34, 764 
(1975).
\bibitem{Siegrist} J. Siegrist {\it et al.}, Phys. Rev. Lett. 36, 700 (1976).
\bibitem{Brandelik} R. Brandelik {\it et al.}, Phys. Lett. 76 B, 361 (1978).
\bibitem{Zhukova} V. Zhukova {\it et al.}, Phys. Rev. D 97, 012002 (2018).
\bibitem{Ablikim1} M. Ablikim {\it et al.}, Phys. Rev. D 101, 012008 (2020).
\bibitem{Aaij} R. Aaij {\it et al.}, Phys. Rev. D 102, 112003 (2020).
\bibitem{Ablikim2} M. Ablikim {\it et al.}, Phys. Rev. D 102, 112009 (2020).
\bibitem{Ablikim3} M. Ablikim {\it et al.}, Phys. Rev. D 104, 092001 (2021).
\bibitem{Ablikim4} M. Ablikim {\it et al.}, Phys. Rev. D 104, 112009 (2021).
\bibitem{Ablikim5} M. Ablikim {\it et al.}, J. High Energy Phys. 07, 064 
(2022).
\bibitem{HLC} L.-K. Hao, K.-Y. Liu, and K.-T. Chao, Phys. Lett. B 546, 216 
(2002).
\bibitem{PGK} M. Piotrowska, F. Giacosa, and P. Kovacs, Eur. Phys. J. C 79, 98
(2019).
\bibitem{BIO} M. Bayar, N. Ikeno, and E. Oset, Eur. Phys. J. C 80, 222 (2020).
\bibitem{GI} S. Godfrey and N. Isgur, Phys. Rev. D 32, 189 (1985).
\bibitem{BGS} T. Barnes, S. Godfrey, and E. S. Swanson, Phys. Rev. D 72, 054026
(2005).
\bibitem{VFV} J. Vijande, F. Fern$\acute {\rm a}$ndez, and A. Valcarce, J.
Phys. G 31, 481 (2005).
\bibitem{OSEF} P. G. Ortega, J. Segovia, D. R. Entem, and F. 
Fern$\acute {\rm a}$ndez, Phys. Lett. B 778, 1 (2018).
\bibitem{wxLXW} W.-X. Li, X.-M. Xu, and H. J. Weber, Eur. Phys. J. C 81, 225 
(2021).
\bibitem{lyLXW} L.-Y. Li, X.-M. Xu, and H. J. Weber, Phys. Rev. D 105, 114025
(2022).
\bibitem{GMKR} H. von Gersdorff, L. McLerran, M. Kataja, and P. V. Ruuskanen,
Phys. Rev. D 34, 794 (1986).
\bibitem{KH} P. F. Kolb and U. Heinz, arXiv:nucl-th/0305084.
\bibitem{NEP} H. Niemi, K. J. Eskola, and R. Paatelainen, Phys. Rev. C 93,
024907 (2016).
\bibitem{ALICE5020CH} S. Acharya {\it et al.}, Phys. Rev. C 101, 044907 (2020).
\bibitem{XW} X.-M. Xu and H. J. Weber, Mod. Phys. Lett. A 35, 2030016 (2020).
\bibitem{CMSJS} E. Colton, E. Malamud, P. E. Schlein, A. D. Johnson, V. J. 
Stenger, and P. G. Wohlmut, Phys. Rev. D 3, 2028 (1971).
\bibitem{CFSW} D. Cohen, T. Ferbel, P. Slattery, and B. Werner, Phys. Rev. D 7,
661 (1973).
\bibitem{LCFMP} M. J. Losty, V. Chaloupka, A. Ferrando, L. Montanet, E. Paul, 
D. Yaffe, A. Zieminski, J. Alitti, B. Gandois, and J. Louie,  Nucl. Phys. B 69,
185 (1974).
\bibitem{PABFF} S. D. Protopopescu, M. Alston-Garnjost, A. Barbaro-Galtieri,
S. M. Flatt$\acute{\rm e}$, J. H. Friedman, T. A. Lasinski, G. R. Lynch, 
M. S. Rabin, and F. T. Solmitz, Phys. Rev. D 7, 1279 (1973).
\bibitem{Hyams} B. Hyams {\it et al.}, Nucl. Phys. B 64, 134 (1973).
\bibitem{EM} P. Estabrooks and A. D. Martin, Nucl. Phys. B 79, 301 (1974).
\bibitem{Srinivasan} V. Srinivasan {\it et al.}, Phys. Rev. D 12, 681 (1975).
\bibitem{FP} C. D. Froggatt and J. L. Petersen, Nucl. Phys. B 129, 89 (1977).
\bibitem{BBMPS} A. A. Bel'kov, S. A. Bunyatov, K. N. Mukhin, O. O. Patarakin, 
V. M. Sidorov, M. M. Sulkovskaya, A. F. Sustavov, and V. A. Yarba, JETP Lett. 
29, 597 (1979).
\bibitem{GKPEY} R. Garc$\acute {\rm i}$a-Mart$\acute {\rm i}$n,
R. Kami$\acute {\rm n}$ski, J. R. Pel$\acute {\rm a}$ez, J. R. de Elvira, and
F. J. Yndur$\acute {\rm a}$in, Phys. Rev. D 83, 074004 (2011).
\bibitem{Rosselet} L. Rosselet {\it et al.}, Phys. Rev. D 15, 574 (1977).
\bibitem{AKMMP} E. A. Alekseeva, A. A. Kartamyshev, V. K. Makar'in, 
K. N. Mukhin, O. O. Patarakin, M. M. Sulkovskaya, A. F. Sustavov, 
L. V. Surkova, and L. A. Chernysheva, Sov. Phys. JETP 55, 591 (1982).
\bibitem{JMTVH} B. Jongejans, R. A. van Meurs, A. G. Tenner, H. Voorthuis,
P. M. Heinen, W. J. Metzger, H. G. J. M. Tiecke, and R. T. van de Walle, Nucl.
Phys. B 67, 381 (1973).
\bibitem{Linglin} D. Linglin {\it et al.}, Nucl. Phys. B 57, 64 (1973).
\bibitem{Mercer} R. Mercer {\it et al.}, Nucl. Phys. B 32, 381 (1971).
\bibitem{Estabrooks} P. Estabrooks, P. K. Carnegie, A. D. Martin, 
W. M. Dunwoodie, T. A. Lasinski, and D. W. G. S. Leith, Nucl. Phys. B 133, 490 
(1978).
\bibitem{Aston} D. Aston {\it et al.}, Nucl. Phys. B 296, 493 (1988).
\bibitem{OO} J. A. Oller and E. Oset, Phys. Rev. D 60, 074023 (1999).
\bibitem{Wong1} C.-Y. Wong, arXiv:1012.0015.
\bibitem{Wong2} C.-Y. Wong, EPJ Web Conf. 71, 00140 (2014).
\bibitem{DHMNS} G. Denicol, U. Heinz, M. Martinez, J. Noronha, and M. 
Strickland, Phys. Rev. D 90, 125026 (2014).
\bibitem{BS} T. Barnes and E. S. Swanson, Phys. Rev. D 46, 131 (1992).
\bibitem{Swanson} E. S. Swanson, Ann. Phys. (N.Y.) 220, 73 (1992).
\bibitem{MM} N. F. Mott and H. S. W. Massey, {\it The theory of Atomic 
Collisions} (Clarendon, Oxford, 1965).
\bibitem{BBS} T. Barnes, N. Black, and E. S. Swanson, Phys. Rev. C 63, 025204
(2001).
\bibitem{WC} C.-Y. Wong and H. W. Crater, Phys. Rev. C 63, 044907 (2001).
\bibitem{BT} W. Buchm\"{u}ller and S.-H. H. Tye, Phys. Rev. D 24, 132 (1981).
\bibitem{KLP} F. Karsch, E. Laermann, and A. Peikert, Nucl. Phys. B 605, 579
(2001).
\bibitem{Xu2002} X.-M. Xu, Nucl. Phys. A 697, 825 (2002).
\bibitem{PDG} M. Tanabashi {\it et al.} (Particle Data Group), Phys. Rev. D 98,
030001 (2018) and 2019 update.
\bibitem{DBGGT} N. B. Durusoy, M. Baubillier, R. George, M. Goldberg, 
A. M. Touchard, N. Armenise, M. T. Fogli-Muciaccia, and A. Silvestri, Phys. 
Lett. B 45, 517 (1973).
\bibitem{Hoogland} W. Hoogland {\it et al.}, Nucl. Phys. B 126, 109 (1977).
\bibitem{ZX} J. Zhou and X.-M. Xu, Phys. Rev. C 85, 064904 (2012).
\bibitem{SXW} Z.-Y. Shen, X.-M. Xu, and H. J. Weber, Phys. Rev. D 94, 034030
(2016).
\bibitem{ZXG} Y.-P. Zhang, X.-M. Xu, and H.-J. Ge, Nucl. Phys. A 832, 112 
(2010).
\bibitem{XMCW} X.-M. Xu, C.-C. Ma, A.-Q. Chen, and H. J. Weber, Phys. Lett.
B 645, 146 (2007).
\bibitem{ALICE5020D} S. Acharya {\it et al.}, J. High Energy Phys. 10, 174
(2018).
\bibitem{CF} F. Cooper and G. Frye, Phys. Rev. D 10, 186 (1974).
\bibitem{ALICE5020KA} S. Acharya {\it et al.}, Phys. Rev. C 106, 034907 (2022).
\bibitem{Satz} H. Satz, arXiv:hep-ph/0602245.
\end{thebibliography}
\end{document}